\documentclass[preprint]{aa} 
\usepackage{graphicx}
\usepackage{txfonts}
\usepackage[T1,hyphens]{url}
\usepackage{hyperref}

\usepackage{tablefootnote}

\usepackage{natbib}
\usepackage[export]{adjustbox}

\bibpunct{(}{)}{;}{a}{}{,} 
\newcommand{\1}{{~\sc i}}
\newcommand{\2}{{~\sc ii}}
\newcommand{\3}{{~\sc iii}}

\newcommand{\kms}{{\,km\,s$^{-1}$}}
\newcommand{\cc}{{\,cm$^{-3}$}} 
\newcommand{\mic}{{\,$\mu$m}}

\usepackage[normalem]{ulem}
\usepackage{xcolor}
\definecolor{colred}{RGB}{255,66,56} 

\usepackage{tikz}
\usepackage[export]{adjustbox}

\usepackage{comment}

\begin{document}

 \title{Physical conditions in the gas phases of the giant H\2\ region LMC-N\,11: II.\ Origin of [C\2] and fraction of CO-dark gas}

           \titlerunning{Origin of [C\2] and fraction of CO-dark gas in LMC-N\,11}
        \authorrunning{Lebouteiller et al.}

   \author{V. Lebouteiller\inst{1,2}, D. Cormier\inst{1}, S.~C. Madden\inst{1}, M. Galametz\inst{1}, S. Hony\inst{3}, Fr\'ed\'eric Galliano\inst{1}, M. Chevance\inst{4}, M.-Y. Lee\inst{5,6}, J. Braine\inst{7}, F.~L. Polles\inst{8}, M. Angel Reque\~na-Torres\inst{9}, R. Indebetouw\inst{10,11}, A. Hughes\inst{12}, N. Abel\inst{13} }
   \institute{$^1$ AIM, CEA, CNRS, Universit\'e Paris-Saclay, Universit\'e Paris Diderot, Sorbonne Paris Cit\'e, F-91191 Gif-sur-Yvette, France \email{vianney.lebouteiller@cea.fr}\\
     $^2$ Department of Physics and Astronomy, University of North Carolina, 290 Phillips Hall CB 3255, Chapel Hill, NC 27599, USA \\
     $^3$ Institut f\"ur theoretische Astrophysik, Zentrum f\"ur Astronomie der Universit\"at Heidelberg, Albert-Ueberle Str.2, 69120 Heidelberg, Germany \\
     $^4$ Astronomisches Rechen-Institut, Zentrum f\"ur Astronomie der Universit\"at Heidelberg, M\"onchhofstra{\ss}e 12-14, 69120 Heidelberg, Germany \\
     $^5$ Korea Astronomy and Space Science Institute, 776 Daedeokdae-ro, 34055 Daejeon, Republic of Korea \\
     $^6$ Max-Planck-Institut f\"ur Radioastronomie, Auf dem H\"ugel 69, 53121 Bonn, Germany\\
     $^7$ Laboratoire d'Astrophysique de Bordeaux, Univ. Bordeaux, CNRS, B18N, all\'ee Geoffroy Saint-Hilaire, F-33615 Pessac, France \\
     $^8$ LERMA, Observatoire de Paris, PSL Research Univ., CNRS, Sorbonne Univ.,
75014 Paris, France \\
     $^9$ Department of Astronomy, University of Maryland, College Park, ND, 20742-2421, USA \\
     $^{10}$ Department of Astronomy, University of Virginia, Charlottesville, VA 22904, USA \\
     $^{11}$ National Radio Astronomy Observatory, 520 Edgemont Road, Charlottesville, VA 22903, USA \\
     $^{12}$ IRAP, Universit\'e de Toulouse, CNRS, UPS, CNES, 31400, Toulouse, France \\
     $^{13}$ University of Cincinnati, Clermont College, 4200 Clermont College Drive, Batavia, OH 45103, USA \\
   }
   
   \date{Received 12/07/2019; accepted 25/10/2019. }

 
  \abstract
  {The ambiguous origin of the [C\2] $158$\mic\ line in the interstellar medium complicates its use for diagnostics concerning the star-formation rate and physical conditions in photodissociation regions. }
   {We investigate the origin of [C\2] in order to measure the total molecular gas content, the fraction of CO-dark H$_2$ gas, and how these parameters are impacted by environmental effects such as stellar feedback. }
   {We observed the giant H\2\ region N\,11 in the Large Magellanic Cloud with SOFIA/GREAT. The [C\2] line is resolved in velocity and compared to H\1\ and CO, using a Bayesian approach to decompose the line profiles. A simple model accounting for collisions in the neutral atomic and molecular gas was used in order to derive the H$_2$ column density traced by C$^+$. }
   {The profile of [C\2] most closely resembles that of CO, but the integrated [C\2] line width lies between that of CO and that of H\1. Using various methods, we find that [C\2] mostly originates from the neutral gas. We show that [C\2] mostly traces the CO-dark H$_2$ gas but there is evidence of a weak contribution from neutral atomic gas preferentially in the faintest components (as opposed to components with low [C\2]/CO or low CO column density). Most of the molecular gas is CO-dark. The CO-dark H$_2$ gas, whose density is typically a few $100$s\,\cc\ and thermal pressure in the range $10^{3.5-5}$\,K\cc, is not always in pressure equilibrium with the neutral atomic gas. The fraction of CO-dark H$_2$ gas decreases with increasing CO column density, with a slope that seems to depend on the impinging radiation field from nearby massive stars. Finally we extend previous measurements of the photoelectric-effect heating efficiency, which we find is constant across regions probed with \textit{Herschel}, with [C\2]  and [O\1] being the main coolants in faint and diffuse, and bright and compact regions, respectively, and with polycyclic aromatic hydrocarbon emission tracing the CO-dark H$_2$ gas heating where [C\2] and [O\1] emit. } 
   {We present an innovative spectral decomposition method that allows  statistical trends to be derived for the molecular gas content using CO, [C\2], and H\1\ profiles. Our study highlights the importance of velocity-resolved photodissociation region (PDR) diagnostics and higher spatial resolution for H\1\ observations as future steps. }


   \maketitle

\section{Introduction}

The [C\2] $158$\mic\ line emits under a variety of conditions in the interstellar medium (ISM) corresponding to the cold and warm neutral medium and warm ionized gas, owing to the relatively low ionization potential to produce C$^+$ ions ($11.3$\,eV) and to the relatively low energy of the $^2{\rm P}_{3/2}$ fine-structure level ($91.3$\,K). In the neutral gas, C$^+$ may exist in the H$^0$ phase but also in the H$_2$ phase, in regions where CO is photodissociated while H$_2$ is self-shielded and shielded by dust, the so-called CO-dark molecular gas (e.g., \citealt{Madden1997a,Grenier2005a,Wolfire2010a}). The [C\2] line has been used to trace the star-formation rate (SFR; e.g., \citealt{DeLooze2014a,Pineda2014a}), to infer physical conditions in photodissociation regions (PDRs), and to calculate the CO-to-H$_2$ conversion factor ($X_{\rm CO}$; e.g., \citealt{Jameson2018a,HerreraCamus2017a,Pineda2017a}). The
growing number of observations in both Galactic and extragalactic environments (with routine detections at $z>6$; e.g., \citealt{Aravena2016a}) has renewed the interest in understanding [C\2] as a diagnostic tool.

It is crucial to provide astrophysical experiments that can isolate the phase contributions to the [C\2] emission toward well-chosen regions and to examine how the origin of [C\2] depends on environmental parameters such as the metallicity or the radiative and mechanical feedback from young stellar clusters. On the one hand, metallicity plays a role through the lower abundance of dust, resulting in less efficient shielding from UV photons and to a larger layer of CO-dark H$_2$ gas in PDRs (\citealt{Madden2018a}, Madden et al.\ in preparation). On the other hand, young massive stars shape the surrounding ISM, thereby modifying the relative filling factors of warm ionized gas and molecular clouds, resulting in a lower ``effective'' extinction on average when measured over the scale of a star-forming region. The age of the molecular cloud in which stars will form is another important parameter, notably because of the H$_2$ formation timescale (e.g., \citealt{Franeck2018a}). 

It is often assumed that most of the [C\2] emission arises from a given dominant ISM phase or alternatively that the relative contributions from the various ISM phases can be recovered from photoionization and photodissociation models (e.g., \citealt{Cormier2015a}). Another method to disentangle the origin of [C\2] is to compare its velocity profile to those of CO and H\1\ $21$\,cm, which are assumed to trace the ``CO-bright'' H$_2$ gas and the H$^0$ gas, respectively, with the remaining [C\2] emission attributed to CO-dark H$_2$ gas (the contribution from the ionized gas being usually determined indirectly for lack of reliable velocity-resolved ionized gas tracers). Using velocity profiles, \cite{Pineda2014a} studied [C\2] in the Milky Way with the \textit{Herschel} GOT C+ survey and found approximately equal contributions ($20-30$\%) from dense PDRs, CO-dark H$_2$ gas, cold atomic gas, and ionized gas. The warm neutral medium does not seem to contribute significantly to the observed [C\2] intensity (see also \citealt{Fahrion2017a}). \cite{Pineda2014a} also find that the extragalactic SFR relationship with gas surface density can be recovered only by combining [C\2] from all the ISM phases, suggesting that SFR determinations using PDR-specific tracers may be difficult to calibrate if the fraction of UV photons absorbed in PDRs varies significantly across objects or within star-forming regions. \cite{Langer2014a} calculated that the fraction of CO-dark H$_2$ gas, $f_{\rm dark}$, in the \textit{Herschel} GOT C+ survey is $\approx75$\%\ in the diffuse clouds and down to $\approx20$\%\ in dense clouds. Toward the Galactic star-forming region M\,17-SW, \cite{PerezBeaupuits2015a} found that about half of the [C\2] traces the CO-dark H$_2$ gas while $\approx36$\%\ originates from the ionized gas.

Metallicity effects and massive stellar feedback effects are conveniently probed through observations of nearby star-forming dwarf galaxies. \cite{Fahrion2017a} examined the ISM at $\approx200$\,pc scales in the nearby star-forming galaxy NGC\,4214 and found that about half of the [C\2] traces the atomic gas and that most ($\approx80$\%) of the H$_2$ gas is CO-dark. The authors also found that $f_{\rm dark}$ is higher in the low-metallicity diffuse region where a super stellar cluster is located, although the
relative influence of metallicity and feedback remains uncertain. Magellanic Clouds allow  smaller spatial scales to be attained by resolving star-forming regions. \cite{Okada2015a} observed N\,159 in the Large Magellanic Cloud (LMC; $\approx1/2$\,Z$_\odot$) and found that the fraction of [C\2] that cannot be associated with CO increases from $\approx20$\%\ around the CO clumps to $\approx50$\%\ in the interclump medium, and that the overall fraction of [C\2] associated with the ionized gas is $\leq15$\%\ (see also \citealt{Okada2019a} for other regions). In LMC-30\,Dor, $90\%$ of [C\2] originate in PDRs and $80-99\%$ of H$_2$ is not traced by CO \citep{Chevance2016a,Chevance2016b}. \cite{RequenaTorres2016a} observed several star-forming regions in the Small Magellanic Cloud (SMC; $\approx1/5$\,Z$_\odot$) and found that most of the H$_2$ gas is traced by [C\2] and not CO.  
Using $54$ lines of sight in the LMC and SMC, \cite{Pineda2017a} find that most of the molecular gas is CO-dark. Overall, the finding that CO-dark H$_2$ gas is predominant in low-metallicity environments is consistent with the picture of CO-emitting regions occupying a smaller filling factor due to the relatively lower dust-to-gas mass ratio, and it is in fact possible to obtain $X_{\rm CO}$ conversion factors close to the Milky Way value in the Magellanic Clouds if filling factor effects are accounted for, as shown by \cite{Pineda2017a}. However, the dependence of $f_{\rm dark}$ gas with metallicity may be indirect, and model results from the \textit{Herschel} Dwarf Galaxy Survey (DGS; \citealt{Madden2013a}) indicate that $f_{\rm dark}$ is mostly a function of the effective cloud extinction, with the latter showing some dependence with metallicity (Madden et al.\ in prep.).

Summarizing these results, [C\2] traces a significant fraction of the H$_2$ gas and the fraction of CO-dark H$_2$ gas increases from dense CO peaks to the diffuse medium (see also \citealt{Mookerjea2016a} in M\,33), and also increases with lower metallicity, and stronger stellar feedback (radiation and/or dynamical and mechanical). The fraction of [C\2] in the ionized gas is usually $\lesssim20$\% as also found in the global analysis of the DGS \citep{Cormier2015a} and in resolved regions such as LMC-30\,Dor \citep{Chevance2016a} and IC\,10 \citep{Polles2019a}. From a kinematics point of view, the [C\2] line is always found to be broader than CO and [C\1] but narrower than H\1, with a profile agreeing better with CO (see also \citealt{Braine2012a,deBlok2016a}). 

\begin{figure*}
  \includegraphics[width=19cm,trim=10 10 0 10]{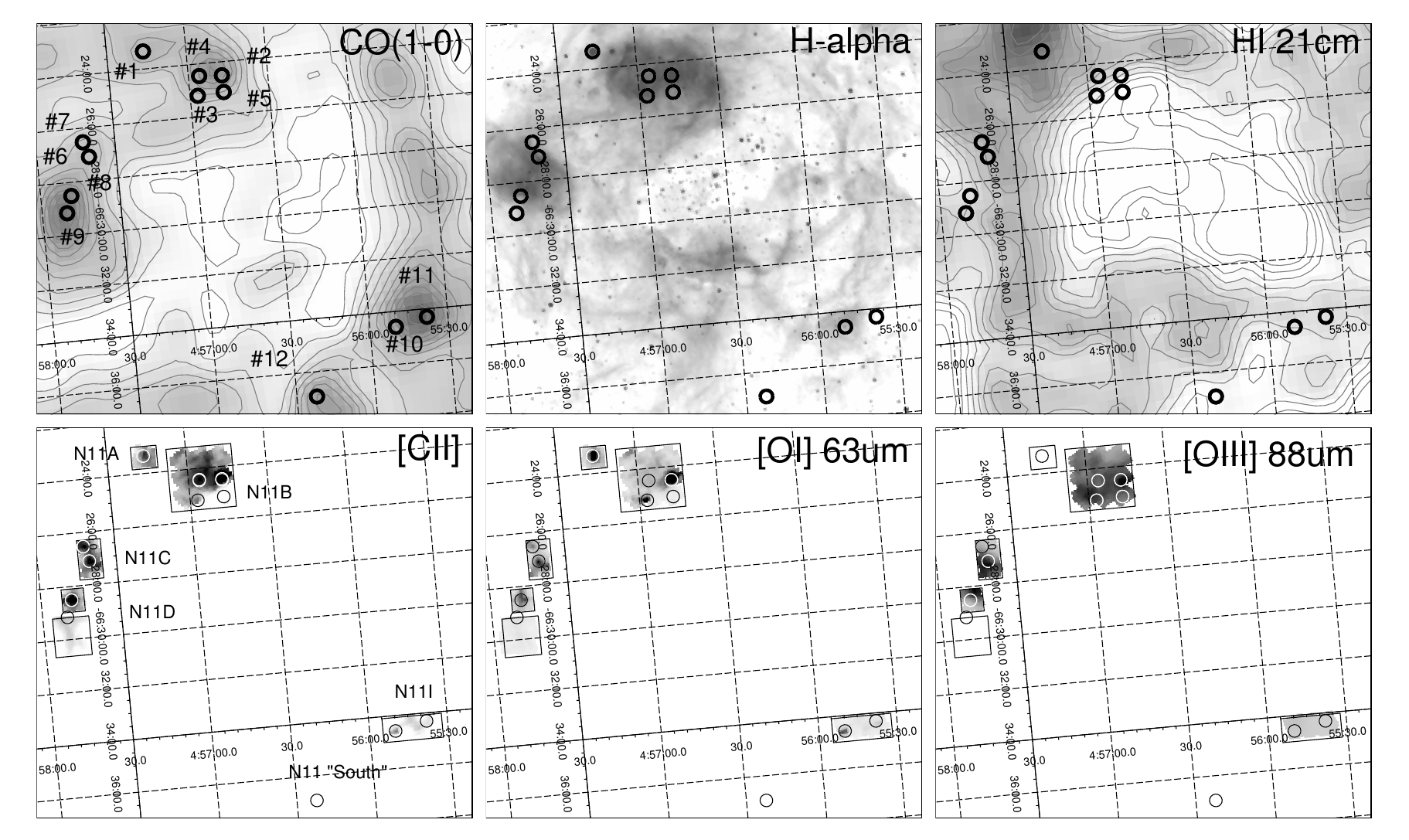} 
  \caption{LMC-N\,11 in CO, H$\alpha$, H\1, [C\2], [O\1], and [O\3]. The SOFIA/GREAT pointings are shown as black/white circles and labelled in the CO panel. The black rectangles show the \textit{Herschel}/PACS coverage. Region names are labelled in the [C\2] panel. GREAT pointings \#1 (N\,11A) and \#9 (part of N\,11D) were not observed in [O\3], and pointing \#12 (N\,11 ``South'') was not observed with PACS. }\label{fig:hicoha} 
\end{figure*}

In a first publication \citep{Lebouteiller2012b}, we investigated N\,11B, part of the second largest ($\approx150$\,pc in diameter) giant H\2\ region N\,11 in the LMC after 30\,Dor. We found a remarkable correlation between the total cooling rate traced by [C\2]+[O\1] and the polycyclic aromatic hydrocarbon (PAH) mid-infrared emission, suggesting that [C\2] emission is predominantly originating from PDRs with a uniform photoelectric-effect heating efficiency. In the present study, we examine the velocity structure of [C\2] in N\,11B and other regions within N\,11 obtained with the GREAT instrument \citep{Heyminck2012a} onboard the SOFIA telescope \citep{Young2012a}. LMC-N\,11 has been studied in detail especially at infrared and submillimeter wavelengths (e.g., \citealt{Israel2011a,Herrera2013a,Galametz2016a}) and has been fully mapped in CO(1-0) \citep{Israel2003a,Wong2011a} and H\1\ $21$\,cm \citep{Kim2003a} (Fig.\,\ref{fig:hicoha}). The N\,11 region was chosen in order to access various environments (e.g., PDRs, quiescent CO clouds, H\2\ regions, ultracompact H\2\ regions; see e.g., \citealt{Lebouteiller2012b,Galametz2016a}) which were expected to result in distinctive [C\2] velocity profiles.

The objectives are (i) to measure the quantity of CO-dark H$_2$ gas traced by [C\2] and the fraction of molecular gas that is CO-dark, (ii) to identify potential [C\2] components associated with atomic gas, and (iii) to probe the influence of the environment (in particular stellar feedback). A specific focus is given in the present study on the velocity profile decomposition method. The comparison between the [C\2], CO, and H\1\ spectral profiles is complex and methods usually determine average properties of the spectral profile along various lines of sight or else use profile fitting starting with the tracers with relatively simple velocity structure and adding velocity components as needed for the more complex ones. Here we use a statistical approach with as few assumptions as possible on the number of components and their properties. 
The observations are described in Section\,\ref{sec:observations}. We derive velocity-integrated properties in Section\,\ref{sec:integrated}. The profile decomposition and the associated results concerning the physical properties of the components are presented in Section\,\ref{sec:indiv}.

\section{Observations}\label{sec:observations}

The observations used in this study are summarized in Table\,\ref{tab:sumobs}. Below we describe the details of each set of observations.

\begin{table}
\caption{Summary of observations used in this study}\label{tab:sumobs}
\begin{tabular}{llll}
\hline
\hline
Instrument & Tracer & LSF$^{\rm a}$ FWHM   & PSF$^{\rm b}$ FWHM               \\
           &        & [\kms]     & [\arcsec]              \\
\hline
SOFIA/GREAT & [C\2], [N\2]      & $1.2$      & $\approx14.4$, $\approx19.1$     \\
  \textit{Herschel}/PACS & [C\2], [N\2] & $250$ & $12$ \\
MAGMA       & CO(1-0)           & $0.53$     & $45$       \\
ALMA        & CO(1-0)           &  $0.1$          & $2$           \\
ATCA+Parkes & H\1\ 21\,cm       & $1.6$      & $60$       \\
VLT/GIRAFFE & H$\alpha$, [Ne\3] & $17$, $15$ & $\approx1$ \\
\hline
\end{tabular}\\
\tablefoottext{a}{Line spread function full width at half maximum (spectral resolution). }
\tablefoottext{b}{Point spread function full width at half maximum (spatial resolution). }
\end{table}

\subsection{SOFIA/GREAT}\label{sec:obs_great}

Twelve pointings were observed within LMC-N\,11 (Table\,\ref{tab:obs_great}; Fig.\,\ref{fig:hicoha}) with SOFIA/GREAT in [C\2] $158$\mic\ and [N\2] $205$\mic\ as part of program $01\_0030$ (PI Lebouteiller). Observations were conducted on 2013, July 19 and 28 deploying from Christchurch, New Zealand.

\begin{table*}
  \caption{SOFIA/GREAT pointing observations }\label{tab:obs_great}
  \begin{tabular}{llllll}
    \hline
    \hline
    Pointing & RA  & DEC & On-source exposure time & $T_{\rm mb}$ rms & Description \\
     &  (J2000) & (J2000) & (min) & (mK) &  \\
    \hline
    \#1 & 4:57:16.2 & -66:23:20.2 & 2.5 & 139  & N\,11\,A $-$ compact H\2\ region $-$ [CII]+CO peak \\
    \#2 & 4:56:47.3 & -66:24:30.8 & 2.5 & 143 & N\,11\,B $-$ PDR $-$ [C\2] bright \\
    \#3 & 4:56:57.1 & -66:25:12.0 & 5.0 & 101 & N\,11\,B $-$ ultracompact H\2\ region $-$ [C\2] bright \\ 
    \#4 & 4:56:55.9 & -66:24:27.7 & 3.1  & 105  & N\,11\,B $-$ center $-$ CO peak \\
    \#5 & 4:56:47.2 & -66:25:09.0 & 14.0  & 60  & N\,11\,B $-$ ionized gas \\ 
    \#6 & 4:57:40.3 & -66:27:06.5 & 3.8 & 92 & N\,11\,C $-$ [C\2] peak \\ 
    \#7 & 4:57:42.1 & -66:26:32.2 & 5.0 & 96  & N\,11\,C $-$ CO peak   \\ 
    \#8 & 4:57:48.4 & -66:28:30.9 & 2.5 & 152 & N\,11\,C $-$ southern [C\2] peak \\ 
    \#9 & 4:57:50.6 & -66:29:09.0 & 7.5 & 80  & N\,11\,D $-$ CO peak \\ 
    \#10 & 4:55:50.1 & -66:34:35.0 & 5.0 & 90 & N\,11\,I $-$ [C\2] peak \\ 
    \#11 & 4:55:38.0 & -66:34:18.4 & 2.5 & 126 & N\,11\,I $-$ CO peak \\  
    \#12 & 4:56:22.4 & -66:36:56.2 & 2.5 & 136 & N\,11 south $-$ CO peak \\ 
    \hline
  \end{tabular}
\end{table*}

Most pointings were previously identified as CO-bright peaks in the
Magellanic MOPRA Assessment (MAGMA) survey \citep{Wong2011a} or as
[C\2] peaks in the \textit{Herschel}/PACS maps
(Fig.\,\ref{fig:hicoha}). A few additional pointings were included,
notably toward the young stellar cluster LH\,10 in N\,11B where the gas is mostly ionized, resulting in bright [O\3]
$88$\mic\ and $24$\mic\ emission
\citep{Lebouteiller2012b}. Overall, the pointings span various
environments including PDRs (e.g., \#2), a quiescent CO cloud (\#12), an ultracompact H\2\
region (\#3), stellar clusters (\#5), among others (Table\,\ref{tab:obs_great}).

The [C\2] line was observed in the GREAT L\#2 channel while [N\2] was observed in the L\#1 channel. We used X(A)FFT spectrometer back-ends. The [C\2] line was detected toward all pointings. The [N\2] line was observed only toward pointings \#1 to \#6 but was not detected. Some scans were affected by emission in the offset field\footnote{Contamination in the offset was checked during the observations (Sect.\,\ref{sec:obs_great}) but only strong contaminations could be identified. } for the observation of N\,11\,I (\#11). The chopper amplitude was changed during the observations and the contaminated scans were not used for the final release.

The half power beam width (HPBW) is around $\approx14-15\arcsec$ for the [C\2] observation and $\approx19-20\arcsec$ for [N\2]. The exact value of the GREAT HPBW during the observation is unfortunately unknown. We use $\theta=14.4\arcsec$ as a tentative HPBW, following the SOFIA/GREAT Observation Planning (Version 9, April 29, 2016\footnote{\url{http://www3.mpifr-bonn.mpg.de/div/submmtech/heterodyne/great/GREAT_calibration.html}}).

\begin{figure*}
\includegraphics[width=4.5cm,height=7.8cm,trim=10 30 0 10,clip]{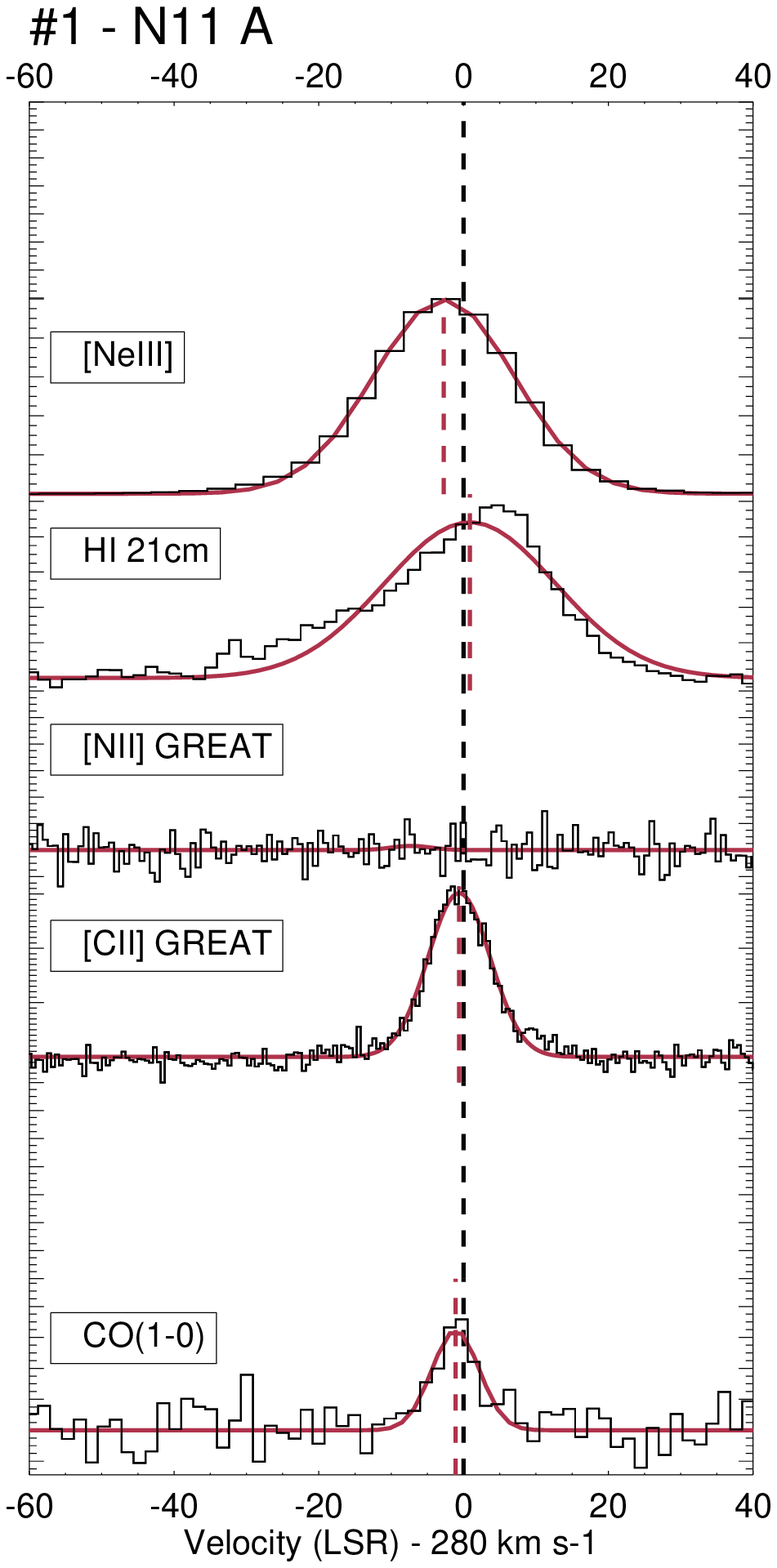}
\includegraphics[width=4.5cm,height=7.8cm,trim=10 30 0 10,clip]{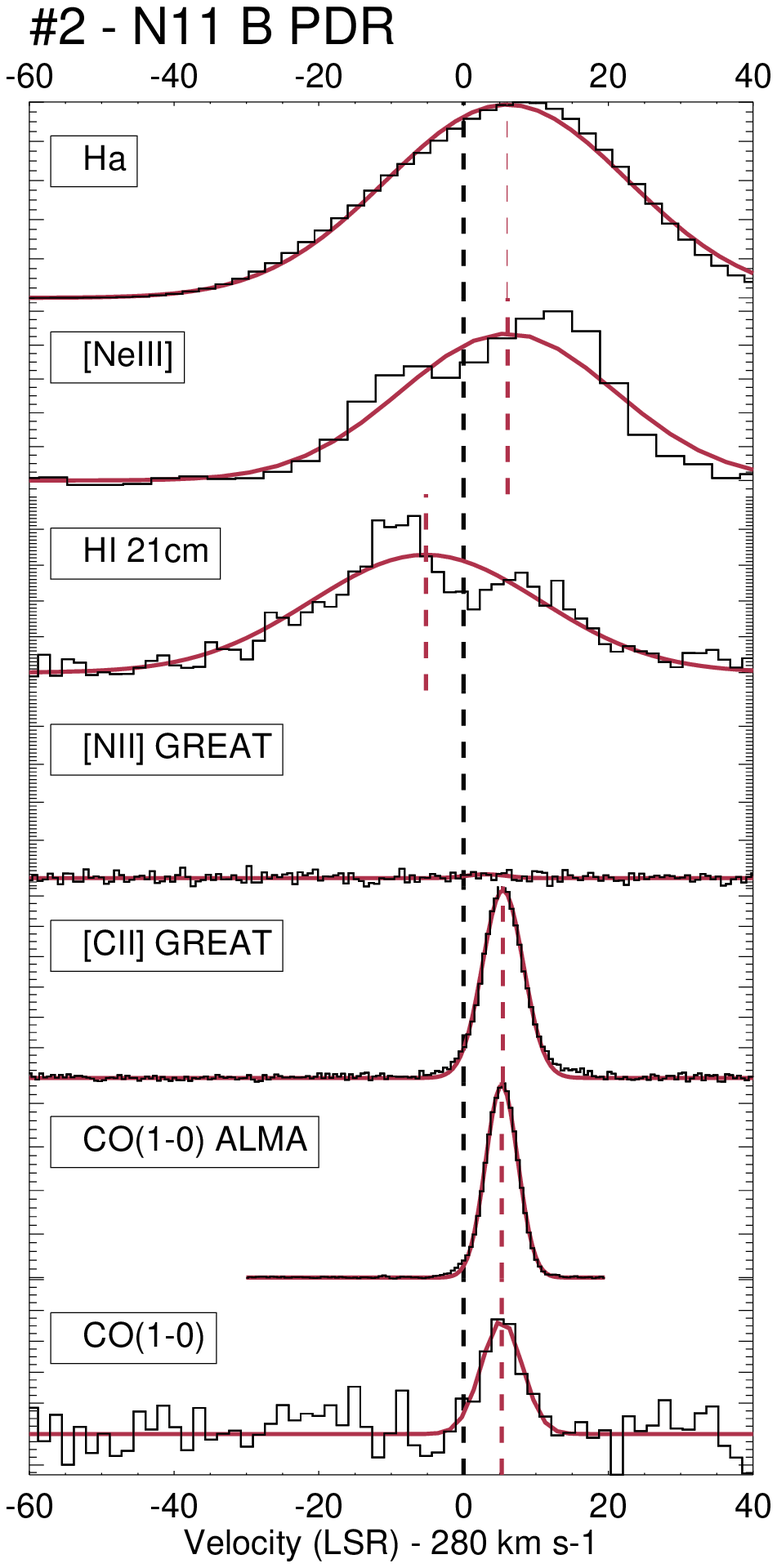}
\includegraphics[width=4.5cm,height=7.8cm,trim=10 30 0 10,clip]{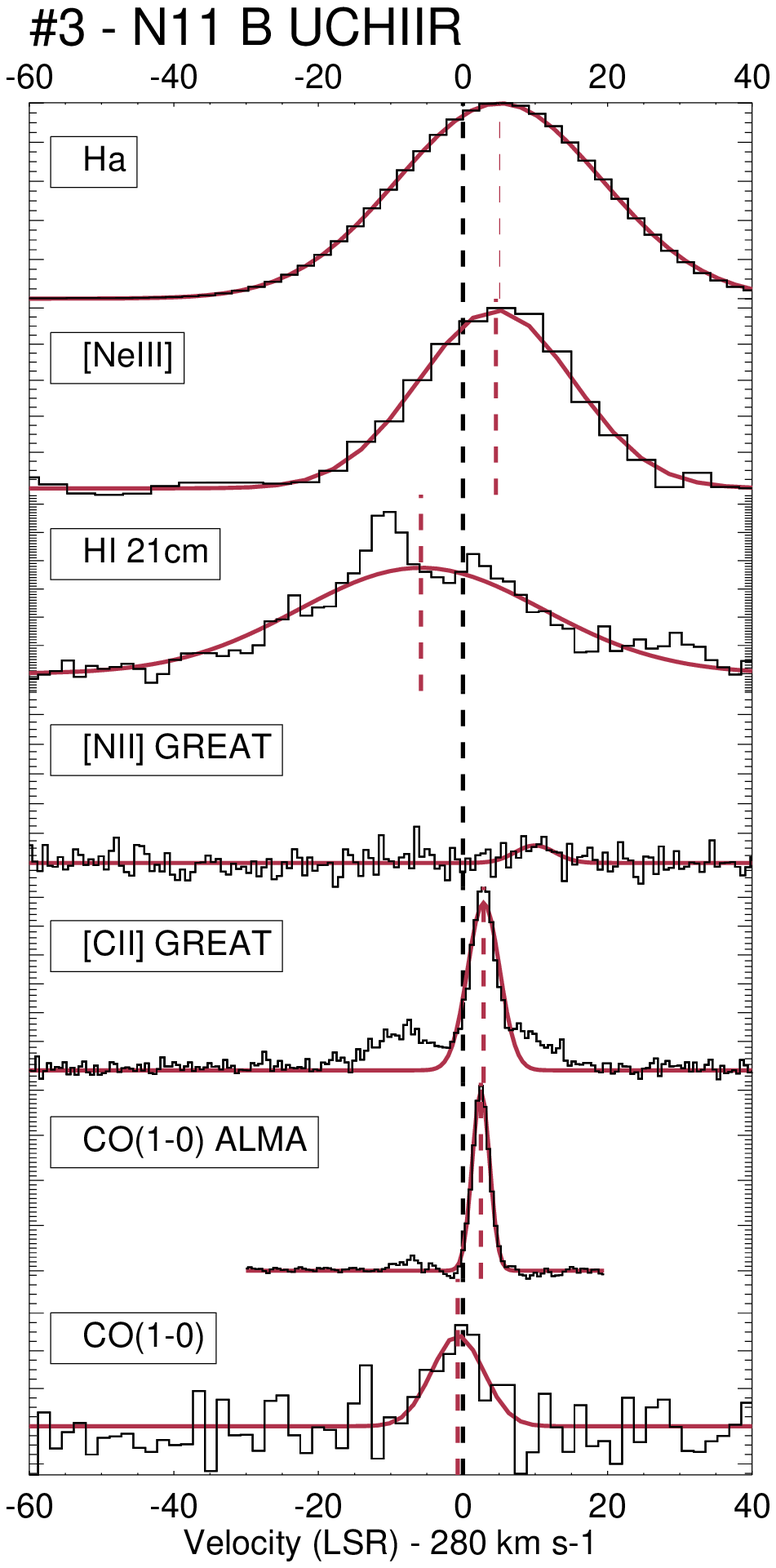}
\includegraphics[width=4.5cm,height=7.8cm,trim=10 30 0 10,clip]{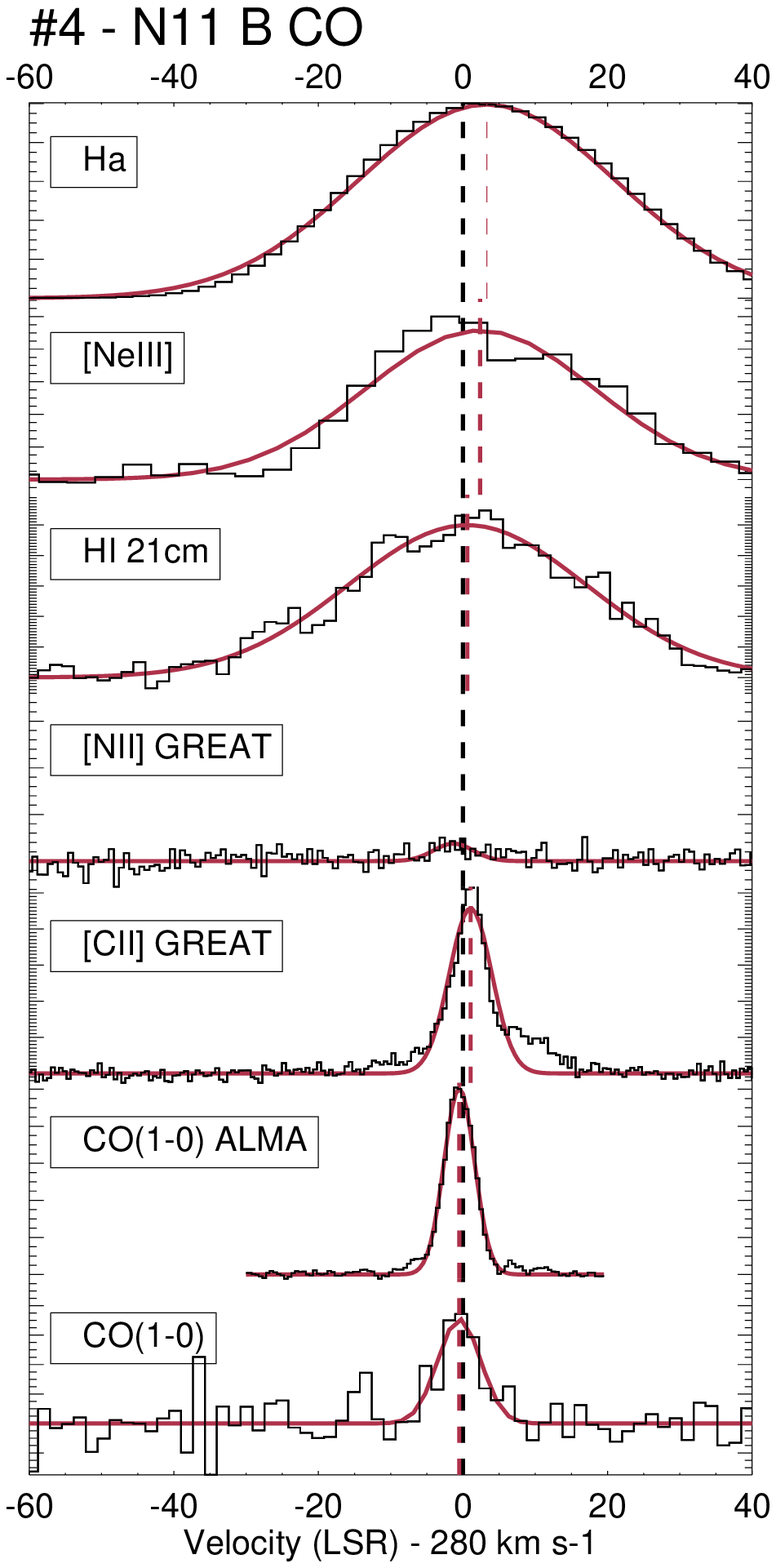}
\includegraphics[width=4.5cm,height=7.8cm,trim=10 30 0 10,clip]{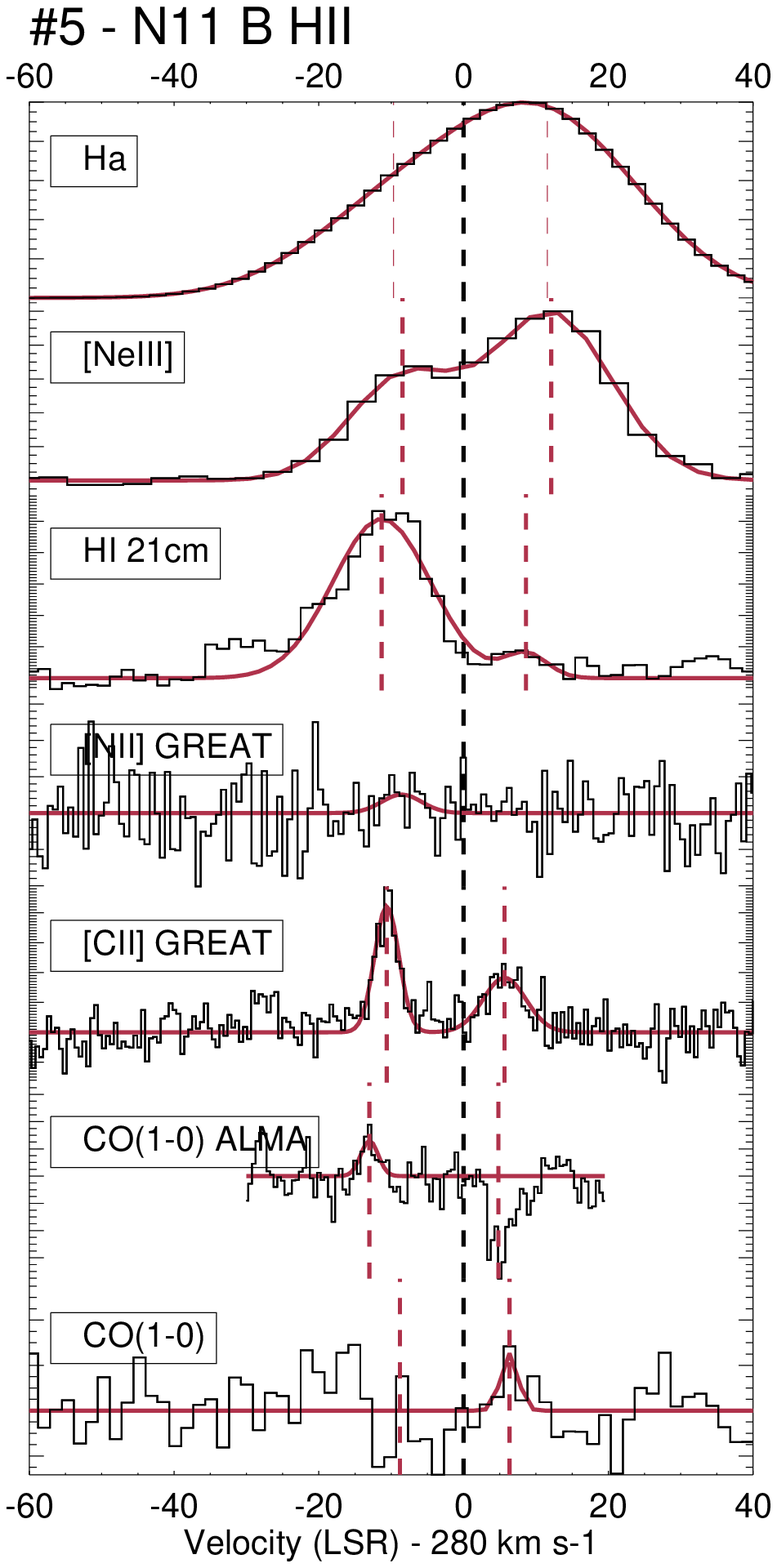}
\includegraphics[width=4.5cm,height=7.8cm,trim=10 30 0 10,clip]{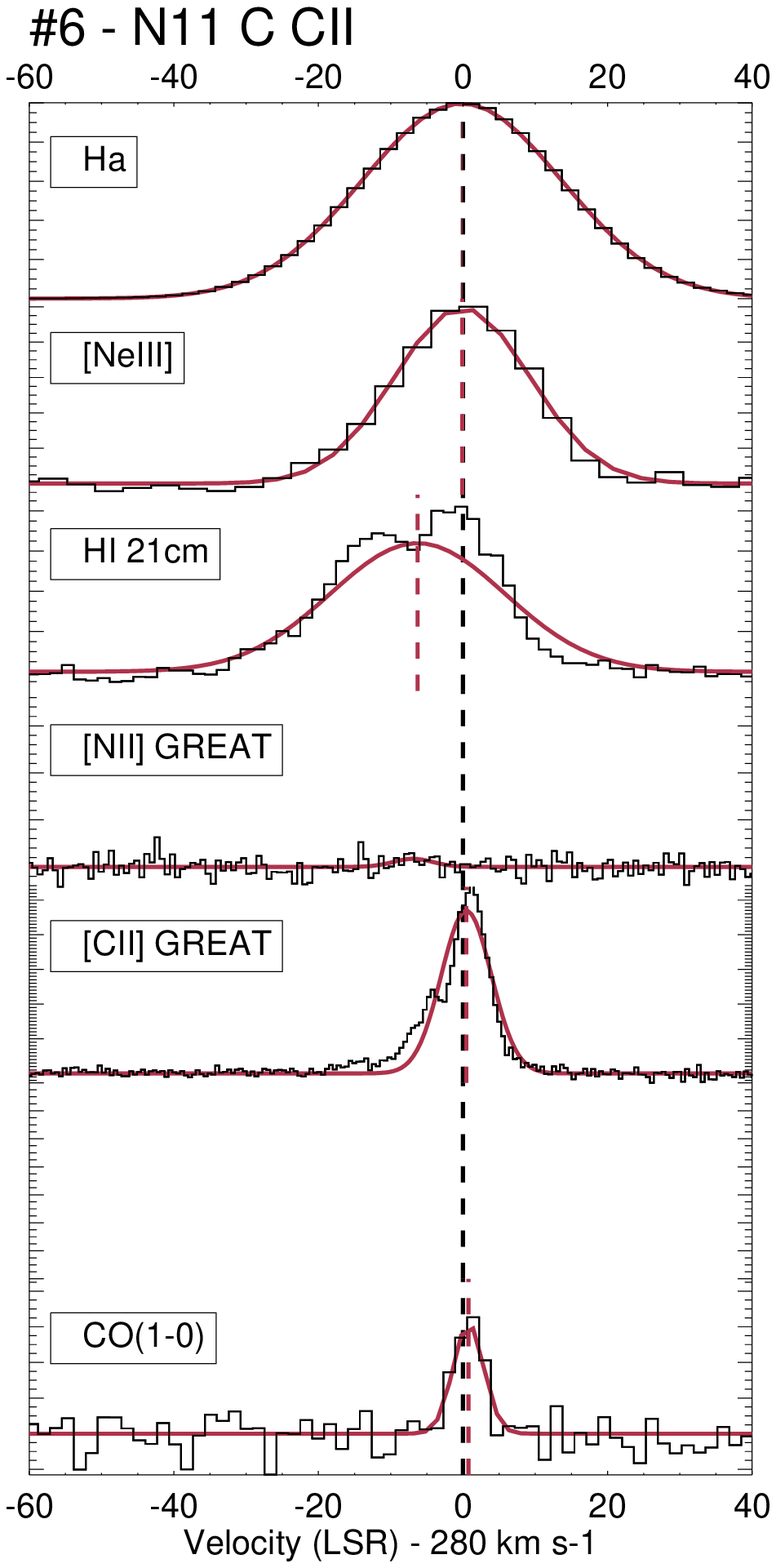}
\includegraphics[width=4.5cm,height=7.8cm,trim=10 30 0 10,clip]{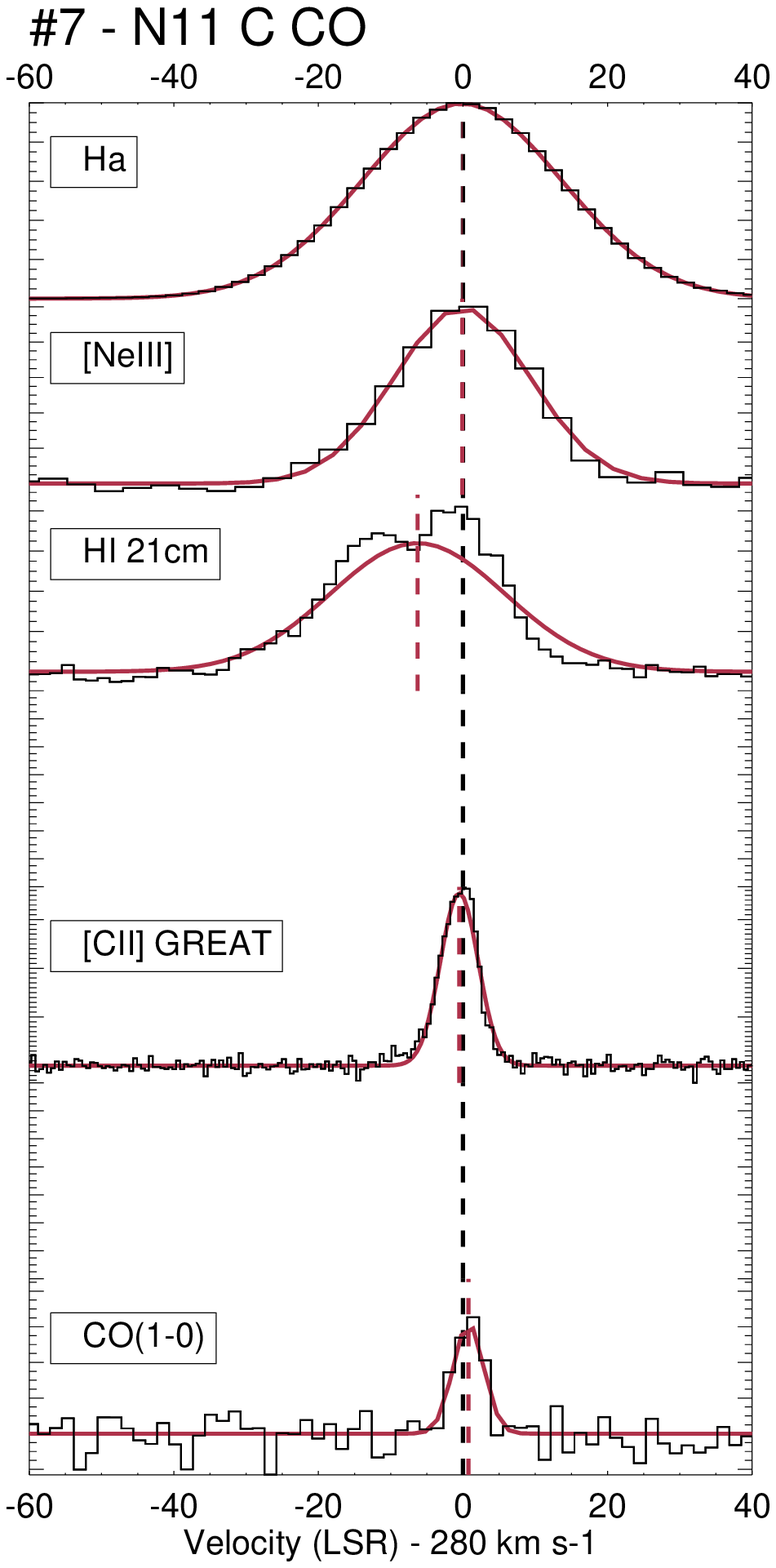}
\includegraphics[width=4.5cm,height=7.8cm,trim=10 30 0 10,clip]{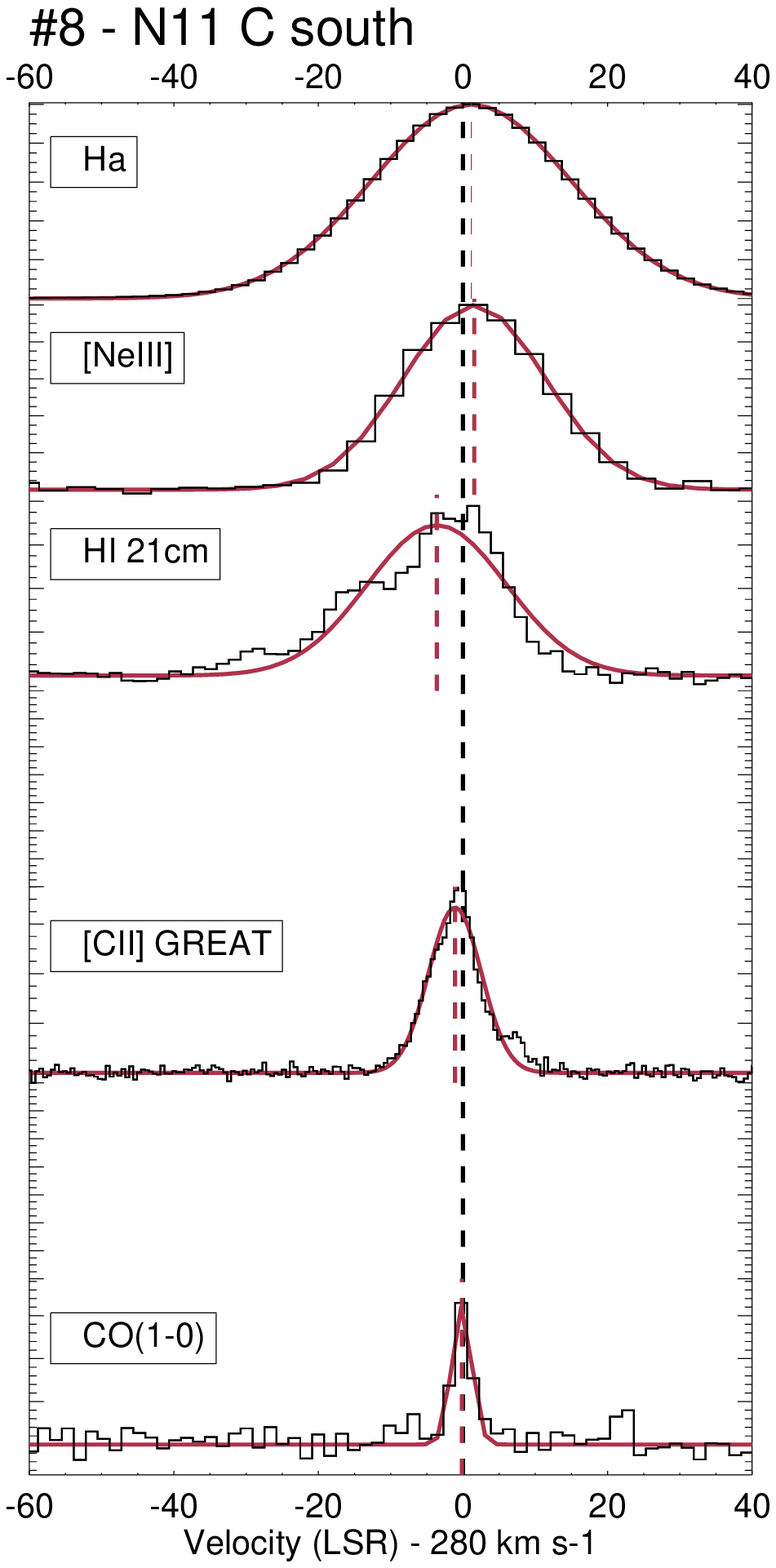}
\includegraphics[width=4.5cm,height=7.8cm,trim=10 0 0 10,clip]{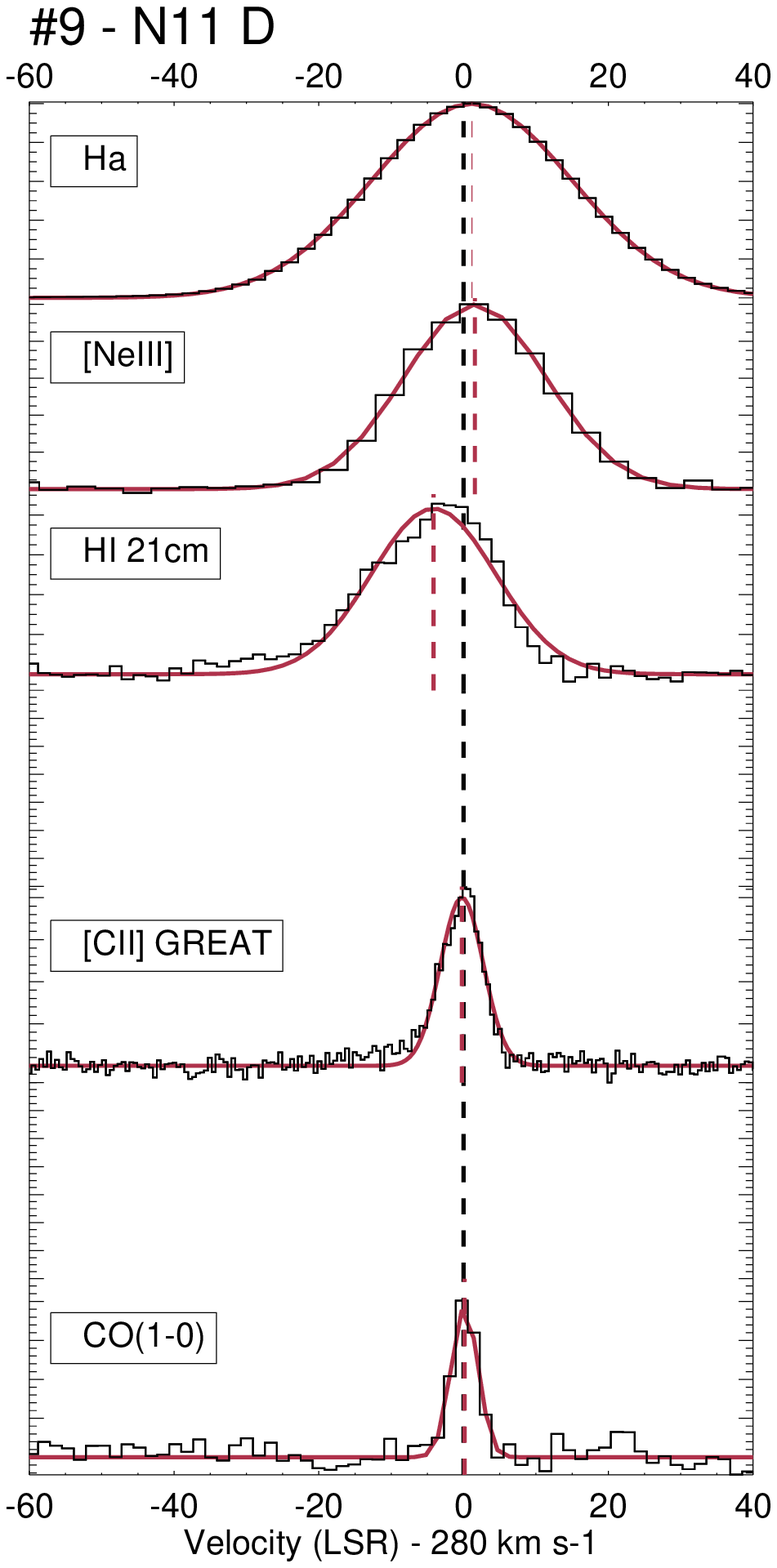}
\includegraphics[width=4.5cm,height=7.8cm,trim=10 0 0 10,clip]{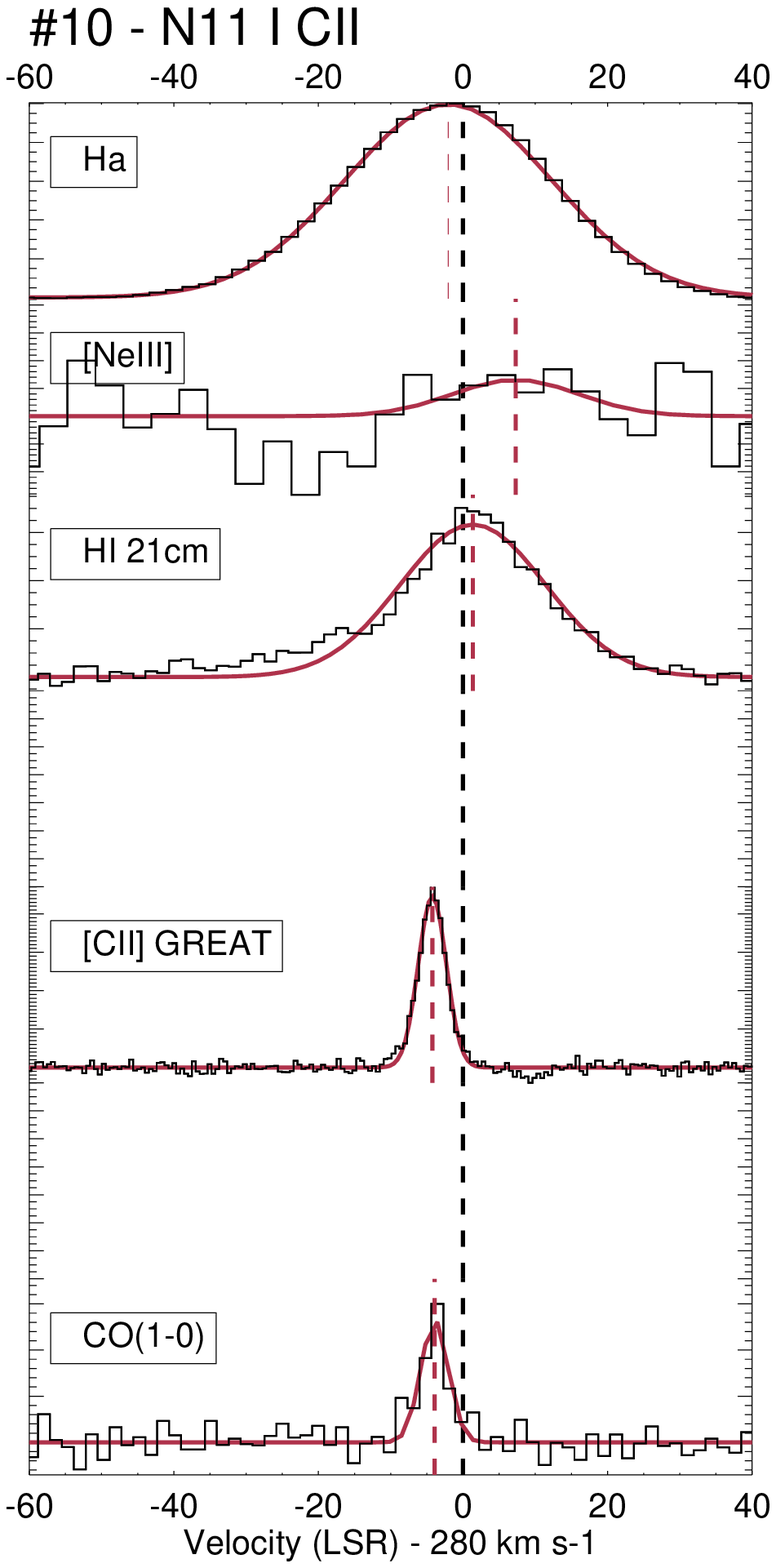}
\includegraphics[width=4.5cm,height=7.8cm,trim=10 0 0 10,clip]{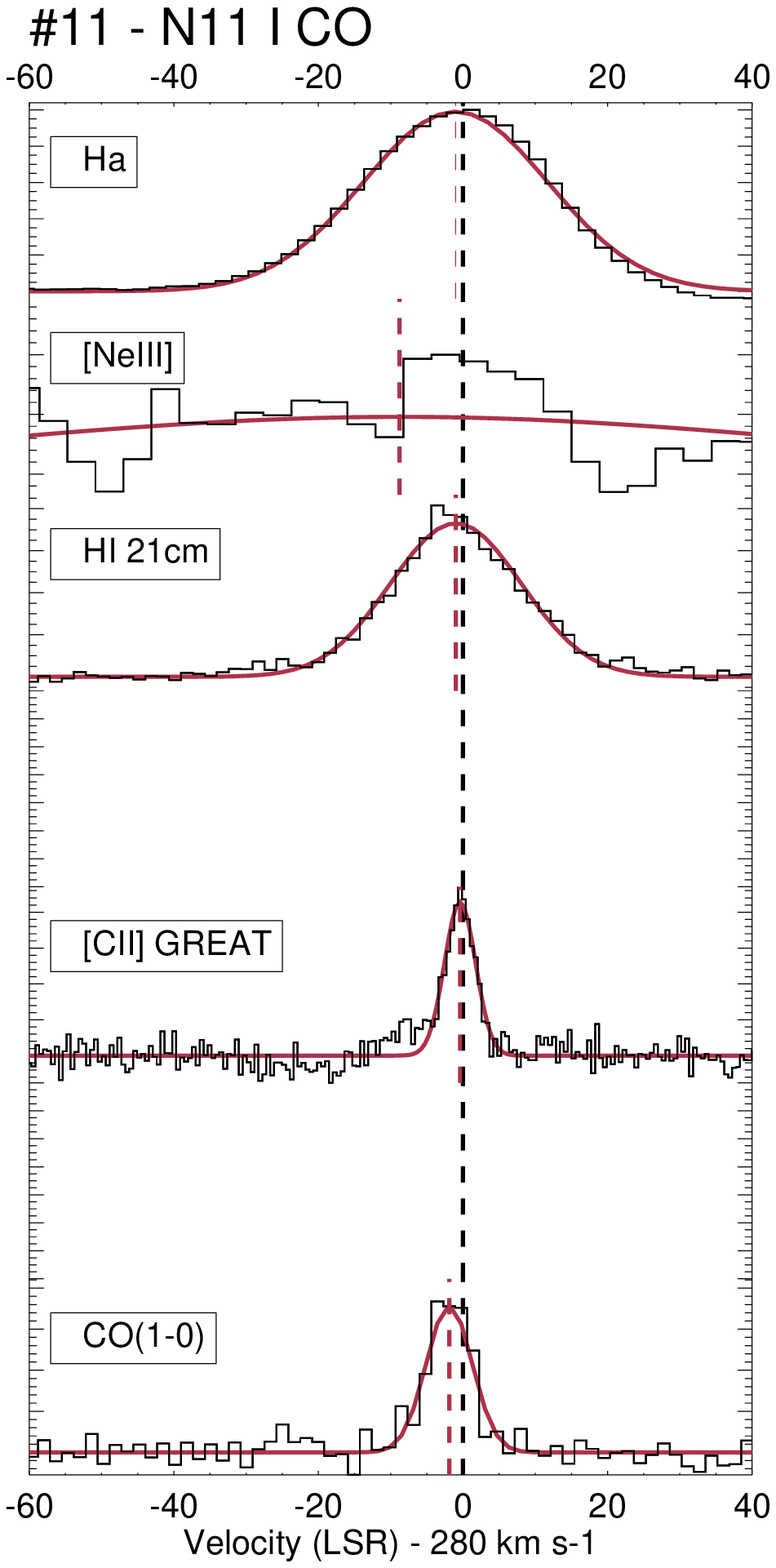}
\includegraphics[width=4.5cm,height=7.8cm,trim=10 0 0 10,clip]{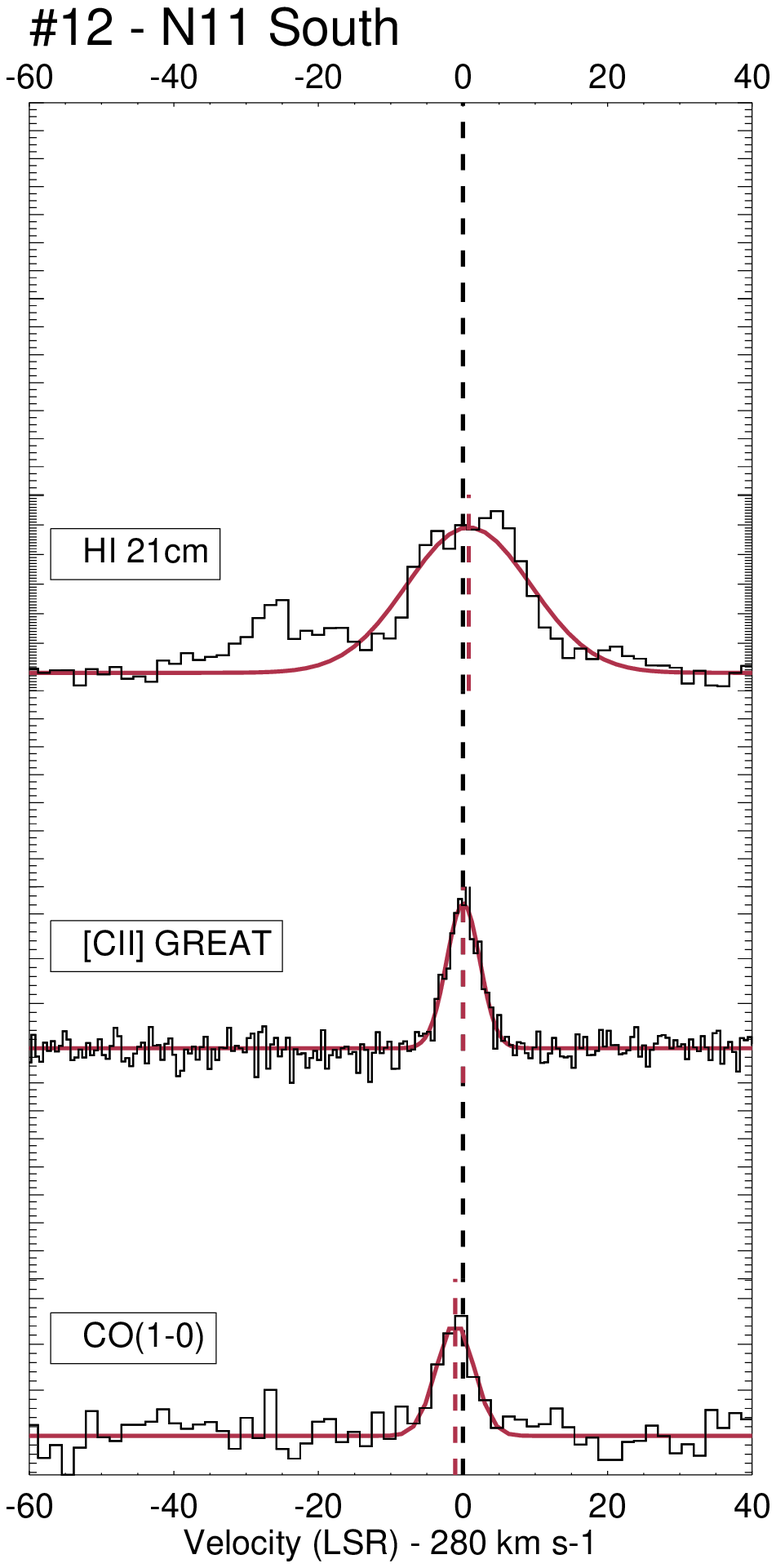}
\caption{Spectra of the main spectral lines used in this study for each GREAT pointing. From top to bottom in each panel: H$\alpha$, [Ne\3] $3868$\AA, H\1 $21$\,cm, [N\2] $205$\mic, [C\2] $158$\mic, and CO(1-0). The histogram shows the data and the red curve shows the Gaussian fit. For this plot, only one component is considered except for pointing \#5. Red dashed vertical lines show the Gaussian fit centroids while black dashed vertical lines show the zero velocity. The spectra are shifted arbitrarily along the y-axis for display purposes. }\label{fig:decomposition_simple}
\end{figure*}

The data were calibrated following \cite{Heyminck2012a} and \cite{Guan2012a}. The reduced spectra shown in this study are in local standard of rest (LSR) velocity (Fig.\,\ref{fig:decomposition_simple}). Spectra were generated from the level 3 GREAT product using the \href{http://www.iram.fr/IRAMFR/GILDAS}{GILDAS/CLASS} package. A baseline of first order was subtracted. Calibration from observed counts to main beam temperature $T_{\rm mb}$ was calculated with a beam efficiency $0.67$ for L1 and $0.65$ for L2. The spectral resolution is $0.15$\kms\ for passband L\#2 and $0.20$\kms\ for L\#1, but spectra were rebinned to obtain a channel width of $1.2$\kms\ to increase the signal-to-noise ratio. For the conversion to Janskys we use the following equation: $S[{\rm Jy}] = 721 T_{\rm mb}[{\rm K}]$ (GREAT science team; private communication). The flux calibration uncertainty is about $10\%$ \citep{Guan2012a}. The final spectra are shown in Figure\,\ref{fig:decomposition_simple}.

\subsection{\textit{Herschel}/PACS}\label{sec:obs_pacs}

\textit{Herschel}/PACS observations were taken as part of the SHINING key program (KPGT\_esturm\_1, PI E.\ Sturm). Small maps were performed with the unchopped scan mode, for regions LMC-N\,11 A, B, C, D, and I in [C\2], [O\1] $63$\mic, and [O\3] $88$\mic\ \citep{Cormier2015a}. The region N\,11\,B was also observed in [O\1] $145$\mic, [N\2] $122$\mic, [N\2] $205$\mic, and [N\3] $57$\mic\ \citep{Lebouteiller2012b}. The region N\,11\,``South'' is the only GREAT pointing (\#12) with no PACS spectroscopy. All PACS lines appear unresolved given the spectral resolution ($55-320$\kms; \citealt{Lebouteiller2012b}).

The comparison between the [C\2] fluxes measured from GREAT individual pointings and from the PACS map is contingent upon the knowledge of the GREAT beam profile and the flux calibration for both instruments. 
We use $\theta=14.4\arcsec$ for the GREAT HPBW (see Sect.\,\ref{sec:obs_great}).
For a Gaussian beam profile, the beam solid angle is then $1.13\,\theta^2$, corresponding to a $\approx17\arcsec$ diameter. 
Therefore, we convolve the PACS map to a resolution of $14.4\arcsec$ and integrate the flux in an aperture of $17\arcsec$ diameter. The same steps are performed for the [N\2] observations. 

\begin{table}
\caption{SOFIA/GREAT and \textit{Herschel}/PACS fluxes. }
\label{tab:compa}
\begin{tabular}{l|ll|ll}
\hline
\hline
Pointing  & \multicolumn{2}{c|}{[C\2]} & \multicolumn{2}{c}{[N\2] $205$\mic} \\
    & \multicolumn{2}{c|}{$\times10^{-15}$\,W\,m$^{-2}$} & \multicolumn{2}{c}{$\times10^{-17}$\,W\,m$^{-2}$} \\
   & PACS  & GREAT$^{\rm a}$ & PACS  & GREAT$^{\rm a}$ \\ 
\hline
\#1 & 1.4  &  2.1 &  ... &  $<5.4$ \\  
\#2 & 3.3 & 5.8 & $<1.3$ &  $<3.6$ \\  
\#3  & 1.1 & 1.7 & $<1.0$ & $<1.8$ \\  
\#4  & 2.2  & 2.7 & $<1.7$ & $<3.1$ \\  
\#5  &  0.4  & $0.11, 0.09^{\rm b}$ & $<1.5$ & $<1.1, <1.9^{\rm b}$ \\  
\#6  &  2.1  & 2.9 &  ... & $<3.8$ \\  
\#7  &  1.9  & 1.6 &  ... &  ...\\  
\#8  &  2.5  & 4.0 &  ... &  ...\\  
\#9  &  $\approx0.7$   & 1.0 &  ... &  ...\\  
\#10  &  0.9  & 1.4 &  ...  &  ...\\  
\#11  & 0.5 & 0.9 &  ...  &  ...\\  
  \#12  &  ...   & 0.6 &  ... &  ... \\
  \hline
\end{tabular}\\
\tablefoottext{a}{Velocity-integrated values. For [N\2], whose velocity structure is unknown, we calculate upper limits using the [C\2] velocity profile. }
\tablefoottext{b}{The first value corresponds to the $-10.6$\kms\ component, the second value to the $5.7$\kms\ component. }
\tablefoot{PACS fluxes are integrated in a $17.2\arcsec$ aperture (see text). The values reported for GREAT (collected over a $17\arcsec$ beam) are integrated over the velocities (i.e., not from a fit).  } 
\end{table}

The fluxes measured with PACS and GREAT are shown in Table\,\ref{tab:compa} for [C\2] and [N\2]\,$205$\mic. The agreement is good overall for [C\2], although the PACS fluxes seem to be lower by a factor of $1.5$ on average (see also \citealt{Fahrion2017a} and \citealt{Schneider2018a}). A better agreement could be reached if the effective GREAT HPBW were somewhat larger ($\approx20\arcsec$ solid angle). Furthermore, the PACS spectrometer data are calibrated for extended emission, so if the emission is indeed extended, the derived flux will scale with the aperture size; if the emission is point-like however, the total flux of the point source is enclosed in an area somewhat larger than the FWHM of the instrument, and the enclosed energy could be typically underestimated by $10-15\%$
\footnote{The PACS Spectrometer Calibration Document v3.0 (7-July-2016)  \url{http://herschel.esac.esa.int/twiki/bin/view/Public/PacsCalibrationWeb}} (see also \citealt{Fahrion2017a}).

\subsection{H\1\ $21$\,cm}\label{sec:obs_hi}

The H\1\ $21$\,cm data cube is taken from the ATCA+Parkes observations by \cite{Kim2003a}, which is, at the time of publication, the highest spatial resolution ($60\arcsec$) H\1\ survey of the LMC. The pixel size is $40\arcsec$ and the velocity channels are $1.6$\kms. The column density is calculated using a conversion factor for optically thin gas $1.823\times10^{18}$\,K\kms\,cm$^{-2}$ \citep{Dickey1990a}.

\begin{figure}
  \includegraphics[width=9.5cm,clip,trim=0 0 0 0]{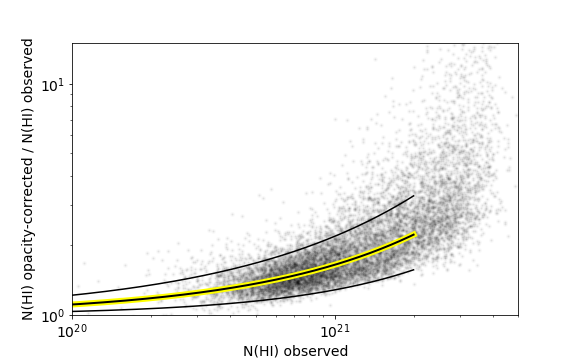}
  \caption{Ratio of the opacity-corrected \citep{Braun2012a} to\ observed H$^0$ column densities across the N\,11 region. The observed H$^0$ column density for individual velocity components is always $<2\times10^{21}$\,cm$^{-2}$ (see Sect.\,\ref{sec:indiv_fh2}).  }\label{fig:nhopacity}
\end{figure}

Optically thick H\1\ may be significant and an important contributor to the dark neutral medium (DNM) gas as proposed by some authors (e.g., \citealt{Fukui2015b}). We account for optically thick H\1\ by using the results of \cite{Braun2012a} for the LMC. The opacity correction is based on a method described in \cite{Braun2009a} which computes the temperature as well as the turbulent broadening on scales of $100$\,pc assuming a spatially resolved isothermal feature (i.e., single temperature along the line of sight). The total column density is then calculated using the profile fit parameters and a correction for the residual emission assumed to be optically thin. To estimate the factor by which we should correct the H$^0$ column density from \cite{Kim2003a}, we use the smooth variation of the ratio between the velocity-integrated opacity-corrected and -uncorrected (optically thin assumption) H$^0$ column densities. Figure\,\ref{fig:nhopacity} shows this ratio for the N\,11 data points. The dispersion in the ratio is partly due to the opacity correction method (e.g., not considering self-absorption) and to the fact that multiple components exist along the line of sight with different opacities. 
For the H$^0$ column densities measured in individual components (Sect.\,\ref{sec:indiv}), the correction due to optically thick H\1\ is less than a factor of two (Fig.\,\ref{fig:nhopacity}). We consider this correction as an upper limit since other methods to infer the fraction of optically thick H\1\ can result in much lower correction factors (e.g., \citealt{Lee2015a,Nguyen2018a,Murray2018a}). 

\subsection{CO (1-0)}\label{sec:obs_co}

The CO(1-0) observations are taken from the MAGMA survey described in \cite{Wong2011a}. Updated reduction is explained in \cite{Wong2017a}. The original spatial resolution is $45\arcsec$ and we use a $15\arcsec$ pixel size grid. Velocity channels are $0.53$\kms. The column density is calculated using a conversion factor\footnote{We use the notation $X'$ to distinguish it from the factor $X$ including the CO-dark H$_2$ gas contribution.} $X'_{\rm CO}=2\times10^{20}$\,cm$^{-2}$\,(K\,\kms)$^{-1}$, which is the fiducial value for resolved CO clumps without the contribution of the CO-dark H$_2$ gas between CO clumps as discussed in \cite{RomanDuval2014a}. In Table\,\ref{tab:cds} we provide the CO intensity, the H$_2$ column density using $X'_{\rm CO}$, $N({\rm H_2|CO})$, and the molecular gas fraction $f({\rm H}_2{\rm |CO})$ defined as $2 \times N({\rm H}_2{\rm |CO})/[N({\rm H}^0)+2 \times N({\rm H}_2{\rm |CO})]$ (i.e., ignoring the CO-dark H$_2$ gas).

\begin{table}
\caption{CO and H$^0$ column densities }\label{tab:cds}
\begin{tabular}{lllll}
\hline
\hline
\# & $N({\rm H}^0)$ & $I$(CO) & $N({\rm H_2|CO})$ & $f({\rm H_2|CO})$ \\ 
   & [$\times10^{21}$\,cm$^{-2}$]  & [K\kms] & [$\times10^{21}$\,cm$^{-2}$]  \\
\hline
\#1 & $5.0$ & $5.3$ & $3.2$ & $0.6$ \\
\#2 & $2.4$  & $5.2$ & $3.1$ & $0.7$  \\
\#3 & $3.1$ & $4.3$ & $2.6$ & $0.6$ \\
\#4 & $3.7$ & $5.3$ & $3.2$  & $0.6$ \\
\#5 & $2.0$ & $<0.8$ & $<0.5$  & $<0.3$ \\
\#6 & $3.7$ & $3.3$ & $2.0$  & $0.5$ \\
\#7 & $3.7$  & $3.3$ & $2.0$ & $0.5$  \\
\#8 & $3.1$ & $5.6$ & $3.3$ & $0.7$ \\
\#9 & $3.3$ & $9.0$ & $5.4$ & $0.8$ \\
\#10 & $3.2$  & $6.3$ & $3.8$ & $0.7$ \\
\#11 & $3.2$ & $14.3$ & $8.6$ & $0.8$ \\
\#12 & $2.4$ & $8.0$ & $4.8$ & $0.8$ \\
\hline
\end{tabular}
\tablefoot{The H$_2$ column density $N({\rm H_2|CO})$ is calculated using $I({\rm CO})$ from the MAGMA survey. We assume $X'_{\rm CO}=2\times10^{20}$\,K\kms\,cm$^{-2}$ (Sect.\,\ref{sec:obs_co}). For this calculation we use the H\1\ and CO observations at a spatial resolution of $60\arcsec$.  }
\end{table}

We also obtained ALMA band-3 observations of N\,11B (programs 2012.1.00532.S and 2013.1.00556.S; PI Lebouteiller). The mosaic covers the area observed with \textit{Herschel}/PACS map (Section\,\ref{sec:obs_pacs}). The spatial resolution is $2.2$\arcsec, with a maximum recoverable scale of $12$\arcsec\ and $47$\arcsec\ for the $12$- and $7$-m arrays respectively. The velocity resolution is $0.64$\kms. The line width ranges between $\approx2$ and $\approx5$\kms. The $7$-m, $12$-m, and Total Power observations were combined for the final data cube. The velocity profiles for the broadest lines are asymmetric, suggesting the presence of multiple components (Fig.\,\ref{fig:decomposition_simple}). 

\begin{figure*}
\begin{minipage}[T]{0.5\textwidth}
    \includegraphics[height=23cm,trim=10 10 5 10,clip]{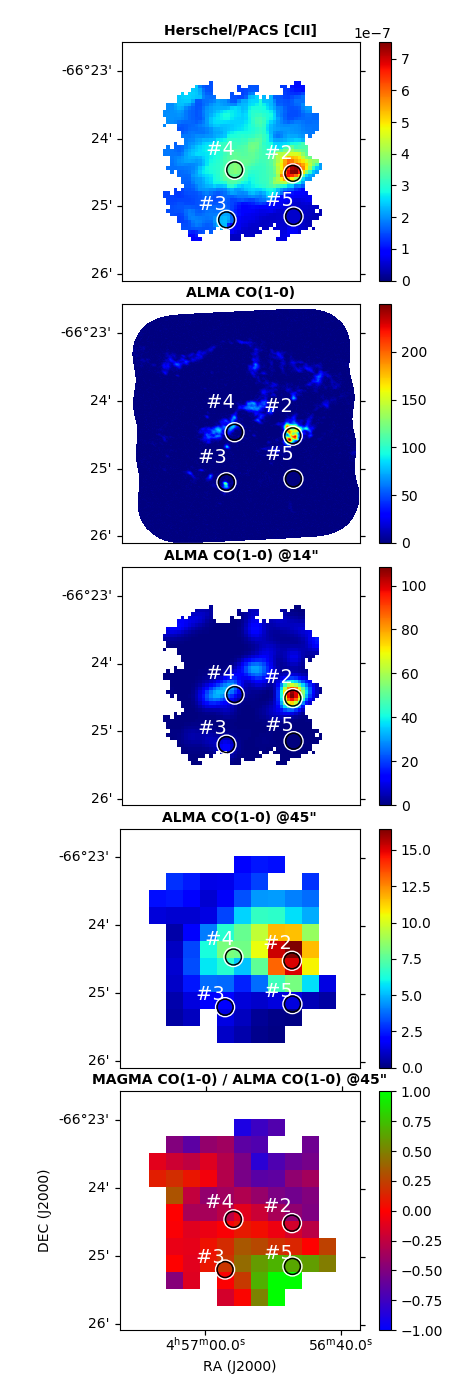}
\end{minipage}\hfill 
\begin{minipage}[T]{0.5\textwidth}
\includegraphics[width=7.4cm,trim=0 47 0 0]{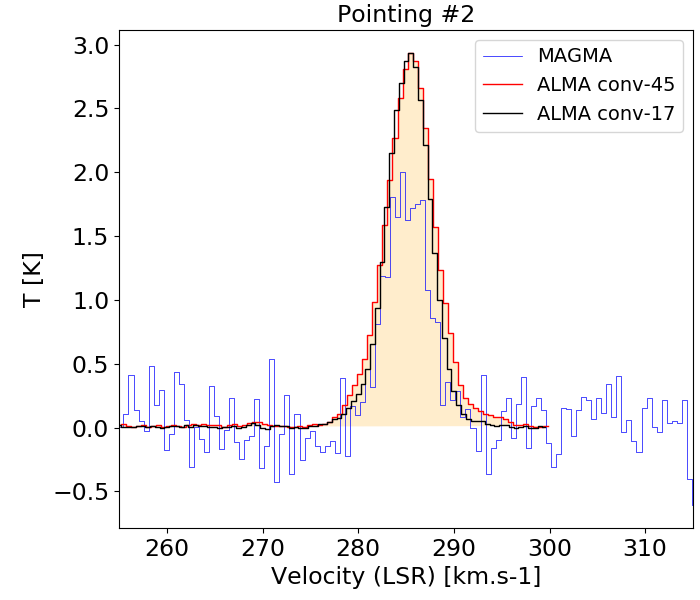}
\includegraphics[width=7.4cm,trim=0 47 0 0]{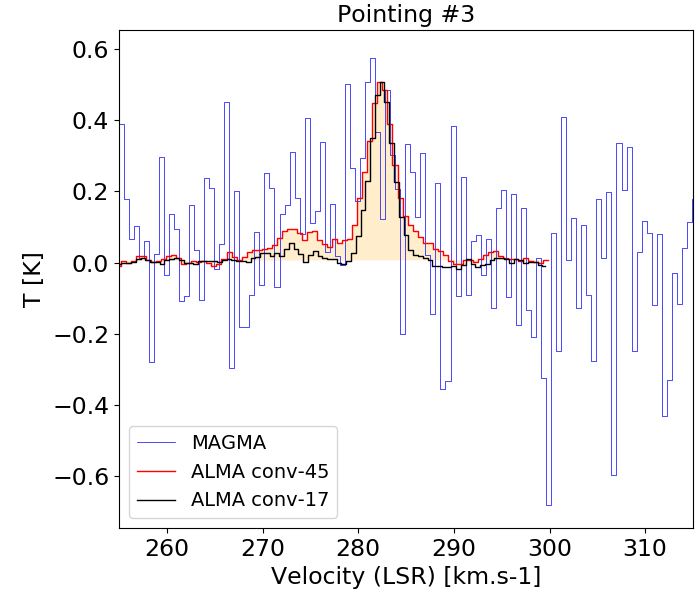}
\includegraphics[width=7.4cm,trim=0 47 0 0]{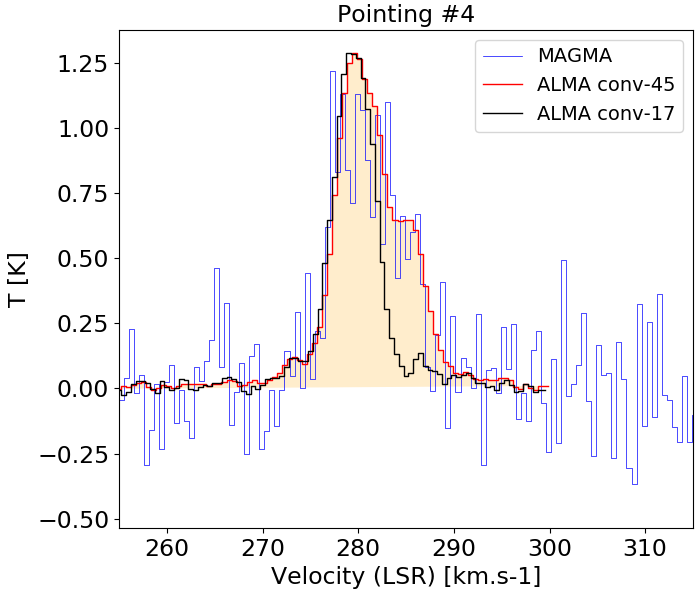}
\includegraphics[width=7.4cm,trim=0 0 0 0]{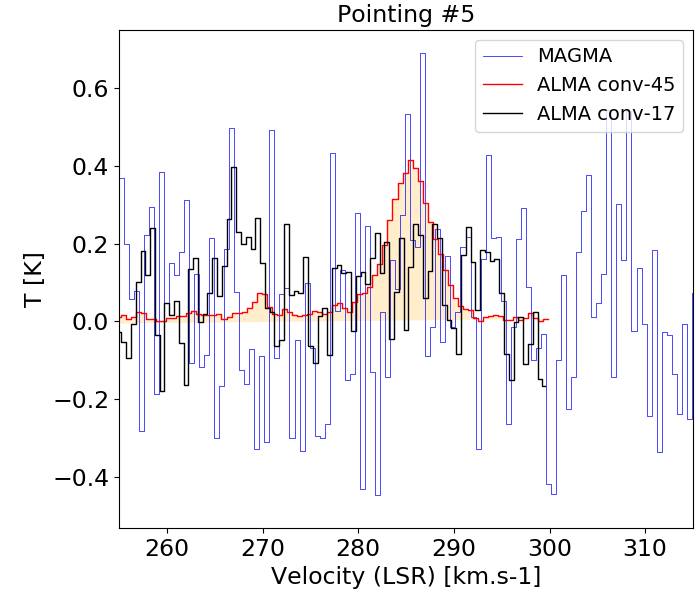}
\end{minipage}
\caption{Comparison between ALMA, MAGMA, and \textit{Herschel} data. Left column, from top to bottom: \textit{Herschel}/PACS [C\2] map of N\,11B, the ALMA CO(1-0) map (original resolution, convolved to $14.4\arcsec$ and projected on PACS [C\2] grid, and convolved to $45\arcsec$ and projected on MAGMA grid), and the ratio ($\log$ units) of CO MAGMA/ALMA. The GREAT pointings are overlaid as $14.4\arcsec$ diameter circles. Right column: MAGMA spectrum (blue histogram) and the ALMA spectra calculated for a $45\arcsec$ resolution (red with shade) and $17\arcsec$ resolution (black) for the four pointings in N\,11B. }\label{fig:spectracoalma}
\end{figure*}

Figure\,\ref{fig:spectracoalma} shows the original ALMA map and the projections on the \textit{Herschel}/PACS [C\2] and MAGMA CO(1-0) grids for comparison.
While pointings \#2 and \#3 coincide well with compact CO peaks observed with ALMA, pointing \#4 is somewhat offset by a few arcseconds with respect to the ALMA clump. Given the GREAT beam size ($14.4\arcsec$; Sect.\,\ref{sec:obs_great}), part of the spatially offset CO cloud emission is expected to contribute to the GREAT pointing \#4.

Figure\,\ref{fig:spectracoalma} also shows the comparison between MAGMA and ALMA spectra. Overall, there is good agreement between the two datasets over a $45\arcsec$ resolution, with the ALMA spectra showing a drastic S/N improvement. We also show the ALMA spectra calculated for a $14.4\arcsec$ resolution (i.e., similar to GREAT). On the one hand, ALMA spectra at $14.4\arcsec$ and $45\arcsec$ resolution are similar for pointing \#2, confirming that most of the emission originates from a cloud smaller than $14.4\arcsec$ with no significant contamination from nearby clouds. On the other hand, the ALMA spectra toward pointings \#3 and \#4 show some differences with spatial resolution, indicating that nearby clouds with different properties (intensity, central velocity, line width) contribute to the spectrum calculated over a $45\arcsec$ beam for these pointings. Pointing \#5 shows no CO emission with ALMA over $14.4\arcsec$ resolution but emission is seen for the $45\arcsec$ resolution which is likely arising from the main PDR complex corresponding to pointing \#2. In the following, the ALMA spectra are used when available in order to compare to the [C\2] spectral profiles. The detailed analysis of the ALMA map is deferred to a future work.

\subsection{VLT/GIRAFFE}\label{sec:obs_opt}

Optical spectroscopy is used to examine tracers of the ionized gas. We use archival ESO/VLT/U2 GIRAFFE observations with the MEDUSA fiber system feeding component. GIRAFFE is a medium-high ($R=5\,500-65\,100$) resolution spectrograph in the $3700-9500$\,\AA\ wavelength range. The MEDUSA fibers allow up to $132$ separate objects (including sky fibres) to be observed in one go. Each fiber has an aperture of $1.2$\arcsec on the sky. 

The sky-subtracted spectra from program 171.D-0237 were downloaded from the \href{http://giraffe-archive.obspm.fr}{GEPI GIRAFFE archive}. We examined in particular H$\alpha$ observed with the HR14a grating with $R=18\,000$ ($\approx17$\kms), and [Ne\3] $3868$\AA\ observed with the HR2 grating with $R=19\,600$ ($\approx15$\kms). Some details on the dataset can be found in \cite{TorresFlores2015a}.

Most GREAT positions could be associated with a nearby GIRAFFE fiber position, except \#12. When several field datasets were available for a given GREAT pointing, we selected only those with the best seeing and S/N. The [Ne\3] and H$\alpha$ spectra are shown in Figure\,\ref{fig:decomposition_simple}. The H$\alpha$ line is well detected toward all pointings, while [Ne\3] is detected everywhere except in N\,11\,D and I (GREAT pointings \#9 and \#10-11 resp.).

\section{Integrated properties along the lines of sight}\label{sec:integrated}

In this section we examine the kinematic properties (radial velocity and line width) integrated along the lines of sight. Results using individual velocity components are discussed in Sect.\,\ref{sec:indiv}.

\subsection{Comparison of the spectral  profiles}\label{sec:kine_neutral}

The CO(1-0) MAGMA spectra can all be fitted reasonably well with a single resolved (i.e., wider than the spectral resolution; Fig.\,\ref{fig:decomposition_simple}) component, with FWHM $\approx3-8$\kms\ for the lines with the best S/N (see Table\,\ref{tab:props_int}). The CO(1-0) line was also observed with ALMA in the N\,11B region (pointings \#2 through \#5), which once convolved to $45\arcsec$ resolution shows good agreement with the MAGMA data (Sect.\,\ref{sec:obs_co}) but also highlights some asymmetry in the line profile (Fig.\,\ref{fig:spectracoalma}). 

Most [C\2] GREAT spectra can also be fitted with a single resolved component, although a few pointings show evidence of multiple components (most notably \#3, \#4, \#5, \#6, and \#8) or an asymmetric profile (e.g., \#7, \#9). The FWHM of [C\2] measured with a single wide component ($\approx4-10$\kms) is always either similar to or larger than the CO FWHM. The H\1\ profile is always much broader than the CO and [C\2] profiles, with FWHM in the range $\approx16-40$\kms.
Similar results were found in various star-forming regions within nearby galaxies \citep{Braine2012a,deBlok2016a,RequenaTorres2016a,Okada2015a,Fahrion2017a} .

We verified that the difference in the profile line width between H\1\ and CO is not due to the spatial resolution by matching both datasets. However, the  wider H\1\ profile as compared to [C\2] is likely driven by the difference in spatial resolution. The H\1\ spectra, taken with $60\arcsec$ resolution (Sect.\,\ref{sec:obs_hi}), include the emission of clouds outside the GREAT beam.
The profile decomposition will allow us to mitigate the biases due to different beam sizes (Sect.\,\ref{sec:indiv}). 
Whether from the spectral profiles (Fig.\,\ref{fig:decomposition_simple}) or from the images (Fig.\,\ref{fig:spectracoalma}), there is no evidence of CO emission not associated with [C\2].

Since [N\2] was not detected with GREAT (Section\,\ref{sec:obs_great}), we attempted to use optical tracers (Sect.\,\ref{sec:obs_opt}) to describe the kinematics of the ionized gas. The observed FWHM of [Ne\3] is $\approx22$\kms, while it is $\approx30-40$\kms\ for H$\alpha$. The difference between the two line widths is mostly due to the thermal broadening. The thermal broadening (FWHM) in the ionized gas is
\begin{equation}
\Delta\nu_D / \nu_0 = 2 \sqrt{\ln 2} \sqrt{2kT/mc^2},
\end{equation}
with $T$ the temperature and $m$ the ionic mass. For a temperature of $10\,000$\,K \citep{ToribioSanCipriano2017a}, we find $\approx22$\kms\ for H$^+$ and $\approx4.8$\kms\ for Ne$^{2+}$. Convolutions with the instrumental resolution (Sect.\,\ref{sec:obs_opt}) and with macro-turbulent motions (typically a few \kms) yield $\approx30$\kms\ for H$\alpha$ and $\approx16$\kms\ for [Ne\3]. Both [Ne\3] and H$\alpha$ thus appear somewhat resolved, with a kinematical, nonthermal width of $\approx15-25$\kms, which likely reflects the dispersion of several individual components along the line of sight. The only GIRAFFE pointings showing a noticeable velocity structure in [Ne\3] are the ones associated with pointings \#2, \#4, and \#5 in N\,11\,B. Overall, the spectral resolution in the optical tracers is unfortunately not sufficient to decompose the velocity profiles but the fact that several velocity components may contribute to the observed ionized gas tracer profiles provides a useful constraint for the origin of [C\2] based on its line width (Sect.\,\ref{sec:linewidth}).

\begin{table*}
\caption{Line kinematic properties along lines of sight. }
\label{tab:props_int}
\begin{tabular}{l|ll|ll|ll|ll|ll|ll|ll}
\hline
\hline
Pointing  & \multicolumn{6}{c|}{CO} & \multicolumn{2}{c|}{[C\2]} & \multicolumn{2}{c|}{H\1} & \multicolumn{2}{c|}{[Ne\3]}  & \multicolumn{2}{c}{H$\alpha$} \\
  & \multicolumn{2}{c|}{MAGMA} & \multicolumn{2}{c|}{ALMA; $45\arcsec$} & \multicolumn{2}{c|}{ALMA; $14.4\arcsec$} & \multicolumn{2}{c|}{} & \multicolumn{2}{c|}{} & \multicolumn{2}{c|}{}  & \multicolumn{2}{c}{} \\
    & $v^{\rm a}$    & $\Delta{v}^{\rm b}$ & $v$    & $\Delta{v}$   &   $v$    & $\Delta{v}$   &   $v$    & $\Delta{v}$   &   $v$    & $\Delta{v}$   &   $v$    & $\Delta{v}$   &   $v$    & $\Delta{v}$   \\
\hline
\#1 & $-1.1$  & $7.7$ & ... & ...  &... & ...  & $-0.6$ & $10.1$ & ($0.9$) & ($27.9$) & $2.3$ & $22.9$ & $...$ & $...$ \\  
\#2 & $5.3$ & $6.6$ &$4.7$ & $5.0$ & $5.2$ & $5.2$  &  $5.4$ & $6.6$ & ($-5.2$) & ($37.3$) & ($11.1$) & ($34.8$) & $6.0$ & $39.4$ \\  
\#3 & $-0.7$ & $8.5$ & $2.0$ & $2.8$ & $2.5$ & $2.8$  & ($2.8$) & ($5.2$) & ($-5.8$) & ($39.8$) & $9.6$ & $25.1$ & $5.1$ & $33.4$   \\  
\#4 & $-0.5$ & $7.0$ &$-0.1$ & $4.9$ & $-0.4$ & $5.1$  &   $1.1$ & $6.9$ & $0.6$ & $38.9$  & ($7.4$) & ($37.4$) & $3.3$ & $42.0$   \\ 
\#5 & $...$ &... & ...  & $...$ & ... & ... & $-10.6$ & $4.0$  & $-11.3$ & $15.9$  & $-3.4$ & $7.4$ & $-9.7$ & $12.4$  \\   
    & $...$ & $...$ &... & ...  & ... & ... & $5.7$ & $7.0$    & $...$ & $...$  & $17.1$ & $7.4$ & $11.6$ & $30.6$  \\  
\#6 & $0.8$ & $5.0$ &... & ...  & ... & ...  &$0.5$ & $8.1$ & ($-6.3$) & ($28.5$)  & $4.9$ & $22.1$ & $-0.1$ & $32.5$  \\    
\#7 & $0.8$ & $5.0$ & ... & ...  & ... & ...  &$-0.5$ & $6.2$ & ($-6.3$) & ($28.5$)  & $4.9$ & $22.1$ & $-0.1$ & $32.5$  \\  
\#8 & $-0.1$ & $3.2$ &  ... & ...  &... & ...  & $-1.1$ & $8.6$ & ($-3.6$) & ($22.7$)  & $6.6$ & $23.1$ & $1.1$ & $32.6$  \\   
\#9 & $0.1$ & $4.3$  & ... & ...  & ... & ...  &$-0.2$ & $7.0$ & $-4.1$ & $20.2$  & $6.6$ & $23.1$ & $1.1$ & $32.6$   \\   
\#10 & $-3.9$ & $4.5$ &  ... & ...  &... & ...  & $-4.2$ & $4.6$ & $1.4$ & $23.7$   & $...$ & $...$ & $-2.0$ & $33.8$   \\  
\#11 & $-1.9$ & $7.5$ & ... & ...  &... & ...  & $-0.4$ & $5.1$ & $-1.0$ & $21.4$  & $...$ & $...$ & $-1.0$ & $30.1$  \\ 
\#12 & $-1.1$ & $6.1$ &  ... & ...  & ... & ...  &$0.0$ & $5.4$ & ($0.8$) & ($20.1$)  & $...$ & $...$ & $...$ & $...$   \\ 
\hline
\end{tabular}
\tablefoot{All values are in \kms. Values between parentheses indicate that the profile is not Gaussian (i.e., either asymmetric and/or multiple components). Only distinct components are considered. The larger FWHM observed for CO with MAGMA as compared to ALMA ($45\arcsec$) is due to the low S/N of the former (Fig.\,\ref{fig:spectracoalma}). }
\tablefoottext{a}{Radial velocities in \kms\ offset from $280$\kms. }
\tablefoottext{b}{Line FWHM in \kms. }
\end{table*}

\subsection{Contribution from the ionized gas to [C\2]}\label{sec:ciiorigin}

The [C\2] line may originate in the neutral (atomic or CO-dark H$_2$) or ionized gas. Unfortunately, no direct comparison of the velocity structure in [C\2] and ionized gas tracers can be performed because the [N\2] $205$\mic\ line was not detected with GREAT (Section\,\ref{sec:obs_great}) and because the spectral resolution for optical tracers is not high enough (Sect.\,\ref{sec:kine_neutral}).

\subsubsection{[C\2] line width}\label{sec:linewidth}

The expected thermal broadening\footnote{For a temperature of $5000$\,K (warm neutral medium), the thermal broadening is $4$\kms, i.e., on the same order as typical macro-turbulence velocity. For the cold neutral medium, macro-turbulence dominates over the thermal broadening. } for C$^+$ in the ionized gas is $\approx6$\kms\ (FWHM for a temperature of $10\,000$\,K). Macro-turbulent motions (typically a few \kms) may increase this value to $\approx10$\kms, but we also keep in mind that the temperature in the WIM, where a significant fraction of C$^+$ could exist, may be colder than $10\,000$\,K. Therefore, we expect [C\2] line widths in the ionized gas for a single velocity component to be in the range $\approx6-10$\kms. 

The observed FWHM for the main [C\2] velocity component is smaller than $\approx6$\kms\ toward pointings \#3, \#10, \#11, \#12, and for one of the two components toward pointing \#5 (Table\,\ref{tab:props_int}). We conclude that [C\2] is necessarily associated with the neutral gas for these components. This conclusion is strengthened if assuming the main [C\2] velocity component in reality corresponds to multiple, narrower components. For some of the other pointings, the [C\2] profile indeed shows multiple blended components (\#4, \#6) or asymmetric profiles (\#7, \#8, \#9), suggesting that the presence of multiple components drives the large total width (Sect.\,\ref{sec:kine_neutral}). The presence of multiple components in the ionized gas is also suggested by the wide [Ne\3] profiles in the optical (Sect.\,\ref{sec:kine_neutral}).

Pointing \#5 is particularly interesting as it shows two well separated [C\2] components, with FWHM of $4.0$\kms\ and $7.0$\kms\ for the low- and high-velocity components respectively. While the FWHM of the low-velocity component is compatible with an origin of [C\2] in the neutral gas, the high-velocity component may come from the ionized gas provided it corresponds to a single cloud.

In summary, the observed FWHMs for [C\2], together with the fact that [C\2] in the ionized gas may originate from more than one component, suggests that the observed [C\2] components are unlikely to arise from the ionized gas, except maybe for one component toward pointing \#5, and perhaps pointing \#1. The line width of all the other [C\2] components is compatible with an origin in the neutral gas, where macro-turbulence competes with the thermal broadening. Other complementary methods are examined in the following.

\subsubsection{Ionized gas traced by [N\2] $122$\mic\ and $205$\mic\ (pointings \#1 through \#6)}\label{sec:ciiorigin_nii}

\begin{figure}
\begin{tikzpicture}
    \node[anchor=south west,inner sep=0] (image) at (0,0) {\includegraphics[width=9.5cm,height=7cm]{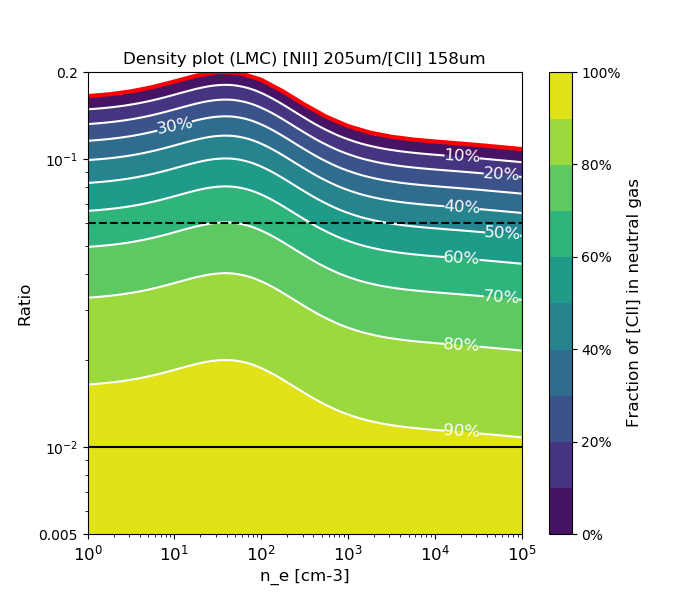}};
  \end{tikzpicture}\\
\begin{tikzpicture}
    \node[anchor=south west,inner sep=0] (image) at (0,0) {\includegraphics[width=9.5cm,height=7cm]{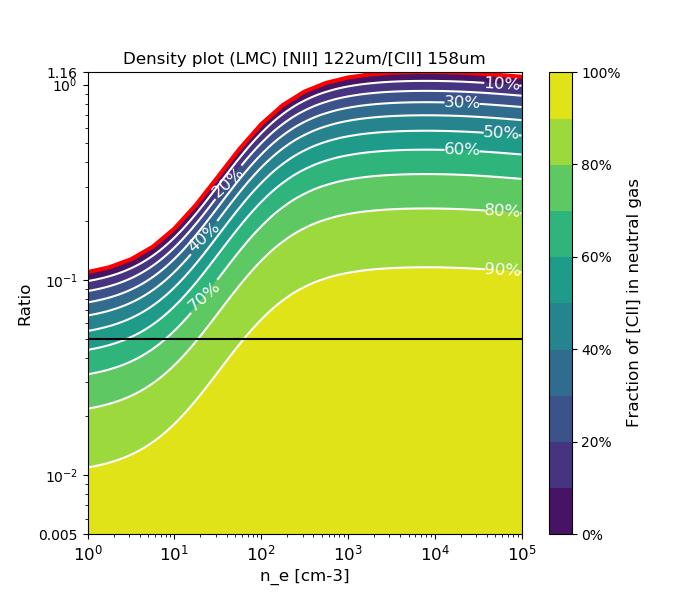}};
  \end{tikzpicture}
\caption{Theoretical [N\2]/[C\2] ratios as a function of density. The ratio values include ionic abundance ionization corrections (see text). The red curve indicates ratio values with [C\2] originating fully in the ionized gas. The solid black lines show the observed upper limits for all pointings observed with \textit{Herschel}/PACS (except \#5 for [N\2] $205$\mic/[C\2] shown as a dashed line). } \label{fig:densityniicii}
\end{figure}

Owing to the energy range required to produce N$^+$ and C$^+$ ions ($14.5-29.6$\,eV and $11.3-24.4$\,eV respectively) and owing to similar critical densities for [C\2] and [N\2] for collisions with $e^-$, [N\2] lines are often used to trace [C\2] emission in the ionized gas (e.g., \citealt{Oberst2006a}). We can then compare the observed [N\2]/[C\2] ratio to the theoretical value in the ionized gas, with any deviation indicating a contribution from [C\2] in the neutral gas. In order to compute the theoretical ratio, we first calculate the ionic abundance ratio N$^+$/C$^+$  using the solar abundance ratio $\log {\rm N/C} = -0.6$ (\citealt{Asplund2009a}; also valid for LMC-30\,Dor \citealt{Pellegrini2011a}) together with the ratio $[$ (N$^+$/N) / (C$^+$/C) $] = 1.26$ computed from MAPPINGS III photoionization grids \citep{Sutherland2013a}. We then need to account for the gas density. While the $205$\mic\ line has a critical density ($180$\cc\ for collisions with e$^-$) close to that of [C\2]\,$158$\mic\ ($\approx30-50$\cc; \citealt{Goldsmith2012a}), the $122$\mic\ line has a larger critical density ($400$\cc), meaning that the [N\2]\,$122$\mic/[C\2] ratio depends somewhat more on density. Figure\,\ref{fig:densityniicii} shows the theoretical [N\2]/[C\2] ratio as a function of density including the ionic abundance correction factor.

Using the upper limits on the [N\2]\,$122$\mic\ and $205$\mic\ lines observed towards N\,11B with \textit{Herschel}/PACS (corresponding to pointings \#2, \#3, \#4, and \#5; Sect.\,\ref{sec:obs_pacs}), we find [N\2]\,$122$\mic/[C\2]$\lesssim0.05$ and [N\2]\,$205$\mic/[C\2]$\lesssim0.01$ except toward \#5 where [N\2]\,$205$\mic/[C\2]$<0.06$ \citep{Lebouteiller2012b}. The comparison with the theoretical ratio in the ionized gas for typical densities between $10^{1-3}$\cc\ (Fig.\,\ref{fig:densityniicii}) indicates that the fraction of velocity-integrated [C\2] originating in the neutral gas is $f_n({\rm [CII]})\gtrsim70\%$ using [N\2]\,$122$\mic\ and $\gtrsim90\%$ using [N\2]\,$205$\mic\ (except for pointing \#5 with $\gtrsim55\%$). Therefore, for pointings \#2, \#3, and \#4, the fraction of [C\2] in the ionized gas is not significant. For pointing \#5 (made of two distinct velocity components), the most stringent constraint is obtained with [N\2]\,$122$\mic/[C\2], with $f_n({\rm [CII]})\gtrsim70\%$ for the sum of both components.
Since the two [C\2] components toward \#5 have similar fluxes, this suggests that the fraction of [C\2] originating in the ionized gas must be relatively small for both. 
The [N\2] $205$\mic\ upper limits calculated from the GREAT spectra (available for \#1 through \#6) do not allow us to improve further the determination of $f_n({\rm [CII]})$. 

\begin{figure}
\includegraphics[width=4.4cm,height=5.2cm,clip]{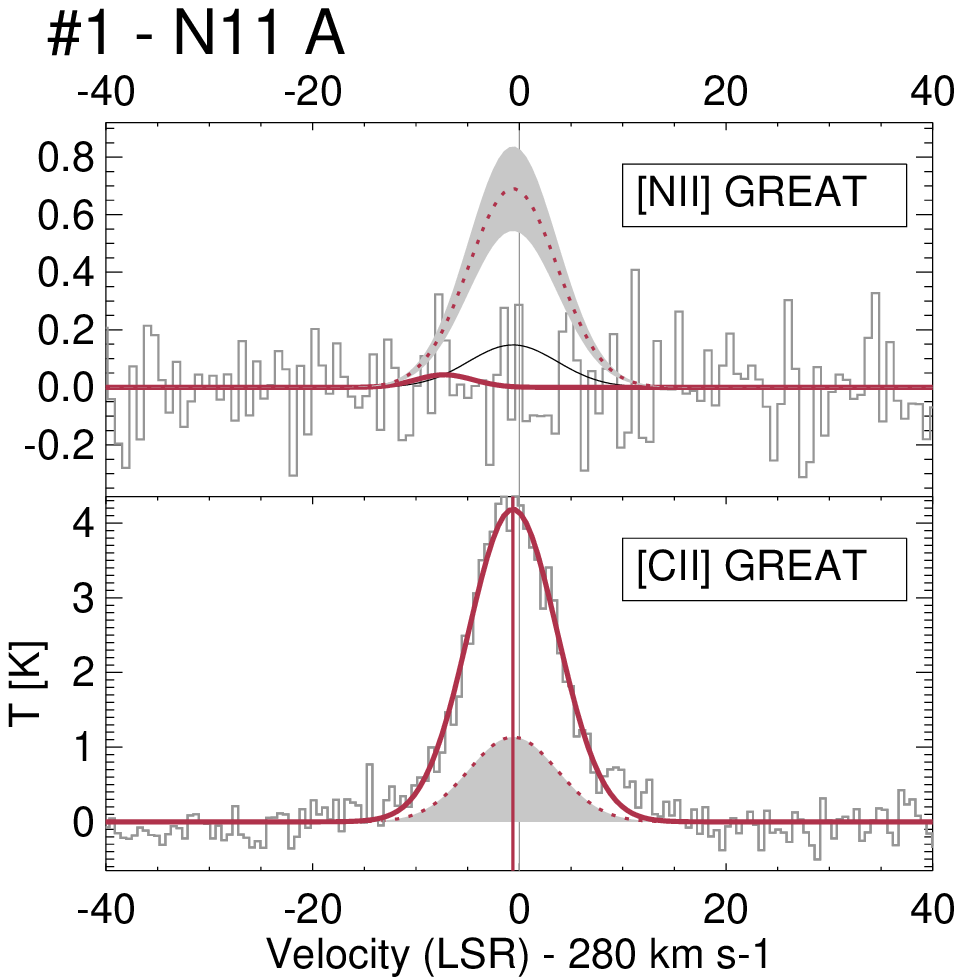}
\includegraphics[width=4.4cm,height=5.2cm,clip]{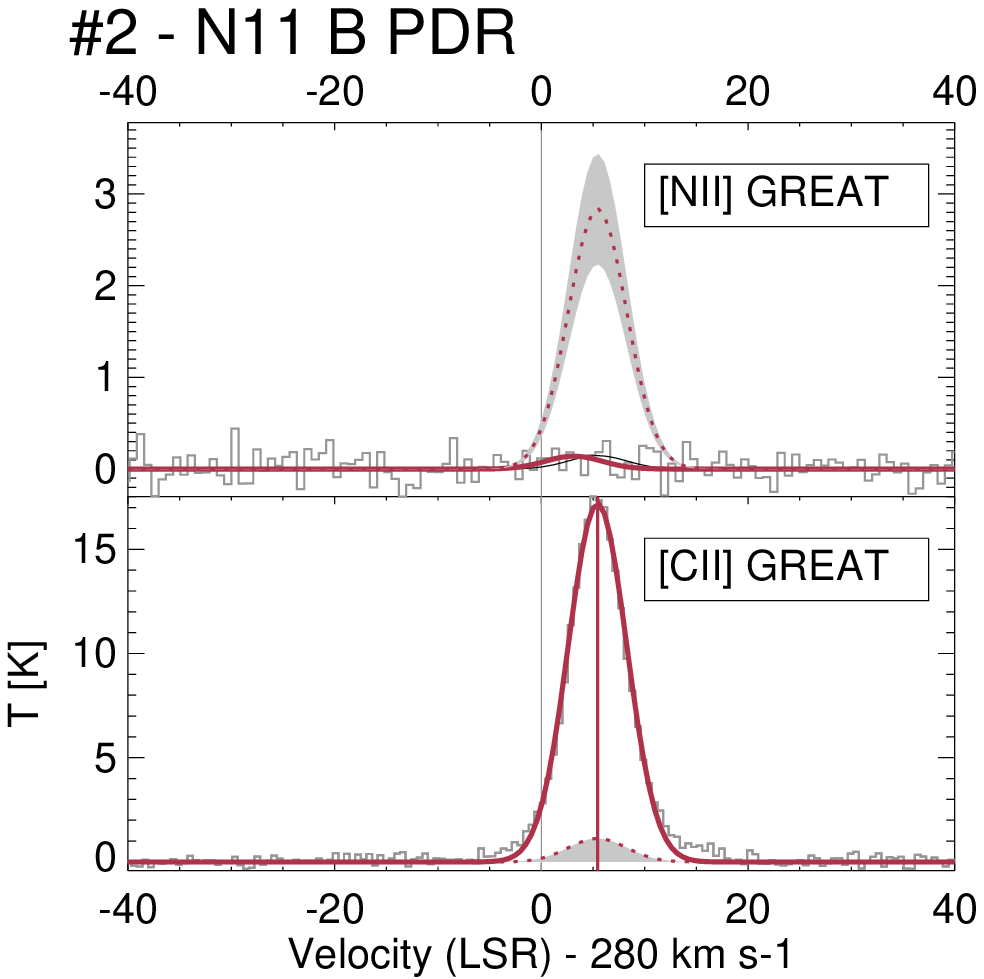}\\
\includegraphics[width=4.4cm,height=5.2cm,clip]{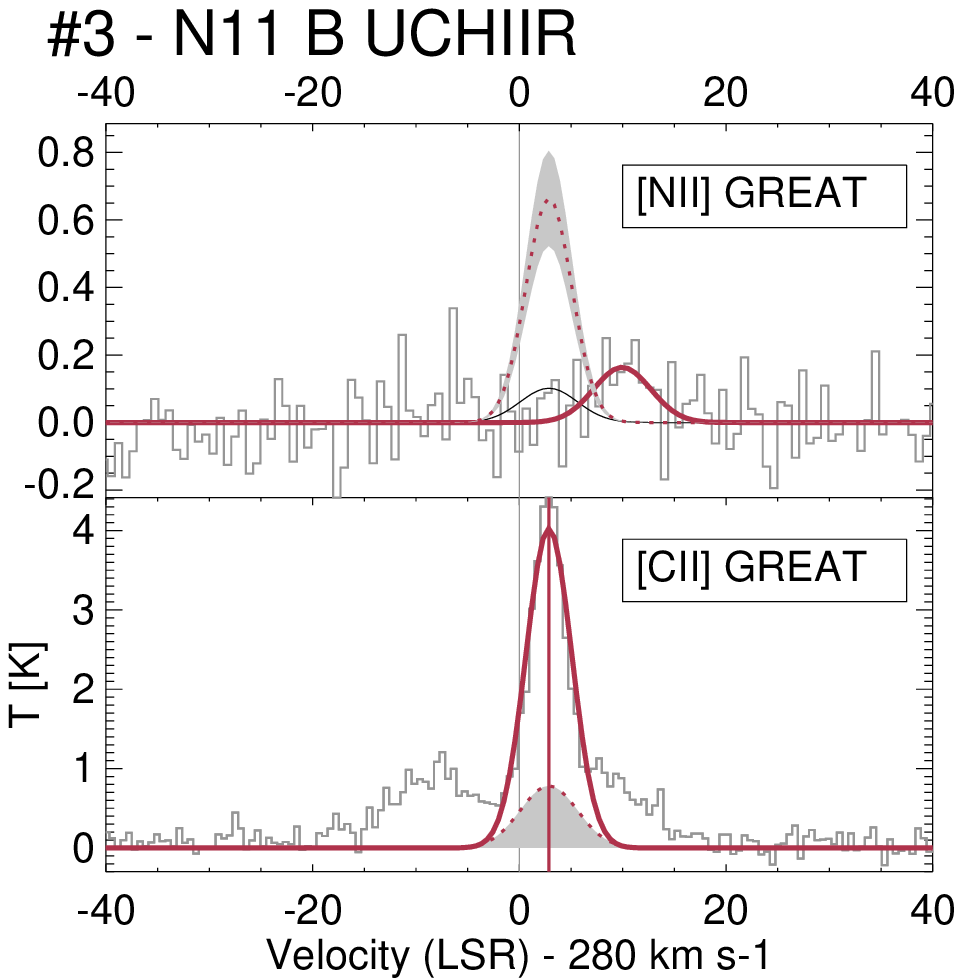}
\includegraphics[width=4.4cm,height=5.2cm,clip]{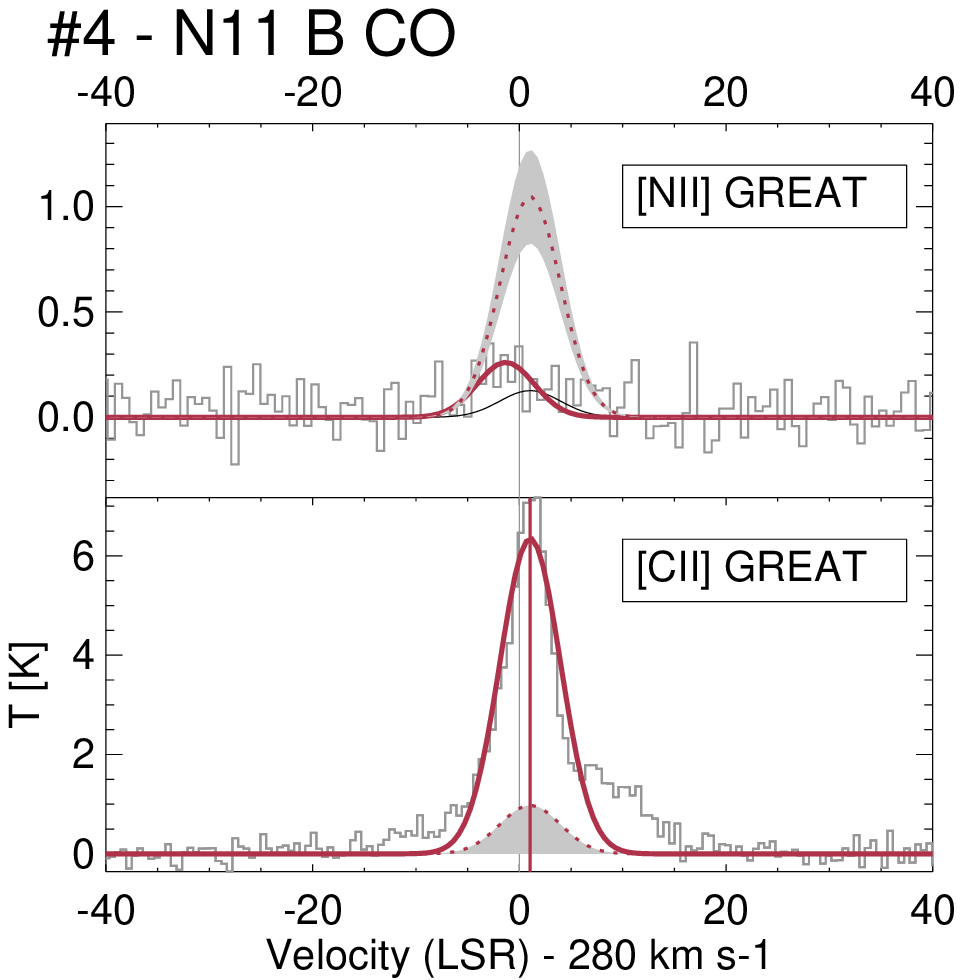}\\
\includegraphics[width=4.4cm,height=5.2cm,clip]{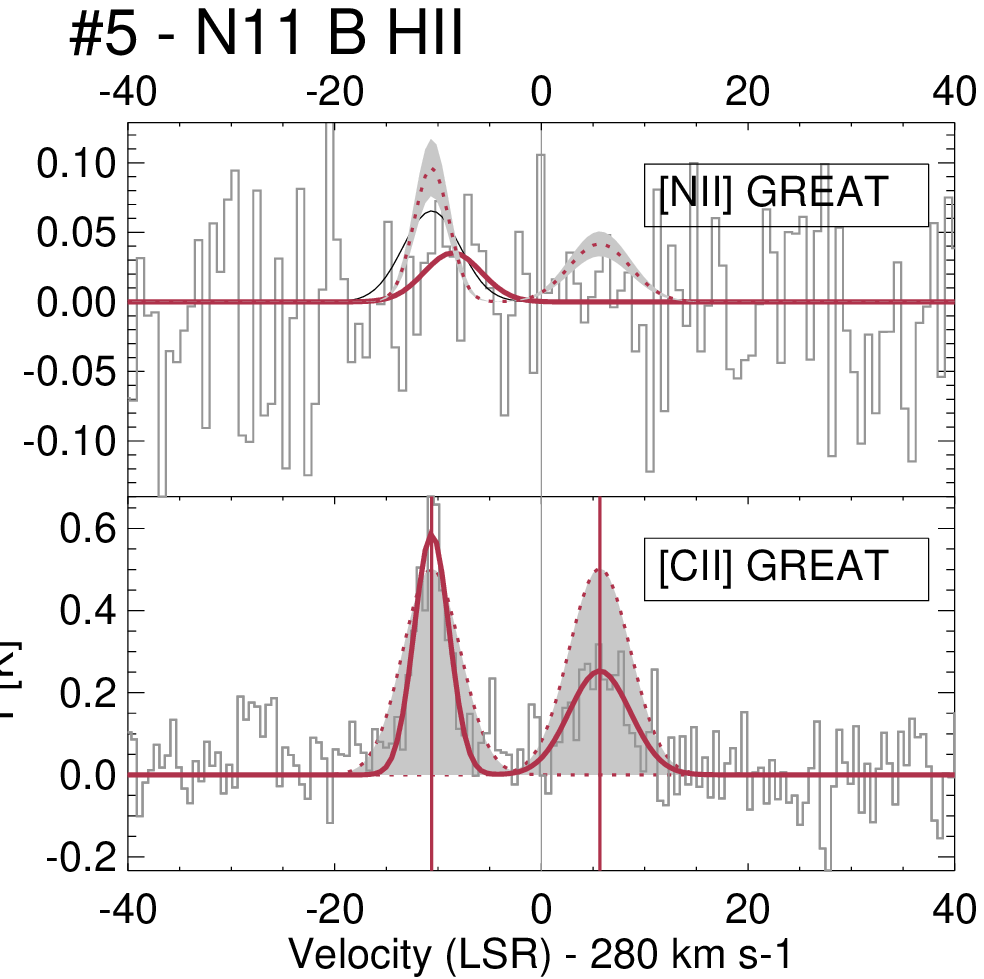}
\includegraphics[width=4.4cm,height=5.2cm,clip]{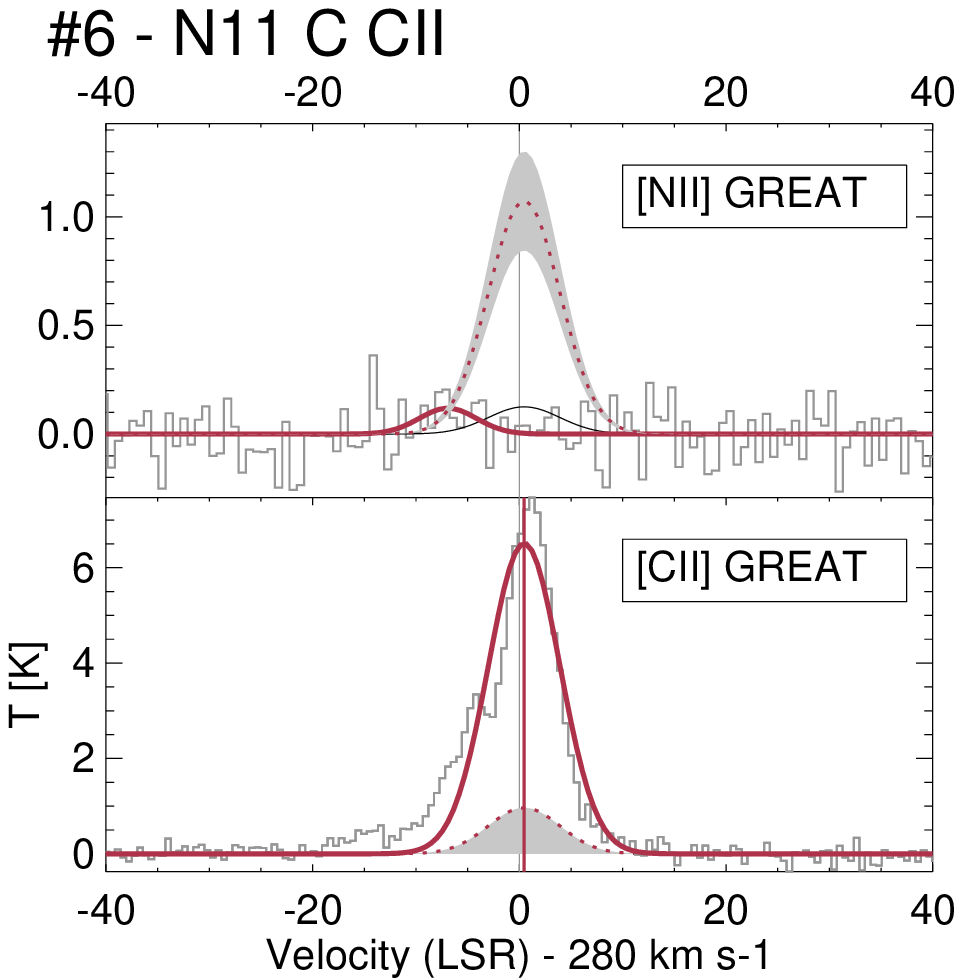}
\caption{Comparison of the GREAT [N\2] and [C\2] spectral profiles for the six pointings with [N\2] observations. The red  solid curve shows for both lines the Gaussian fit to the main velocity components as in Figure\,\ref{fig:decomposition_simple}. For the [N\2] spectrum the black curve corresponds to the RMS flux while the red dotted curve shows the expected [N\2] profile assuming [C\2] arises only from the ionized gas, with the uncertainty shown as the gray area (see text). For the [C\2] spectrum the red dotted curve and shaded area show the profile of [C\2] in the ionized gas calculated using the [N\2] RMS (i.e., the brightest possible [C\2] component arising from the ionized gas). }\label{fig:decomposition_simple_ciinii}
\end{figure}

We can also model the velocity profile of [N\2]\,$205$\mic\ with GREAT assuming [C\2] originates fully from the ionized gas and assuming the same velocity structure (i.e., ignoring potential differences in individual component line widths). Our results are illustrated in Figure\,\ref{fig:decomposition_simple_ciinii} in which we also show the brightest ionized gas [C\2] emission possible using the [N\2]\,$205$\mic\ upper limits. The density is chosen as $100$\cc\ for this test. It is clear that [C\2] originates overall from the neutral gas. For pointing \#5, the simulated [N\2] profile is compatible with the observed profile within uncertainties but we recall that the results using velocity-integrated measurements with \textit{Herschel}/PACS suggest that the [C\2] contribution from ionized gas is not dominant. We can also conclude that relatively faint [C\2] components observed toward \#3, \#4, or \#6 cannot be explained by [C\2] emission in the ionized gas.

\subsubsection{Photoelectric heating efficiency proxy}\label{sec:ciioipah}

Finally, the origin of [C\2] can be examined indirectly by comparing the neutral atomic gas cooling traced by the [C\2] and [O\1] lines to the gas heating. The latter can be traced by far-infrared emission but contamination by warm dust in the ionized phase can become an issue \citep{Lebouteiller2012b}. The PAH emission is another tracer of the neutral gas heating, and it is plausible that PAHs actually dominate the gas heating as compared to very small grains. In LMC-N\,11B, \cite{Lebouteiller2012b} found that the ratio ([C\2]+[O\1])/PAH is uniform, indicating that the photoelectric heating efficiency is fairly constant (see also \citealt{Helou2001a,Croxall2012a,Okada2013a}). The ratio remains constant even toward the stellar cluster LH\,10 (corresponding to pointing \#5) where ionized gas dominates the infrared line emission, suggesting that [C\2] and [O\1] still trace the neutral atomic gas in the foreground and background. 

We now revisit this finding by extending the measurement of ([C\2]+[O\1])/PAH to all PACS maps in LMC-N\,11. Since there is no mid-IR spectra available for all pointings in N\,11, we use the IRAC photometry bands to evaluate the PAH emission. We refer to \cite{Lebouteiller2012b} for the method. In short, we use the \textit{Spitzer}/IRAC photometry data points to identify and flag out stellar emission, to estimate and correct for the dust continuum, in order to provide the emission of PAHs either in the IRAC $5.8$\mic\ band (corresponding to the PAH $6.3$\mic\ feature) or in the IRAC $8.0$\mic\ band (corresponding to the PAH $7.7+8.6$\mic\ features).

\begin{figure}
\includegraphics[width=9cm,height=12cm,clip]{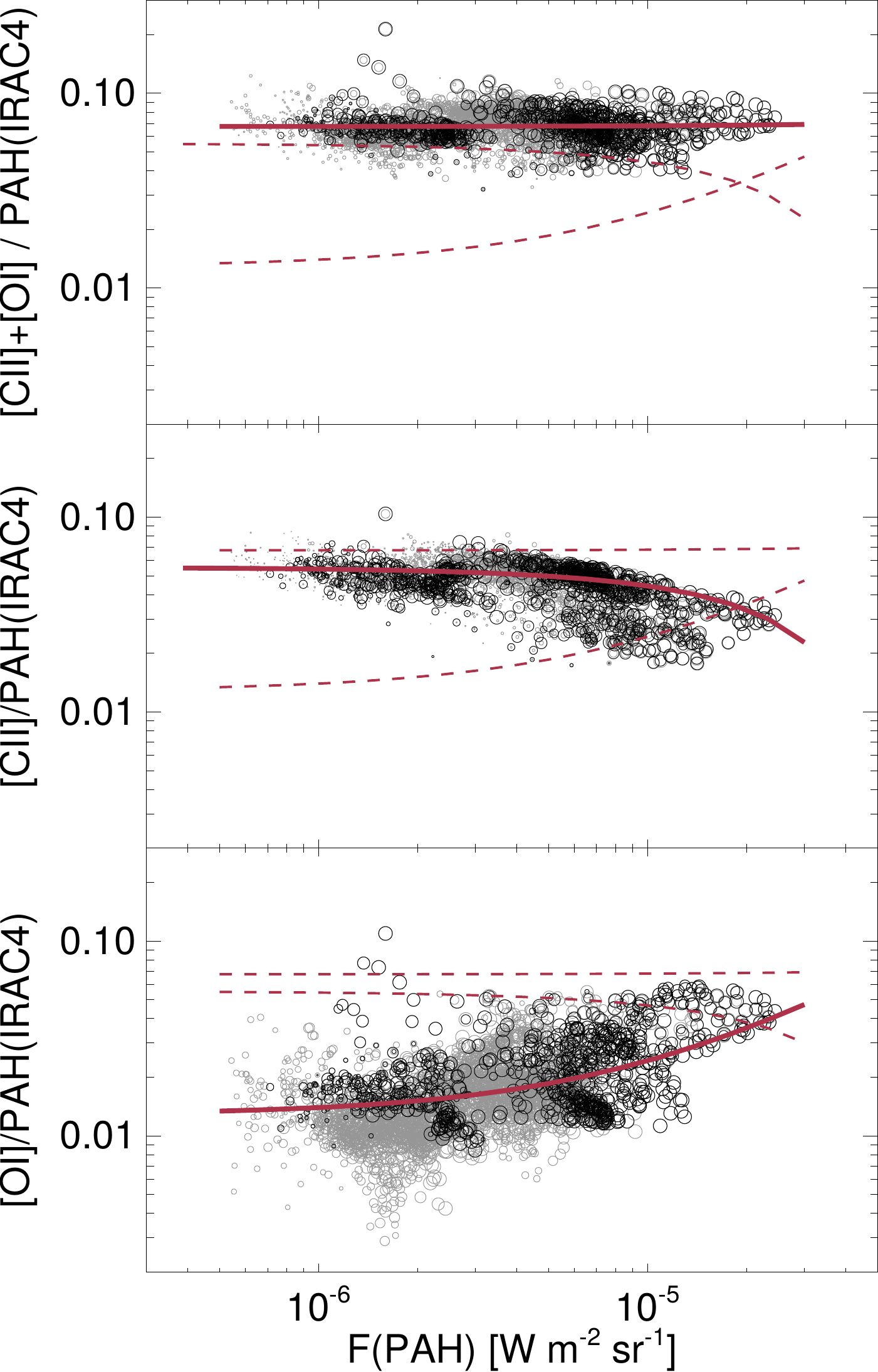}
\caption{Variation of [O\1]/PAH (bottom), [C\2]/PAH (middle), and ([C\2]+[O\1])/PAH (top) across the regions in LMC-N\,11 mapped with PACS. The size of the symbol scales with the signal-to-noise ratio. Black circles correspond to GREAT pointings. The solid line shows the linear regression for the corresponding ratio while the dashed lines show the regression obtained for the two other ratios. }\label{fig:ciioipah}
\end{figure}

Results are shown in Figure\,\ref{fig:ciioipah}, where it can be seen that the cooling rates provided by [C\2] and [O\1] compensate, with [C\2] dominating in the faintest, likely more diffuse, regions, and with [O\1] dominating in the brightest, more compact, regions. The sum [C\2]+[O\1] is however proportional to the PAH emission with no obvious trend with PAH brightness. The remarkably small scatter in ([C\2]+[O\1])/PAH in many regions including PDRs (\#2) suggests that there is no significant extra [C\2] emission arising from the ionized phase. Potential PAH emission in the ionized gas is ignored as it is plausible that they are photodestroyed (see e.g., \citealt{Madden2006a,Lebouteiller2007a,Chastenet2019a}). 

As for the other diagnostics using velocity-integrated values, this result does not preclude the presence of relatively faint [C\2] components that could be associated with the ionized gas. Overall, the results obtained with ([C\2]+[O\1])/PAH, together with those obtained in Sections\,\ref{sec:linewidth} and \ref{sec:ciiorigin_nii}, suggest that most of the [C\2] arises in the neutral gas. We find no velocity component that can be unambiguously associated with ionized gas.

\section{Individual component analysis}\label{sec:indiv}

In this section we attempt to decompose the CO, [C\2], and H\1\ profiles toward the GREAT pointings in order to derive the physical conditions associated with each individual component, with the help of simple two-phase models (neutral atomic and molecular). We assume that the ionized contribution to [C\2] is negligible (Sect.\,\ref{sec:ciiorigin}).

\subsection{Profile decomposition}\label{sec:profdecomp}

We assume that the velocity structure reflects the presence of several individual components, which we refer to as ``clouds'', defined by their radial velocity. We ignore clouds with potentially significant velocity gradient along the line of sight. In other words, all tracers (CO, [C\2], H\1) are assumed to peak at the same velocity for any given cloud.
Some caveats pertain to the different spatial resolutions of the observations. The H\1\ emission in particular may arise from clouds that are not contributing to the GREAT beam (Sect.\,\ref{sec:kine_neutral}). However, this confusion is mitigated when individual velocity components are considered, since these are more likely to originate from a relatively more spatially confined cloud.

The velocity profiles of CO, [C\2], and H\1\ can be decomposed in various ways, for instance by fitting Gaussian components for the tracer with the narrowest profile, typically CO, and using these centroids for the adjustment of other tracers with other velocity components added as needed (e.g., \citealt{Okada2019a}). We decided to use a statistical approach instead, by adjusting the various profiles simultaneously with velocity components not being fixed or inferred from any other specific profile. Any given component is defined by its velocity and line width, while the component intensity is determined independently for each CO, [C\2], and H\1. The sum of the components reproduces the global CO, [C\2], and H\1\ profiles. We also performed decompositions allowing potentially different line widths for each tracer for any given velocity component, but the main results shown in the following remain unchanged. 

We use a Bayesian approach with the Markov Chain Monte Carlo (MCMC) \texttt{PyMC3} code \citep{Salvatier2016a} along with the Metropolis-Hastings sampling algorithm. Parameters are described in Table\,\ref{tab:bayespar}. We identify three parameters that control the number of possible solutions, namely the number of components, the minimum line width, and the minimum separation between components. Although we could narrow down the number of solutions by introducing these parameters within the Bayesian inference (eventually finding a unique converged solution), other degenerate solutions may also be acceptable. Since we consider that we do not have enough observational constraints and priors for these parameters, we chose to test different fixed values:
\begin{itemize}
\item The \textit{number of components} is set to $10$ (minimum to reproduce the H\1\ profile) and $15$ (to allow potential significant blends).
\item The \textit{minimum line width} is set to $1$ and $2$\kms. The velocity profiles in Figure\,\ref{fig:decomposition_simple} show that, when the [C\2] profile is visibly made of more than one component (e.g., \#3, \#4, \#6), the [C\2] component associated with CO emission shows the same line width as the CO component, with FWHM values $\gtrsim3$\kms; Table\,\ref{tab:props_int}). This suggests that macro-turbulence dominates the [C\2] and CO line width. Therefore, we impose the same line width for a given velocity component in each tracer and ignore line width differences due to mean molecular weight. The lowest value of $1$\kms\ is motivated by the typical turbulence measured for interstellar clouds (e.g., \citealt{Welty1994a,Welty1996a}).  
\item The \textit{minimum component separation} is set to $0$\kms (components with same velocity but potentially different line width), $1$\kms, and $2$\kms\ (larger values resulting in unsatisfactory fits). 
\end{itemize}

\begin{figure}
  \includegraphics[width=8.5cm,height=21.5cm,clip,trim=0 0 0 00]{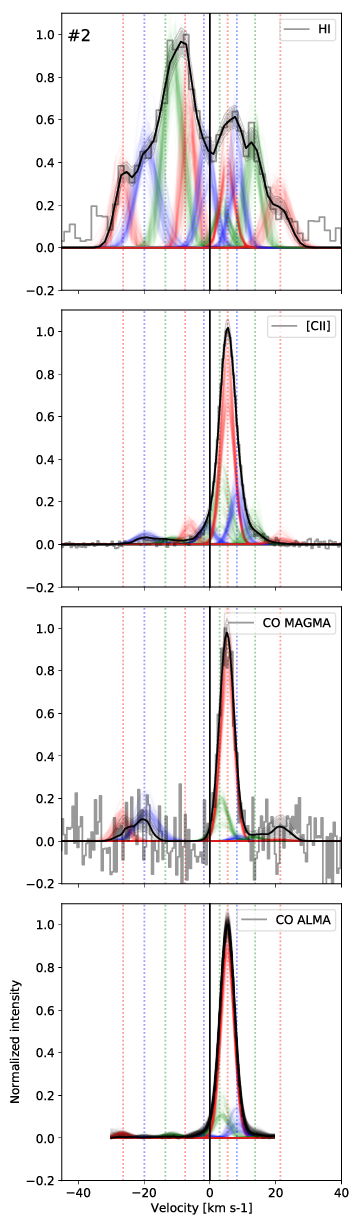}
\caption{Illustration of the profile decomposition method for pointing \#2 for one of the model input parameter sets ($N_{\rm c}=10$; $\sigma_{\rm min}=1$\kms; $\Delta{v}_{\rm min}=1$\kms). The gray histogram shows the data. The red, green, and dark blue curves correspond to different components and the dotted vertical line shows the velocity for each component. The black solid curve shows the total inferred profile. }\label{fig:decompose_pointing4_illustration}
\end{figure}

We note that the converged solution for the decomposition, like any other approach, is not claimed to represent the actual velocity structure, but we hope to infer statistically representative trends by making use of the dispersion in the Markov chain (related to flux uncertainties in the spectra), by considering different models (i.e., with different input parameters), and by considering the various pointings. The MCMC method draws a sequence of random variables corresponding to each parameter in the decomposition. Therefore, there are some excursions, especially in the burn-in phase (first iterations, or "states", in the chain). Once the burn-in phase is finished, the chain enters a high-probability region where the result has converged. The chain still contains a random distribution though, with excursions according to the uncertainties in the parameter. The standard deviation of the Markov chain therefore provides a probability density function. Figure\,\ref{fig:decompose_pointing4_illustration} shows an illustration of the simultaneous fit of all the components. The profile decomposition for the other pointings can be found in Appendix\,\ref{secapp:profiles}.

\begin{table}
  \caption{Parameters for the decomposition method}\label{tab:bayespar}
  \begin{tabular}{lll}
    \hline
    \hline
    Parameter & Type & Prior/value \\
    \hline
    Number of components ($N_{\rm c}$) & fixed & $10, 15$ \\
    Minimum line width ($\sigma_{\rm min}$) & fixed & $1, 2$\kms \\
    Minimum separation (${\Delta}v_{\rm min}$) & fixed & $0, 1, 2$\kms \\
    \hline
    Intensity (Gaussian area) & free, uniform & $>0$ \\
    Line width & free, half-normal & $>\sigma_{\rm min}$ \\
    Separation between components & free, half-uniform & $>{\Delta}v_{\rm min}$ \\
    \hline
  \end{tabular}
\end{table}

\subsection{Model-independent quantities}\label{sec:indiv_fh2}
 
For each velocity component we compile the [C\2] line intensity, the H$^0$ column density, the H$_2$ column density derived from CO, $N({\rm H_2|CO})$, and the resulting CO-traced molecular gas fraction $f({\rm H_2|CO})=2N({\rm H_2|CO})/(2N({\rm H_2|CO})+N({\rm H}^0))$. The CO-dark H$_2$ gas will be accounted for later with the model (Sect.\,\ref{sec:modelcomps}). 

We wish to stress that beam dilution may affect H\1\ and CO observations differently if the respective beam filling factors are different. The beam filling factor for CO is likely much smaller than for H\1, and as a result the value of $f({\rm H_2|CO})$ should correspond to the beam filling factor of fully molecular clouds embedded in a mostly atomic medium rather than to the actual molecular gas fraction of a single cloud filling the beam.

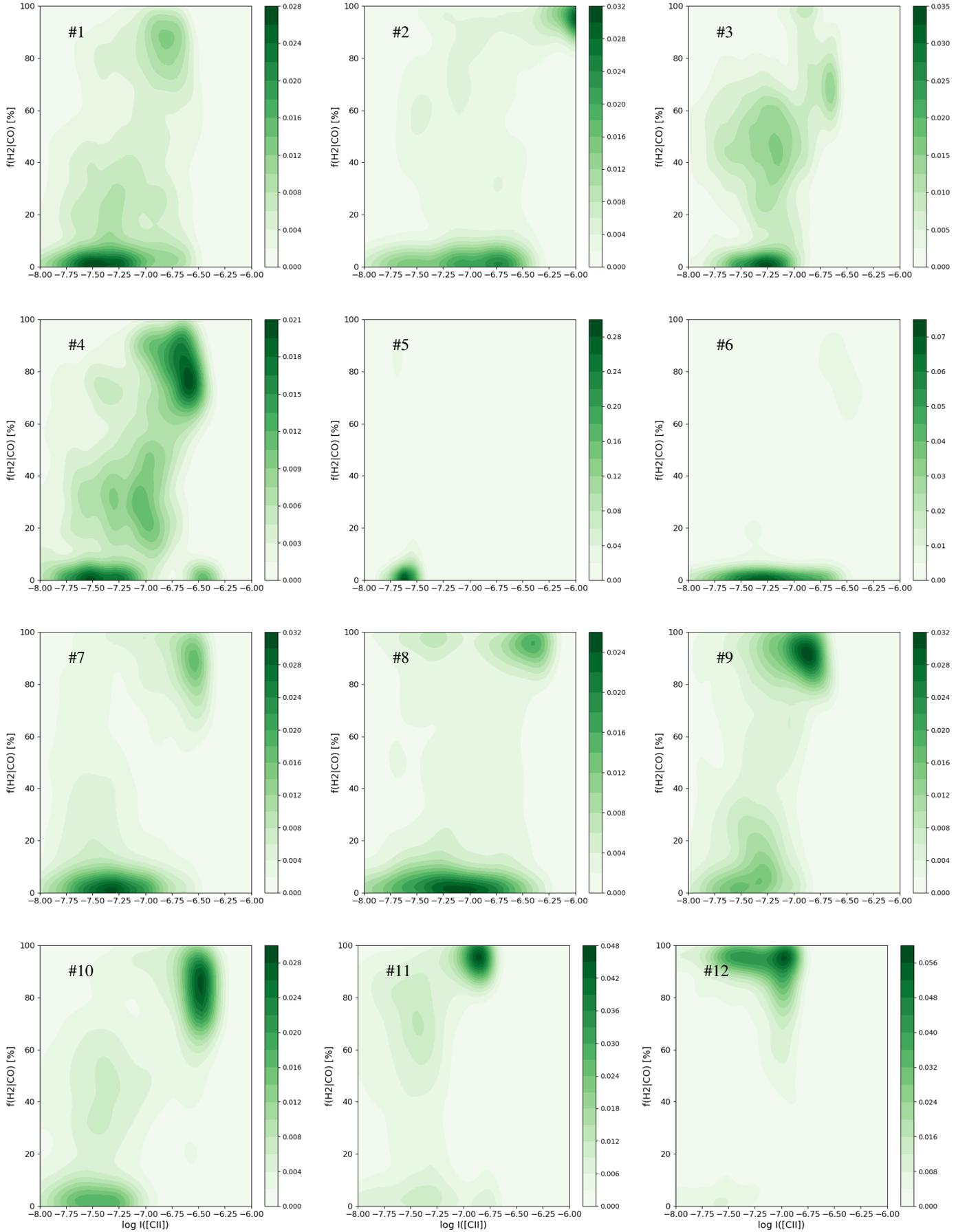
\begin{figure*} 
\begin{tikzpicture}
  \node[anchor=south west,inner sep=0] (image) at (0,0) {\includegraphics[width=6cm,trim=20 20 20 20]{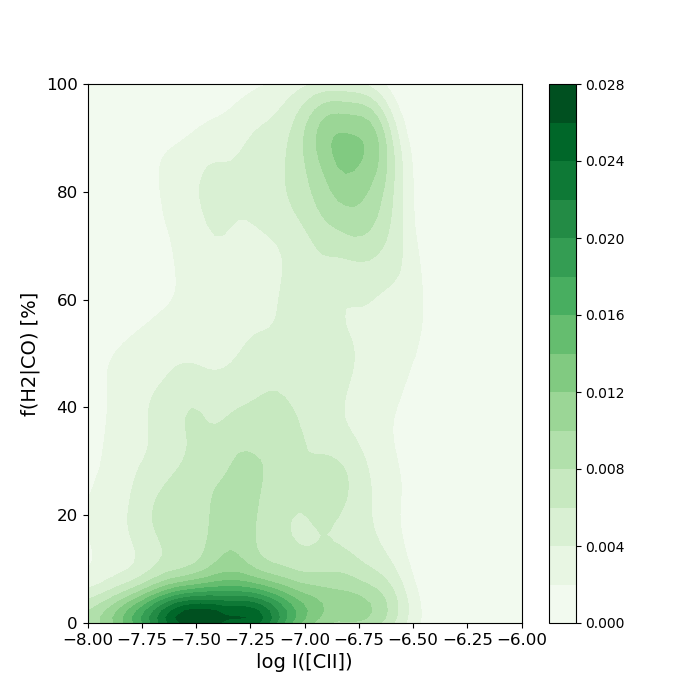}}; \node[right] at (1.,5) {\#1}; 
\end{tikzpicture}
\begin{tikzpicture}
  \node[anchor=south west,inner sep=0] (image) at (0,0) {\includegraphics[width=6cm,trim=20 20 20 20]{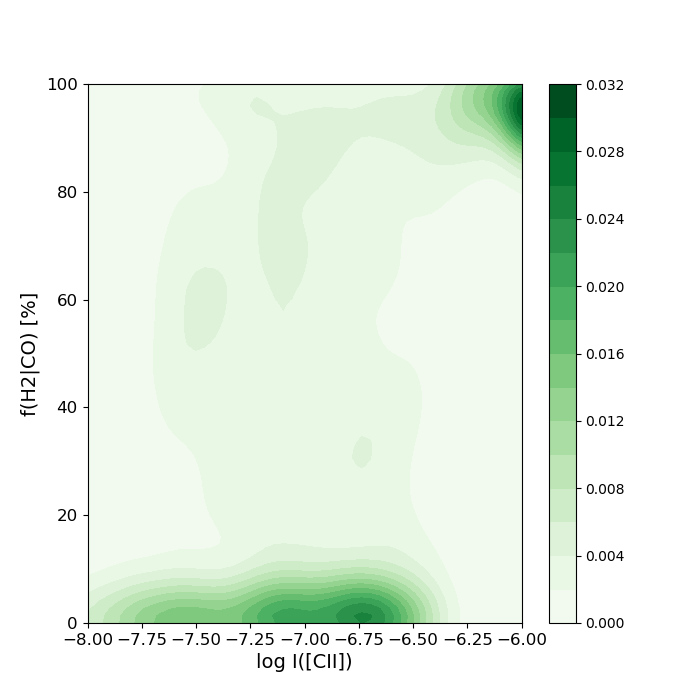}}; \node[right] at (1,5) {\#2}; 
\end{tikzpicture}
\begin{tikzpicture}
  \node[anchor=south west,inner sep=0] (image) at (0,0) {\includegraphics[width=6cm,trim=20 20 20 20]{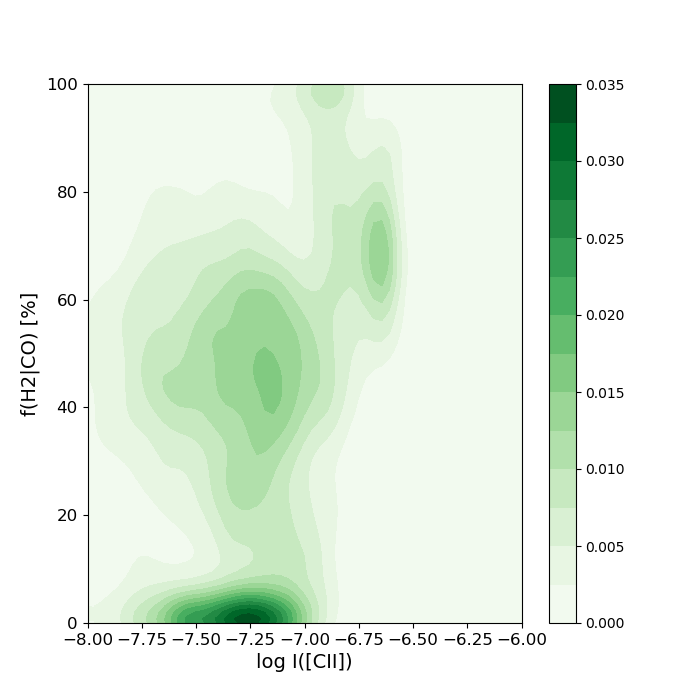}}; \node[right] at (1,5) {\#3}; 
\end{tikzpicture}
\begin{tikzpicture}
  \node[anchor=south west,inner sep=0] (image) at (0,0) {\includegraphics[width=6cm,trim=20 20 20 20]{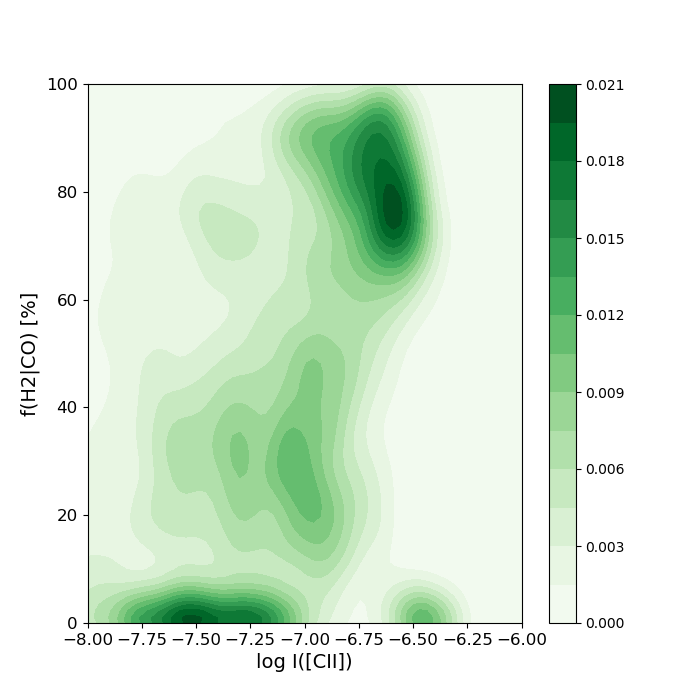}}; \node[right] at (1,5) {\#4};
\end{tikzpicture}
\begin{tikzpicture}
  \node[anchor=south west,inner sep=0] (image) at (0,0) {\includegraphics[width=6cm,trim=20 20 20 20]{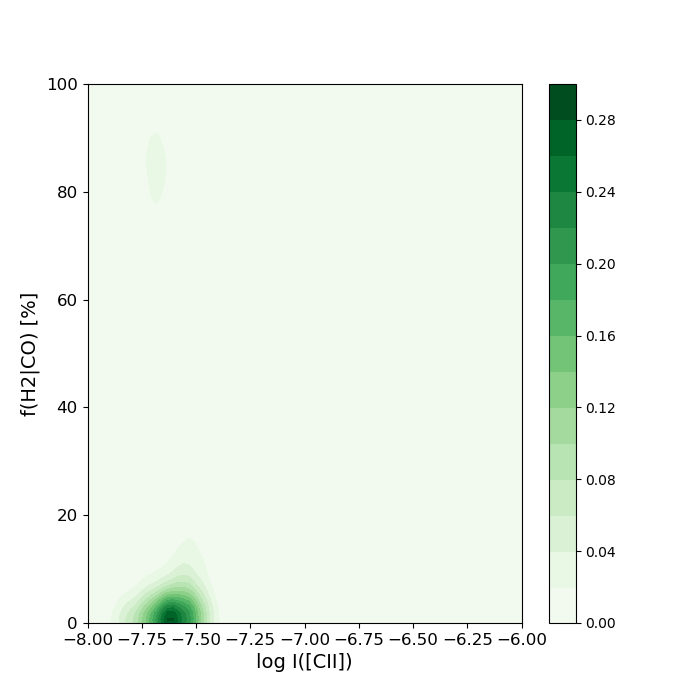}}; \node[right] at (1,5) {\#5};
\end{tikzpicture}
\begin{tikzpicture}
  \node[anchor=south west,inner sep=0] (image) at (0,0) {\includegraphics[width=6cm,trim=20 20 20 20]{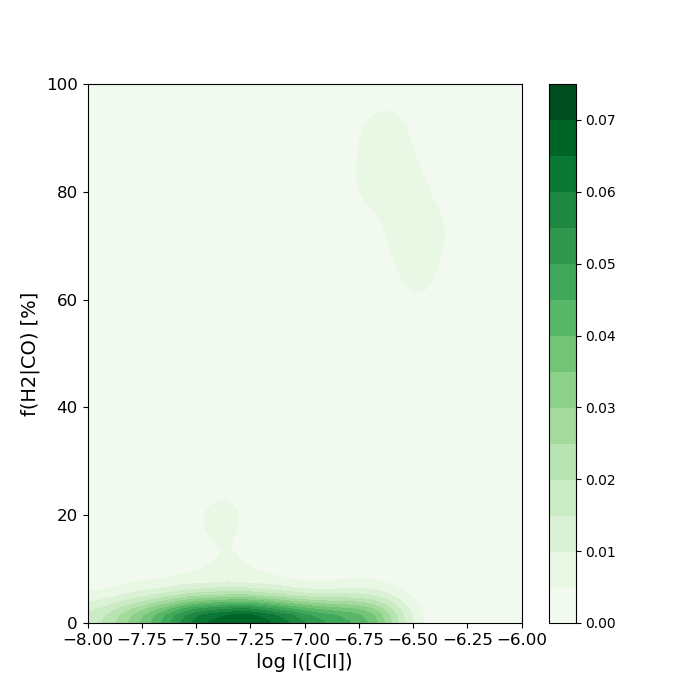}}; \node[right] at (1,5) {\#6}; 
\end{tikzpicture}
\begin{tikzpicture}
  \node[anchor=south west,inner sep=0] (image) at (0,0) {\includegraphics[width=6cm,trim=20 20 20 20]{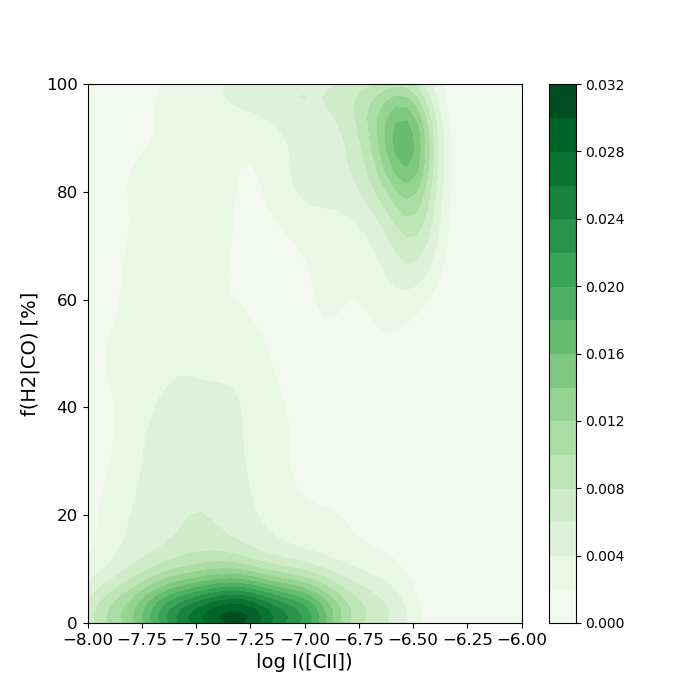}}; \node[right] at (1,5) {\#7}; 
\end{tikzpicture}
\begin{tikzpicture}
  \node[anchor=south west,inner sep=0] (image) at (0,0) {\includegraphics[width=6cm,trim=20 20 20 20]{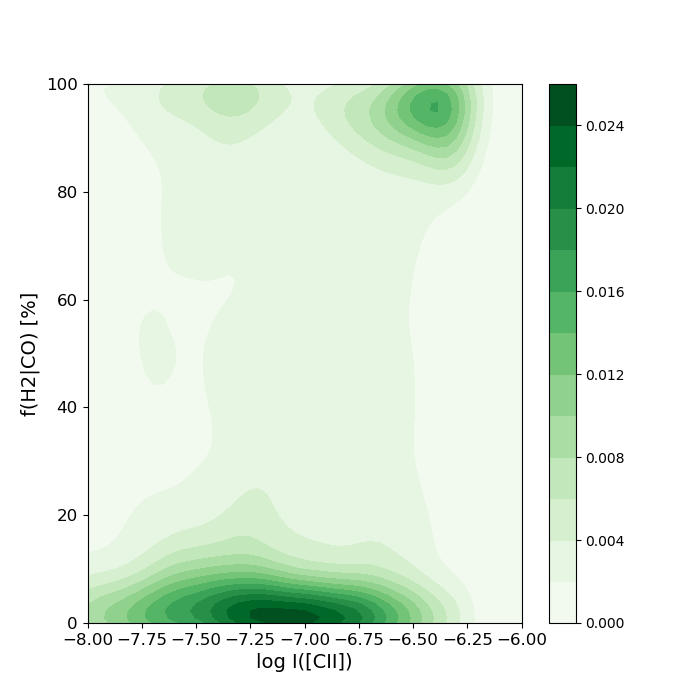}}; \node[right] at (1,5) {\#8}; 
\end{tikzpicture}
\begin{tikzpicture}
  \node[anchor=south west,inner sep=0] (image) at (0,0) {\includegraphics[width=6cm,trim=20 20 20 20]{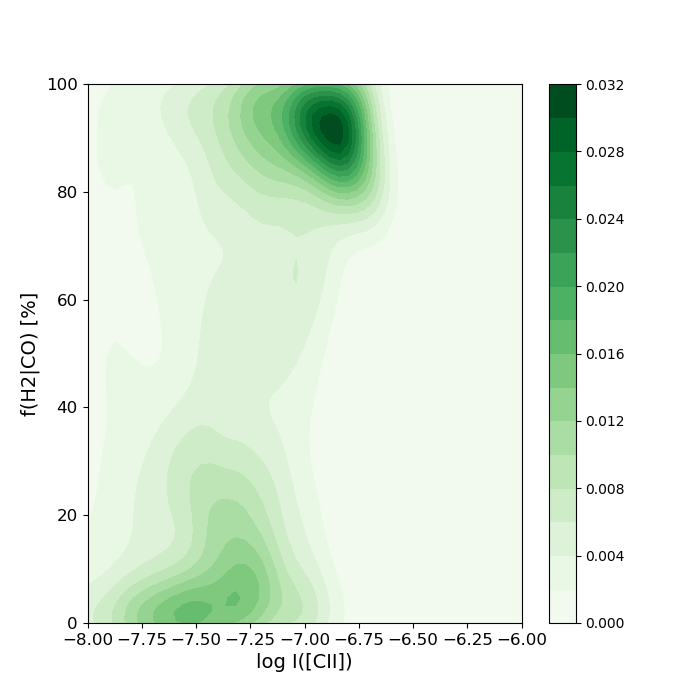}}; \node[right] at (1,5) {\#9}; 
\end{tikzpicture}
\begin{tikzpicture}
  \node[anchor=south west,inner sep=0] (image) at (0,0) {\includegraphics[width=6cm,trim=20 20 20 20]{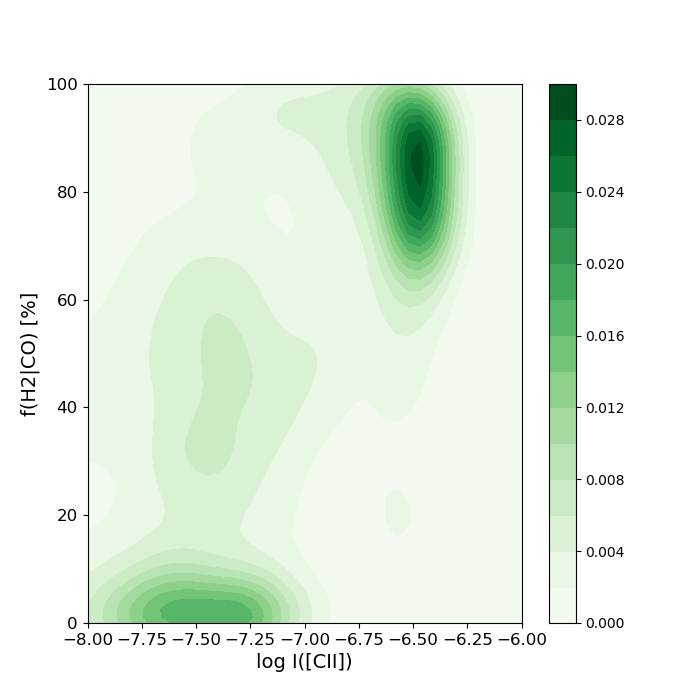}}; \node[right] at (1,5) {\#10}; 
\end{tikzpicture}
\begin{tikzpicture}
  \node[anchor=south west,inner sep=0] (image) at (0,0) {\includegraphics[width=6cm,trim=20 20 20 20]{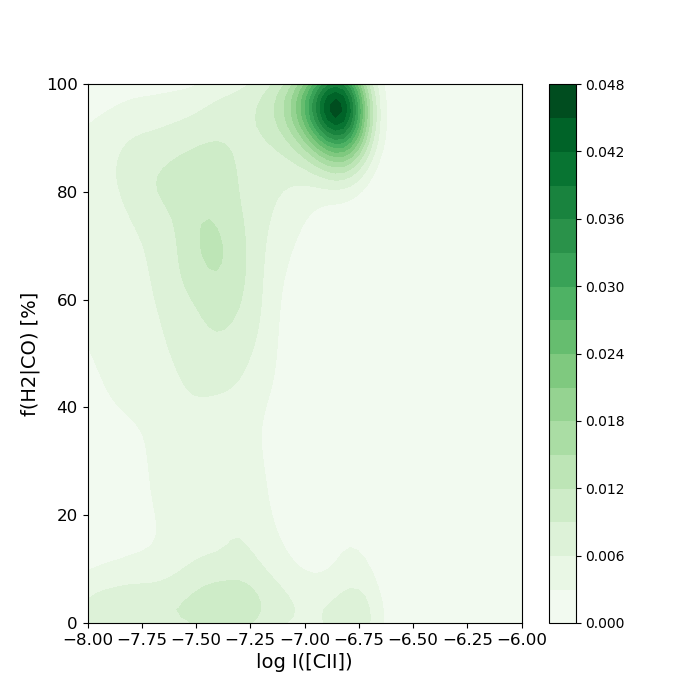}}; \node[right] at (1,5) {\#11}; 
\end{tikzpicture}
\begin{tikzpicture}
  \node[anchor=south west,inner sep=0] (image) at (0,0) {\includegraphics[width=6cm,trim=20 20 20 20]{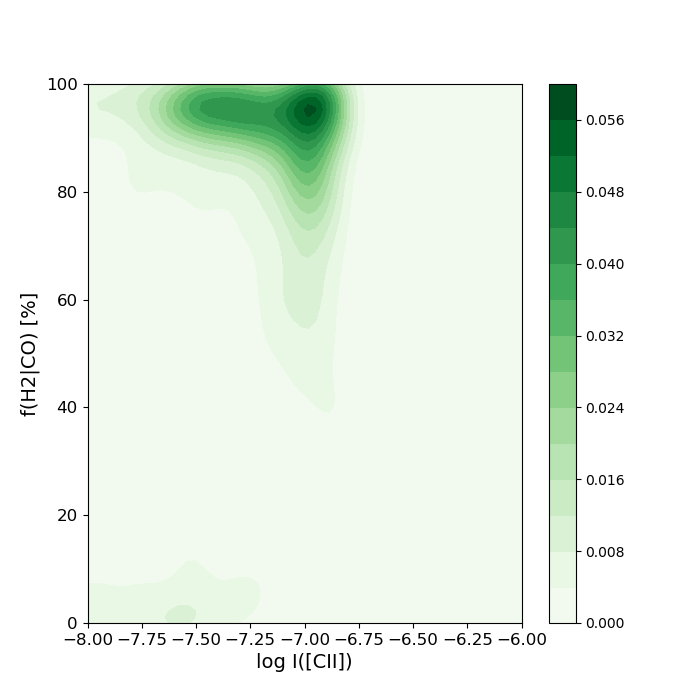}}; \node[right] at (1,5) {\#12};
\end{tikzpicture}
\caption{Bivariate kernel density estimate (non-parametric probability density function) of $I({\rm [CII]})$ vs.\ the molecular gas fraction ignoring CO-dark H$_2$ gas, $f({\rm H_2|CO})$, for all pointings. The shade scales with the density of points.  }\label{fig:resifco_each}
\end{figure*}

The decomposition results are illustrated in Figure\,\ref{fig:resifco_each}, where we show the bivariate kernel density estimate (non-parametric probability density function) of the [C\2] emission and $f({\rm H_2|CO})$. The kernel density estimation makes use of the data cloud corresponding to every velocity component of every pointing for all the elements in the Markov chain, and for all model input parameters (number of components, minimum component line width, and minimum component separation). Hence the kernel density estimate conveniently indicates whether different solutions to the profile decomposition cluster around similar locii in the parameter space.
Most of the components in Figure\,\ref{fig:resifco_each} seem to lie either at $f({\rm H_2|CO})\lesssim10\%$ or $f({\rm H_2|CO})\gtrsim60\%$, suggesting a sharp transition between CO-bright H$_2$ gas and either CO-dark H$_2$ or atomic gas. Components with a low $f({\rm H_2|CO})$ are our best candidates for evidence of significant CO-dark H$_2$ gas amount. 

\subsection{CO-dark gas properties}\label{sec:modelcomps}

\subsubsection{Selection of components}\label{sec:selcomps}

Our models account for the neutral gas only (atomic and molecular). The ionized gas contribution to the integrated line emission is negligible but its contribution to specific velocity components, in particular faint ones, is more difficult to assess (Sect.\,\ref{sec:ciiorigin}).
An upper limit can be calculated for [C\2] in the ionized gas for the pointings with velocity-resolved [N\2]\,$205$\mic\ observations (Sect.\,\ref{sec:ciiorigin_nii}; Fig.\,\ref{fig:decomposition_simple_ciinii}). In the following, we ignore all the [C\2] components below $2\times10^{-8}$\,W\,m$^{-2}$\,sr$^{-1}$ (corresponding to $\approx2\sigma$) to ensure we select [C\2] components arising in the neutral gas. Although such [C\2] components are faint, and although they do not contribute much to the kernel density estimate plots, we prefer to ignore them for clarity and robustness of the final results. This threshold also prevents over-interpretation of potential [C\2] upper limits corresponding to H\1-identified components that do not contribute to the GREAT beam.

\subsubsection{Modeling strategy}\label{sec:modelstrat}

While the fraction of [C\2] associated with the atomic or molecular phase could be derived from the general profile shapes (e.g., \citealt{Okada2019a}), we choose here a different approach making use of the CO and H$^0$ column densities derived for {each} velocity component. Since we lack additional constraints (e.g., [O\1], [C\1], or the infrared luminosity) for the individual velocity components, we cannot derive the physical parameters from PDR models such as the impinging radiation field intensity or the cloud extinction for each individual component. Indirect measurements of such parameters making use of 2D projected quantities are themselves subject to caution (see \citealt{Seifried2019a} for the extinction). Here we rely on a simple agnostic approach to calculate the level population of C$^+$ accounting for collisions with H$^0$, H$_2$, and $e^-$ as a function of gas temperature and density (see details in \citealt{Lebouteiller2013a}). The clouds that are identified thanks to the velocity decomposition methods are expected to contain both atomic gas, molecular gas traced by CO, and molecular gas traced by C$^+$. For a given velocity component, the CO-dark H$_2$ gas may be related to a CO clump in which case the cloud emits both in [C\2] and CO. The CO-bright H$_2$ gas is not considered in the model because we only calculate the properties of [C\2]-emitting gas, and as such, the H$_2$ column density associated with CO is simply calculated using a fiducial $X'_{\rm CO}$ factor (Sect.\,\ref{sec:obs_co}). 

\begin{figure}[h]
\includegraphics[width=9cm,clip]{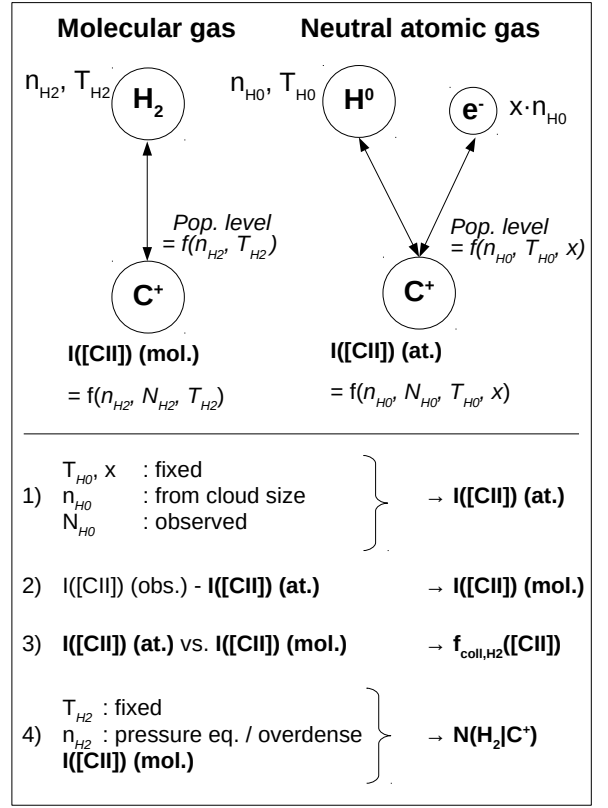}
\caption{Illustration of the model strategy. Two phases are considered for which the level population of C$^+$ is calculated as a function of density $n$, temperature $T$, and the ionization fraction $x$ in the atomic medium. Collisional rates with each partner also depend on temperature. The [C\2] intensity is calculated using the level population and the column density of H$^0$ or H$_2$ and the fraction of C into C$^+$ in each phase (see \citealt{Lebouteiller2013a} for details).  }\label{fig:modelstrat}
\end{figure} 

The strategy goes as follows (illustrated in Figure\,\ref{fig:modelstrat}):
\begin{itemize}
\item Step 1: For each component we calculate a range for the atomic gas number density and the [C\2] emission is estimated in the atomic phase using the atomic gas number density and column density.
\item Step 2: The [C\2] intensity from the molecular gas is inferred from the difference of the estimated [C\2] emission in the atomic phase with the observed [C\2].  
\item Step 3: We compare the [C\2] emission in each phase to compute $f_{\rm coll,H2}({\rm [CII]})$, which is the fraction of [C\2] associated with gas where collisions with H$_2$ dominate, that is, the CO-dark H$_2$ gas.
\item Step 4: From the [C\2] intensity from the molecular gas, we calculate the H$_2$ column density traced by C$^+$, $N({\rm H}_2|{\rm C}^+)$. 
\end{itemize}

For step (1), the number density $n({\rm H}^0)$ in the atomic gas is calculated using the H$^0$ column density and assuming that the individual cloud size along the line of sight lies in the range $1-10$\,pc\footnote{The number density is calculated as $n=N/L$, where $N$ is the column density and $L$ is the cloud size. }. Constraints on the cloud size are mostly given by the H\1\ morphology and typical cloud sizes observed with ALMA in 30\,Dor \citep{Indebetouw2013a} or N\,11 (this study). The inferred density $n({\rm H}^0)$ therefore ranges from a few \cc\ to $\sim10^3$\cc. The temperature in the atomic phase is fixed assuming that all H\1\ components that can be associated in velocity with [C\2] trace the cold rather than the warm neutral medium (CNM, WNM resp.; see also \citealt{Pineda2013a,Pineda2017a}). Accordingly, we estimate a temperature of $100$\,K from [O\1]\,$63$\mic/[C\2] (App.\,\ref{secapp:oicii}). The resulting pressure is then about $10^{2.5-5}$\,K\cc, which is similar to what is found in predominantly atomic gas within \textit{Herschel} KINGFISH galaxies \citep{HerreraCamus2017a} and to other regions in the LMC \citep{Okada2019a} using far-infrared lines. 
The ionization fraction is fixed to a value of $n_e/n_{\rm H}=10^{-4}$, that is, a typical value for a UV-illuminated diffuse gas in which free electrons are provided by ionization of species with an ionization potential below $13.6$\,eV (i.e., no significant ionization from cosmic rays or X-rays).

For step (3), the main free parameter is the H$_2$ column density traced by C$^+$, $N({\rm H}_2|{\rm C}^+)$. We use two constraints for the molecular component: (i) there is no lower limit on the molecular cloud size, but we use the same upper limit as for the atomic gas emission ($10$\,pc), and (ii) the number density in the molecular phase scales by default with that in the atomic phase. For the second hypothesis, the scaling assumes by default thermal pressure equilibrium. The expected temperature of the CNM and of the CO-dark H$_2$ gas is about $10-100$\,K in Milky Way conditions \citep{Glover2016a,Tang2016a,Seifried2019a}. A somewhat warmer temperature is expected in metal-poor environments due to the larger photoelectric-effect heating efficiency (itself due to the low dust content), to the lack of metal coolants, and to the harder interstellar radiation field (ISRF). We assume in the following a temperature of $50$\,K in the molecular phase but our results are not changed significantly if we take $100$\,K. At thermal pressure equilibrium, the density in the H$_2$ gas is therefore twice larger than in the H$^0$ gas, reaching up to $\approx10^3$\cc, on the low end of values found in \cite{Pineda2017a} for a sample of other LMC star-forming regions. We allow the molecular gas to be denser if no solution can be found with a thermal pressure equilibrium hypothesis. This is motivated by the fact that pressure equilibrium might not always be satisfied (e.g., \citealt{VazquezSemadeni2000a,Scoville2013a}) and that thermal pressure may depend on the radiation field intensity and star-formation activity (e.g., \citealt{Ostriker2010a}). Furthermore, the estimated densities correspond to averages along the line of sight while the density structure of the neutral atomic phase and CO-dark H$_2$ gas may differ significantly. 

\noindent From these results, we can then calculate the following quantities for each velocity component:
\begin{itemize}
\item The total H$_2$ column density,
\begin{equation} 
N({\rm H}_2) = N({\rm H}_2|{\rm C}^+) + N({\rm H}_2|{\rm CO}).
\end{equation}
\item The fraction of CO-dark H$_2$ gas, 
\begin{equation}
f_{\rm dark} = \frac{ N({\rm H}_2|{\rm C}^+) }{ N({\rm H}_2) }.
\end{equation}
\item The mass fraction of molecular gas,
\begin{equation}
f({\rm H}_2) = \frac{ 2 \times N({\rm H}_2) }{ 2 \times N({\rm H}_2) + N({\rm H}^0) }.
\end{equation} 
\end{itemize}

\begin{figure}[t]
\includegraphics[width=9cm,clip]{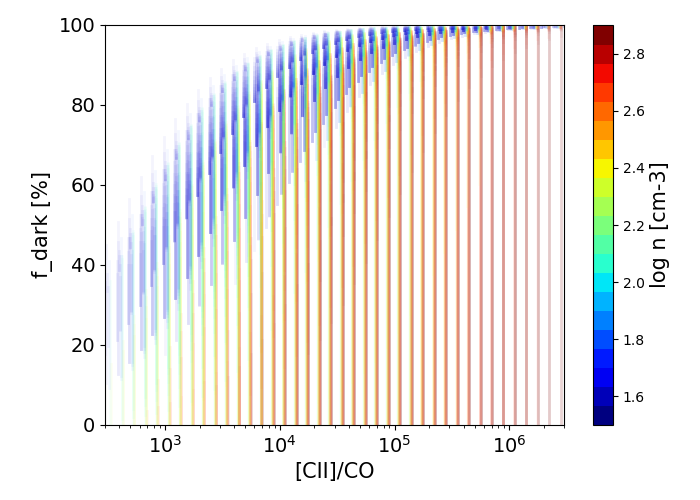}
\caption{Possible model values for $f_{\rm dark}$ as a function of [C\2]/CO and volume density in the neutral atomic phase (color bar). The models shown are those corresponding to the observed ranges for $N({\rm H}^0)=[10^{18}, 10^{21.3}]$\,cm$^{-2}$, $N({\rm H}_2|{\rm CO})=[10^{18}, 10^{21.5}]$\,cm$^{-2}$, and $I([{\rm CII}])=[10^{-8}, 10^{-6}]$\,W\,m$^{-2}$\,sr$^{-1}$. Model results are shown for a temperature in the neutral atomic phase of $100$\,K and $50$\,K in the molecular phase. The striping is due to the incomplete coverage in the model results.  
}\label{fig:model_plots}
\end{figure}

\noindent We note that the mass of CO-dark H$_2$ gas we measure is a lower limit because we consider only the gas traced by C$^+$. In principle however, other tracers with possibly different
properties may also correspond to CO-dark H$_2$  (\citealt{Glover2016a,Clark2019a}). In Figure\,\ref{fig:model_plots} we show the possible model values for $f_{\rm dark}$ as a function of [C\2]/CO for the observed ranges of $N({\rm H}^0)$, $N({\rm H}_2|{\rm CO})$, and $I([{\rm CII}])$. The correlation between $f_{\rm dark}$ and [C\2]/CO is tighter for low densities, corresponding to low H$^0$ column densities (due to the assumed cloud size), and therefore to a low fraction of [C\2] in the neutral atomic gas. For such conditions, the [C\2]/CO ratio is proportional to first order\footnote{The tight relationship between $f_{\rm dark}$ and [C\2]/CO is also due to the assumption that the [C\2] and CO emission originate from a single cloud. The global [C\2]/CO ratio measured for a collection of clouds is not a linear function of the ratios in individual clouds. } to the column density ratio $N({\rm C}^+)/N({\rm CO})$ (e.g., \citealt{Crawford1985a}). As the atomic hydrogen column density increases, so does the possible contribution of the neutral atomic gas to [C\2] (shifting the observed [C\2]/CO to larger values) and so does the number density in the atomic and consequently the molecular gas (leading to lower $f_{\rm dark}$ values). The dependency of $f_{\rm dark}$  with temperature in the atomic gas is relatively small unless the temperature is much larger than $\sim500$\,K. Even then, if we were to choose a temperature of several thousand K (WNM conditions), $f_{\rm dark}$ values would be lower by less than about $10$\%\ in value.

\begin{figure}[t]
\includegraphics[width=9cm,clip]{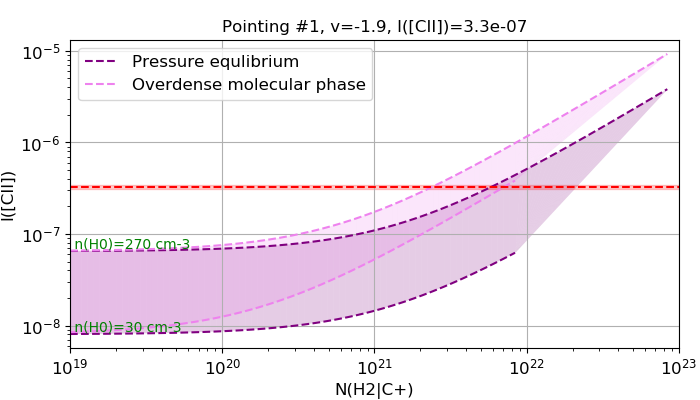}
\includegraphics[width=9cm,clip]{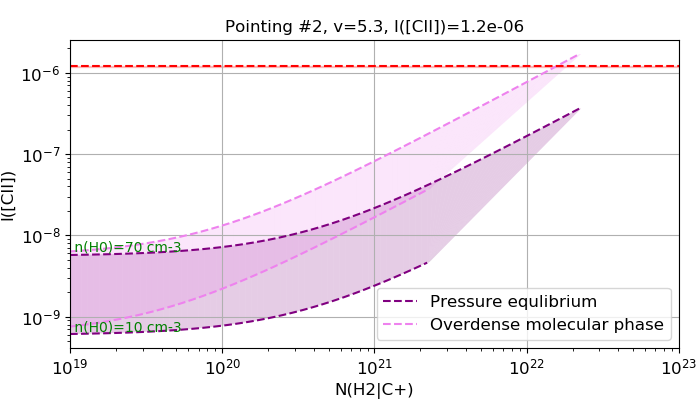} 
\caption{Illustration of model results for one components in pointing \#1 (top) and in \#2 (bottom). The [C\2] surface brightness is calculated as a function of the H$_2$ column density associated with C$^+$. The horizontal red line shows the observed [C\2] surface brightness. The purple and pink shaded areas show the range of model results assuming thermal pressure equilibrium and overdense molecular gas respectively (see text). The bottom and top boundaries are set by the H$^0$ column density together with the cloud size constraint. The right boundary is set by the maximum allowed molecular cloud size. In the top panel the model with thermal pressure equilibrium can reproduce the observations while in the bottom panel, a model with an overdense molecular gas is required. }\label{fig:modelstrat_application}
\end{figure}

\begin{figure*}
\centering

\hfil
\begin{tikzpicture}
  \node[anchor=south west,inner sep=0] (image) at (0,0) {\includegraphics[width=0.35\textwidth]{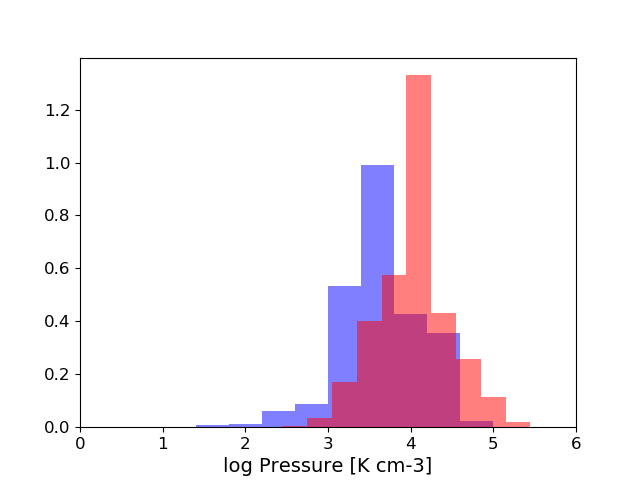}}; \node[right] at (1.05,3.4) {\#1};
\end{tikzpicture}
\hfil
\begin{tikzpicture}
  \node[anchor=south west,inner sep=0] (image) at (0,0) {\includegraphics[width=0.31\textwidth]{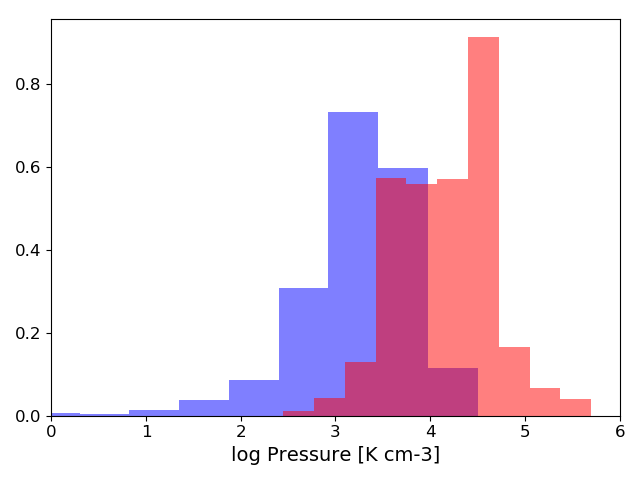}}; \node[right] at (1.05,3.4) {\#2};
  \end{tikzpicture}
\hfil
\begin{tikzpicture}
  \node[anchor=south west,inner sep=0] (image) at (0,0) {\includegraphics[width=0.31\textwidth]{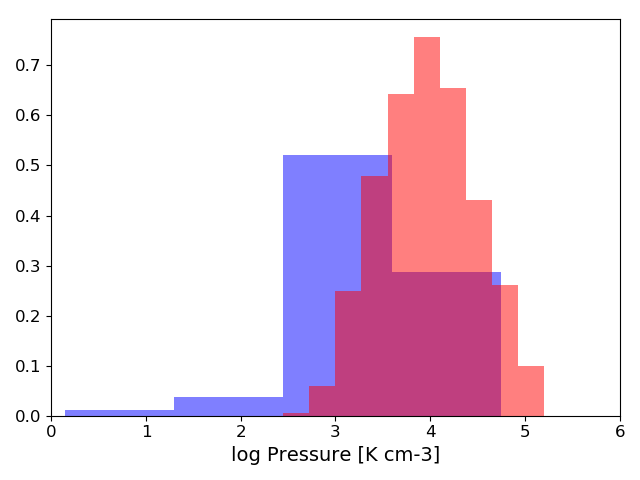}}; \node[right] at (1.05,3.4) {\#3};
  \end{tikzpicture}

\hfil
\begin{tikzpicture}
  \node[anchor=south west,inner sep=0] (image) at (0,0) {\includegraphics[width=0.31\textwidth]{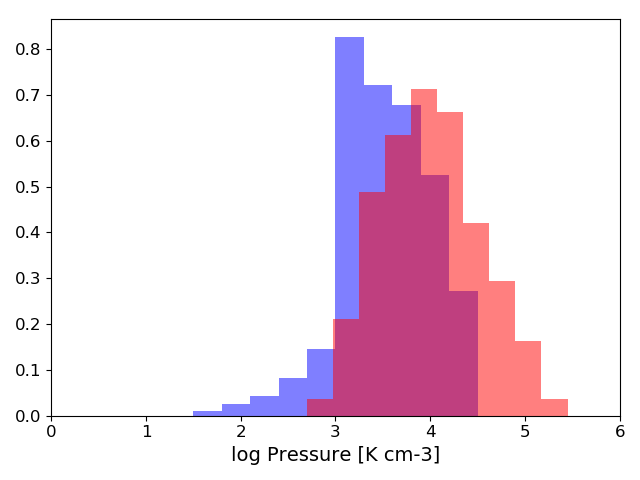}}; \node[right] at (1.05,3.4) {\#4};
  \end{tikzpicture}
\hfil
\begin{tikzpicture}
  \node[anchor=south west,inner sep=0] (image) at (0,0) {\includegraphics[width=0.31\textwidth]{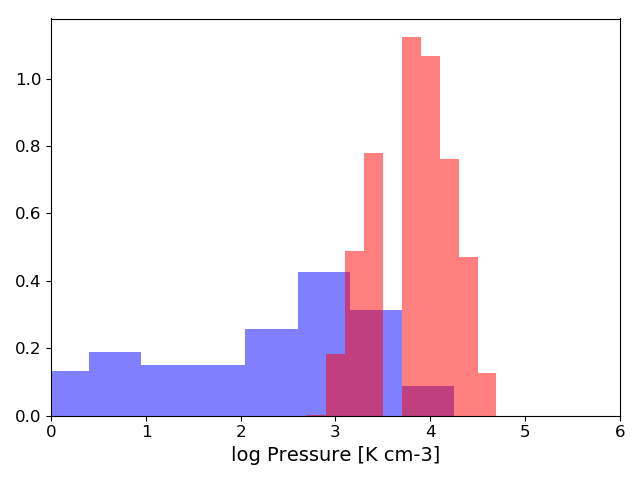}}; \node[right] at (1.05,3.4) {\#5};
  \end{tikzpicture}
\hfil
\begin{tikzpicture}
  \node[anchor=south west,inner sep=0] (image) at (0,0) {\includegraphics[width=0.31\textwidth]{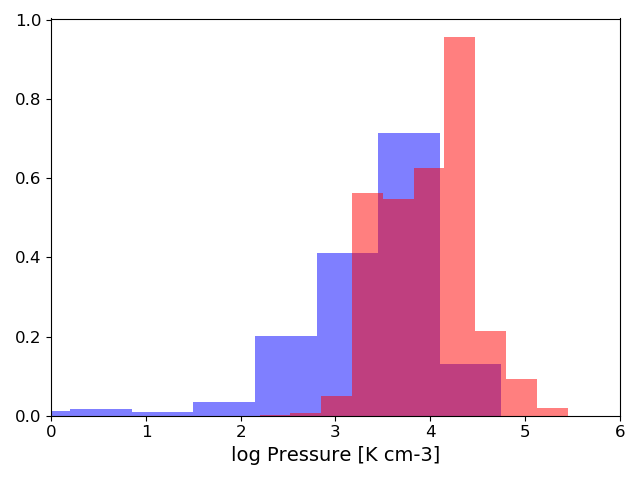}}; \node[right] at (1.05,3.4) {\#6};
  \end{tikzpicture}

\hfil
\begin{tikzpicture}
  \node[anchor=south west,inner sep=0] (image) at (0,0) {\includegraphics[width=0.31\textwidth]{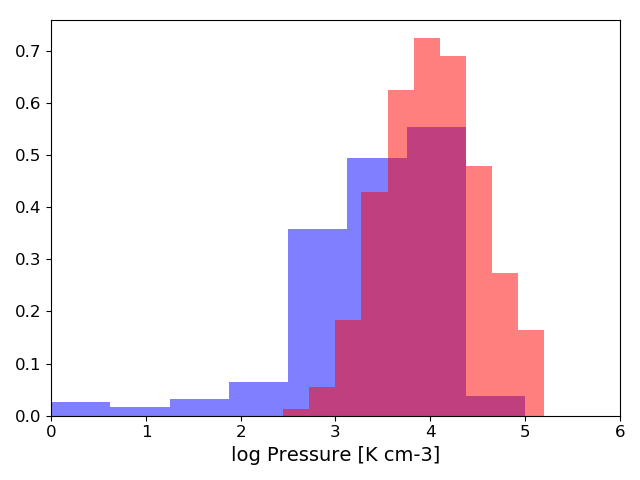}}; \node[right] at (1.05,3.4) {\#7};
  \end{tikzpicture}
\hfil
\begin{tikzpicture}
  \node[anchor=south west,inner sep=0] (image) at (0,0) {\includegraphics[width=0.31\textwidth]{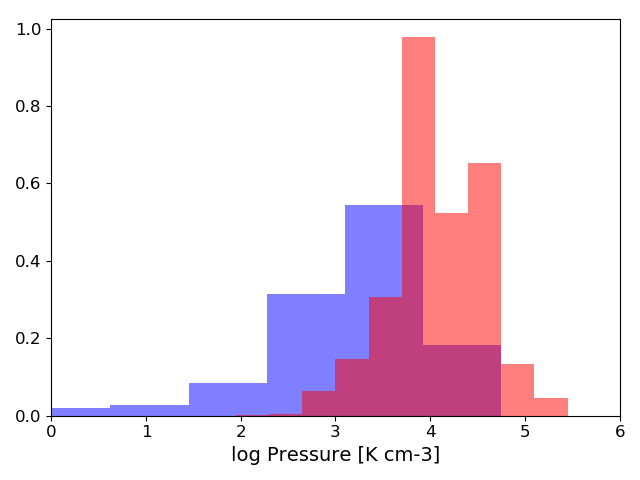}}; \node[right] at (1.05,3.4) {\#8};
  \end{tikzpicture}
\hfil
\begin{tikzpicture}
  \node[anchor=south west,inner sep=0] (image) at (0,0) {\includegraphics[width=0.31\textwidth]{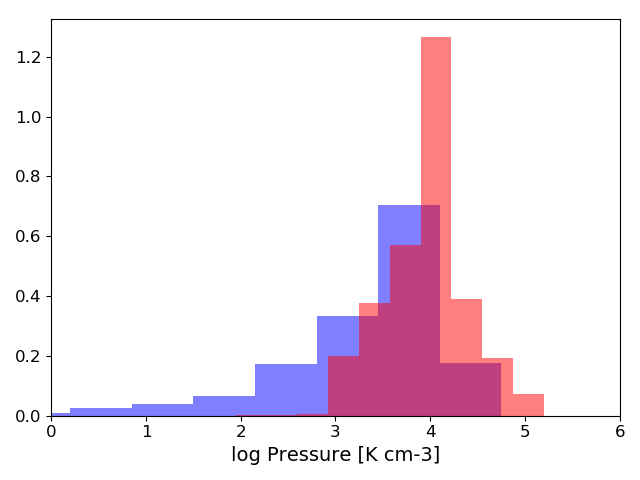}}; \node[right] at (1.05,3.4) {\#9};
  \end{tikzpicture}

\hfil
\begin{tikzpicture}
  \node[anchor=south west,inner sep=0] (image) at (0,0) {\includegraphics[width=0.31\textwidth]{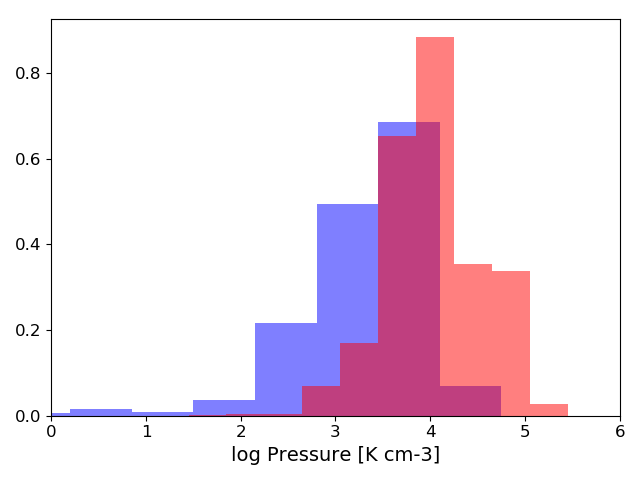}}; \node[right] at (1.05,3.4) {\#10};
  \end{tikzpicture}
\hfil
\begin{tikzpicture}
  \node[anchor=south west,inner sep=0] (image) at (0,0) {\includegraphics[width=0.31\textwidth]{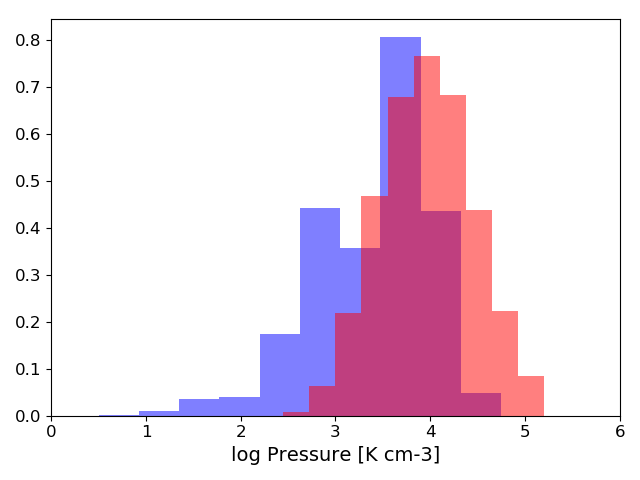}}; \node[right] at (1.05,3.4) {\#11};
  \end{tikzpicture}
\hfil
\begin{tikzpicture}
  \node[anchor=south west,inner sep=0] (image) at (0,0) {\includegraphics[width=0.31\textwidth]{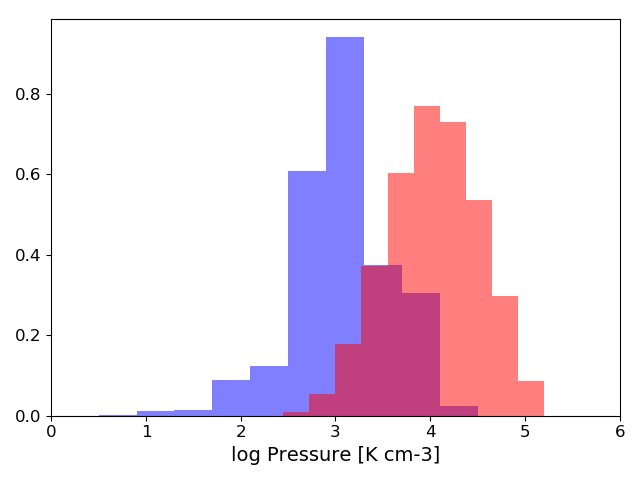}}; \node[right] at (1.05,3.4) {\#12};
  \end{tikzpicture}
\caption{Histogram of the modeled pressure in the neutral atomic (blue) and CO-dark H$_2$ gas (red) for all the velocity components. Results from all decomposition methods are combined (see Table\,\ref{tab:bayespar}). }\label{fig:pressures}
\end{figure*}
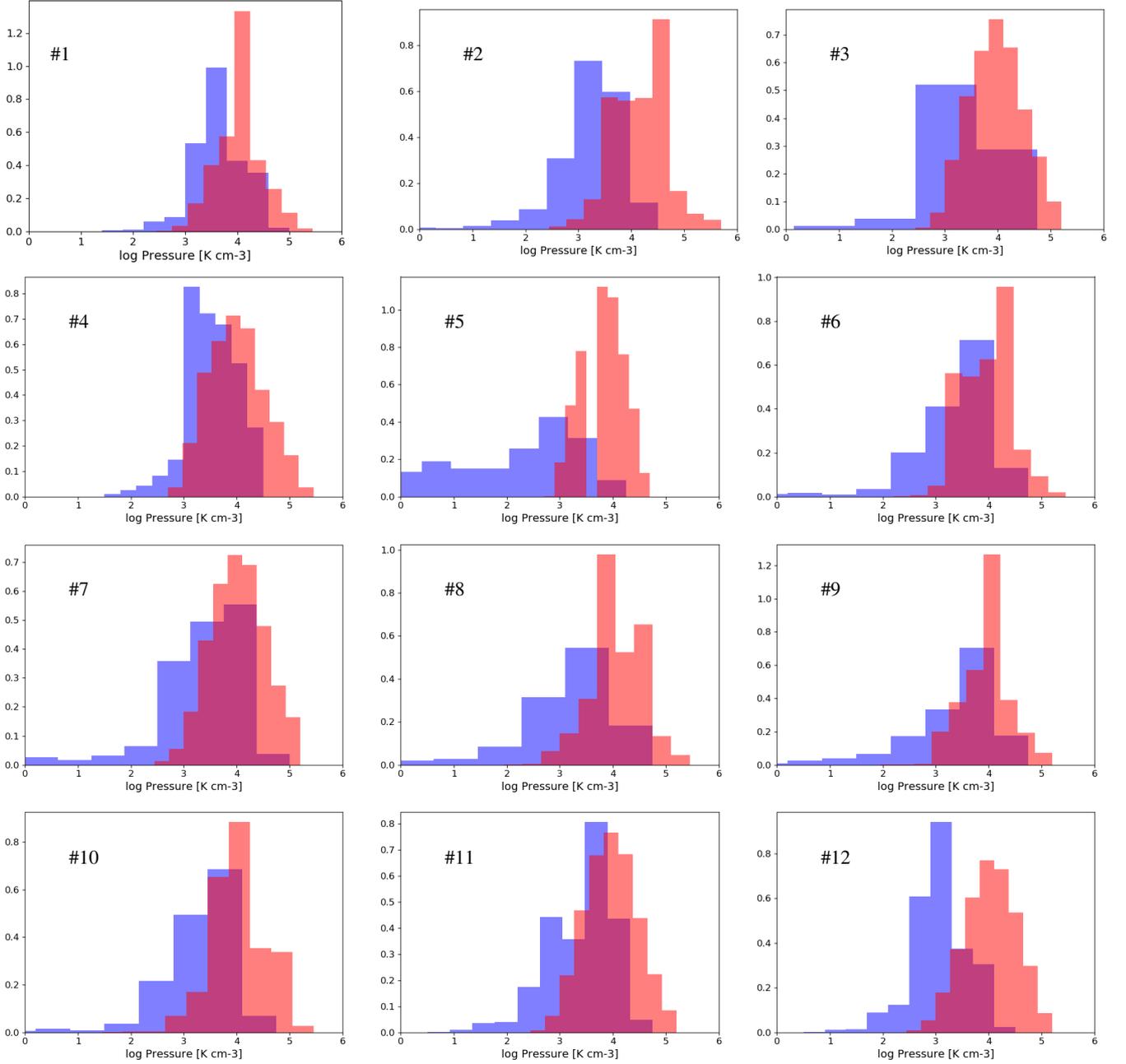

An illustration of the model calculation is shown in Figure\,\ref{fig:modelstrat_application}. In practice, densities in the CO-dark H$_2$ gas lie around $10^{1.5-3.5}$\cc, corresponding to a pressure of $10^{3.5-5}$\,K\cc\ (Fig.\,\ref{fig:pressures}). The total H$_2$ column density $N({\rm H}_2)$ measured this way varies across pointings from $\approx10^{21}$\,cm$^{-2}$ (in \#5) to $\approx2\times10^{22}$\,cm$^{-2}$ (in \#10).

\subsection{Results}

\subsubsection{[C\2] in the neutral atomic gas}\label{sec:ciiorigin2}

\begin{figure*} 
\begin{tikzpicture}
  \node[anchor=south west,inner sep=0] (image) at (0,0) {\includegraphics[width=6cm,trim=20 20 20 20]{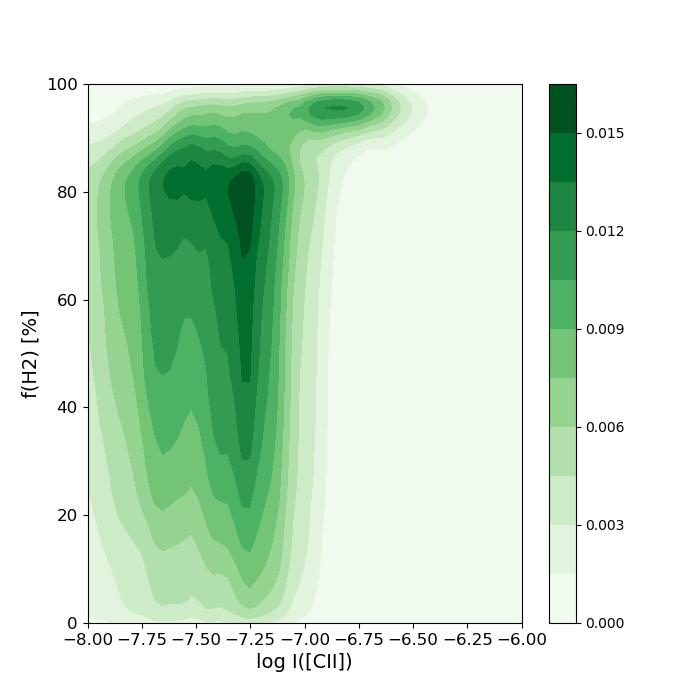}}; \node[right] at (4.,1.5) {\#1}; 
\end{tikzpicture}
\begin{tikzpicture}
  \node[anchor=south west,inner sep=0] (image) at (0,0) {\includegraphics[width=6cm,trim=20 20 20 20]{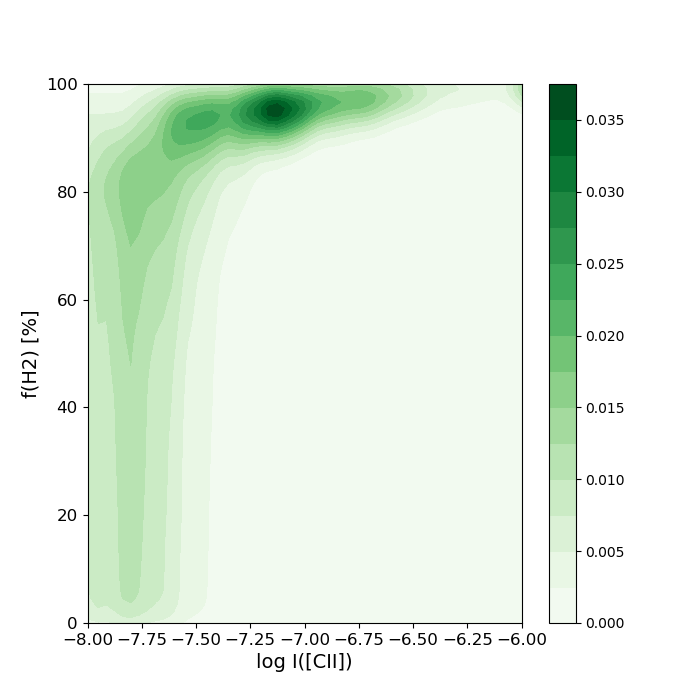}}; \node[right] at (4.,1.5) {\#2}; 
\end{tikzpicture}
\begin{tikzpicture}
  \node[anchor=south west,inner sep=0] (image) at (0,0) {\includegraphics[width=6cm,trim=20 20 20 20]{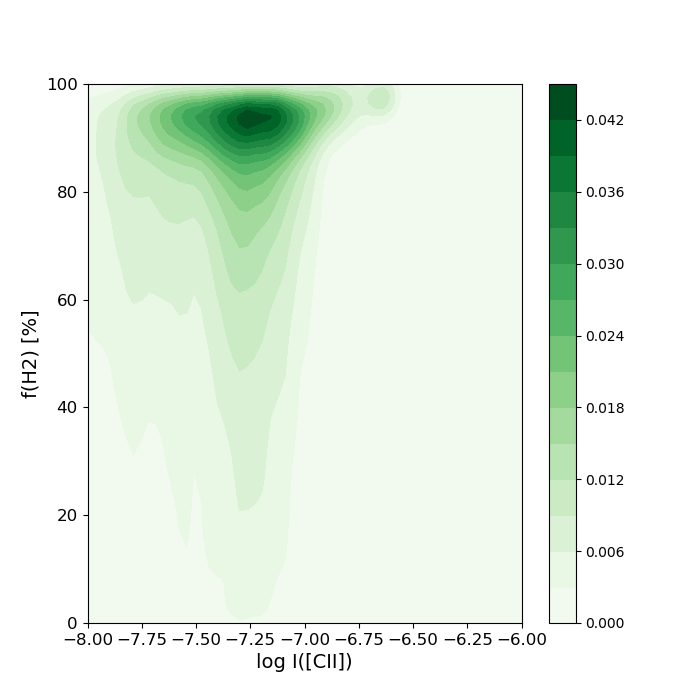}}; \node[right] at (4.,1.5) {\#3}; 
\end{tikzpicture}
\begin{tikzpicture}
  \node[anchor=south west,inner sep=0] (image) at (0,0) {\includegraphics[width=6cm,trim=20 20 20 20]{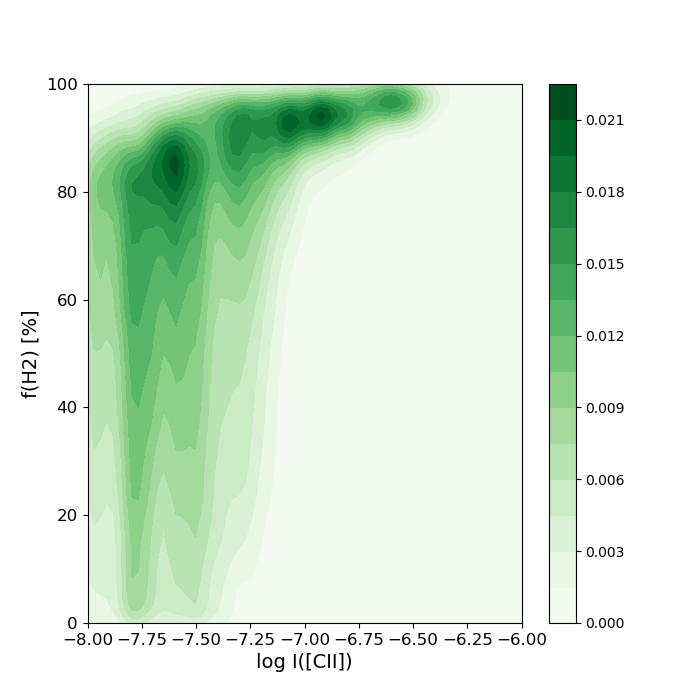}}; \node[right] at (4.,1.5) {\#4};
\end{tikzpicture}
\begin{tikzpicture}
  \node[anchor=south west,inner sep=0] (image) at (0,0) {\includegraphics[width=6cm,trim=20 20 20 20]{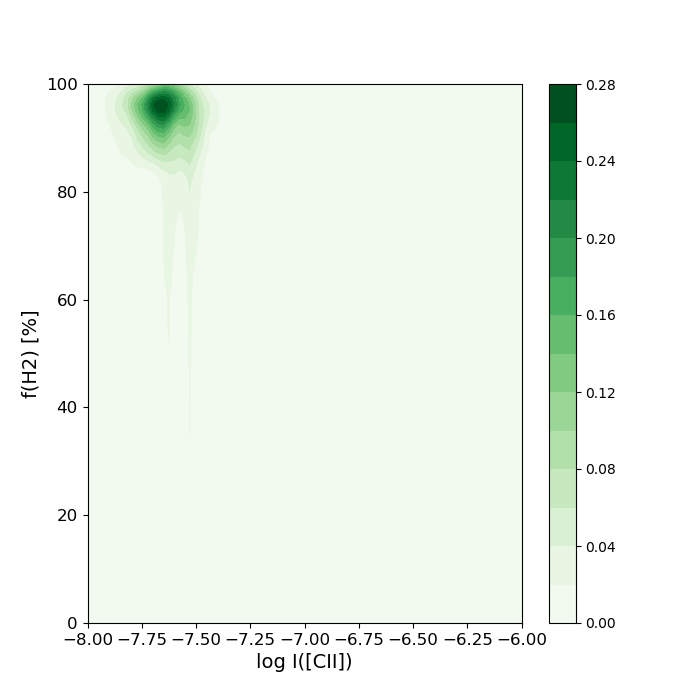}}; \node[right] at (4.,1.5) {\#5};
\end{tikzpicture}
\begin{tikzpicture}
  \node[anchor=south west,inner sep=0] (image) at (0,0) {\includegraphics[width=6cm,trim=20 20 20 20]{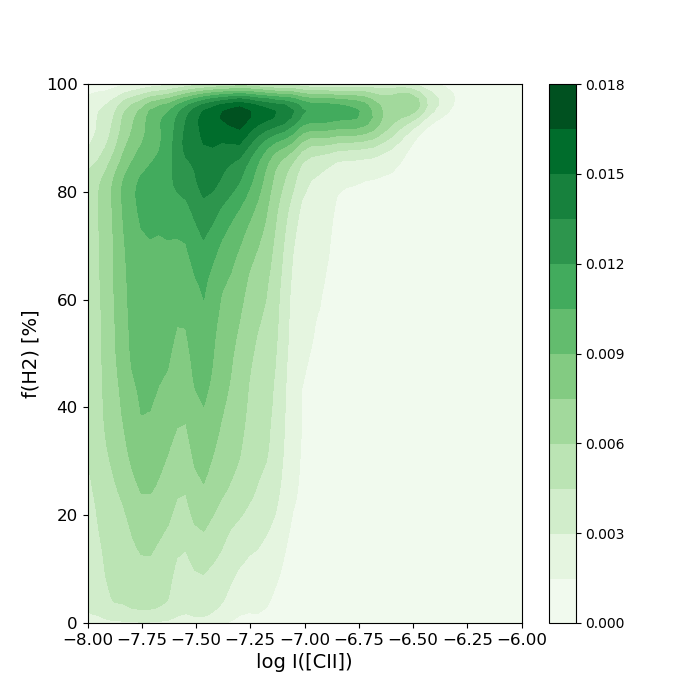}}; \node[right] at (4.,1.5) {\#6}; 
\end{tikzpicture}
\begin{tikzpicture}
  \node[anchor=south west,inner sep=0] (image) at (0,0) {\includegraphics[width=6cm,trim=20 20 20 20]{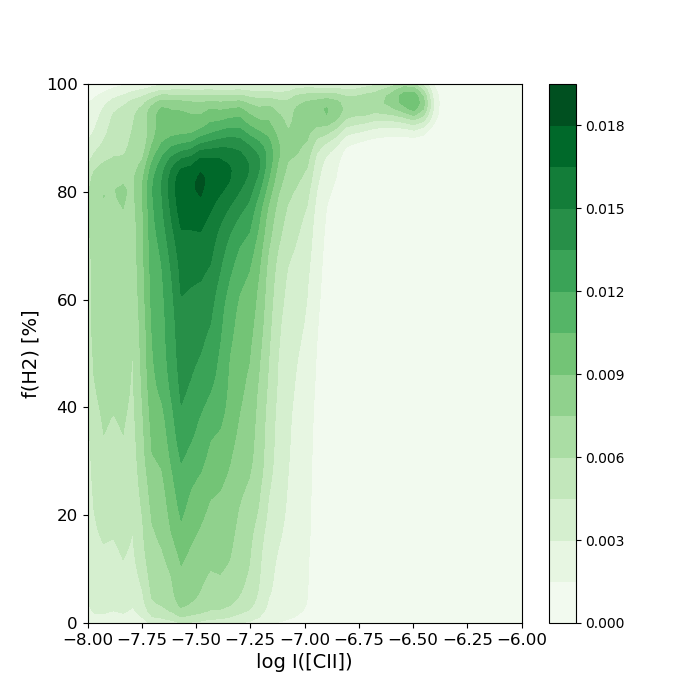}}; \node[right] at (4.,1.5) {\#7}; 
\end{tikzpicture}
\begin{tikzpicture}
  \node[anchor=south west,inner sep=0] (image) at (0,0) {\includegraphics[width=6cm,trim=20 20 20 20]{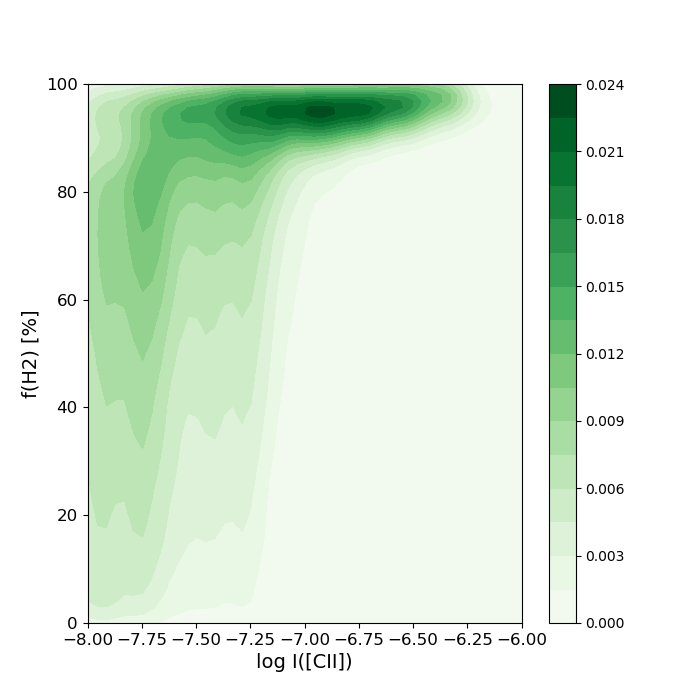}}; \node[right] at (4.,1.5) {\#8}; 
\end{tikzpicture}
\begin{tikzpicture}
  \node[anchor=south west,inner sep=0] (image) at (0,0) {\includegraphics[width=6cm,trim=20 20 20 20]{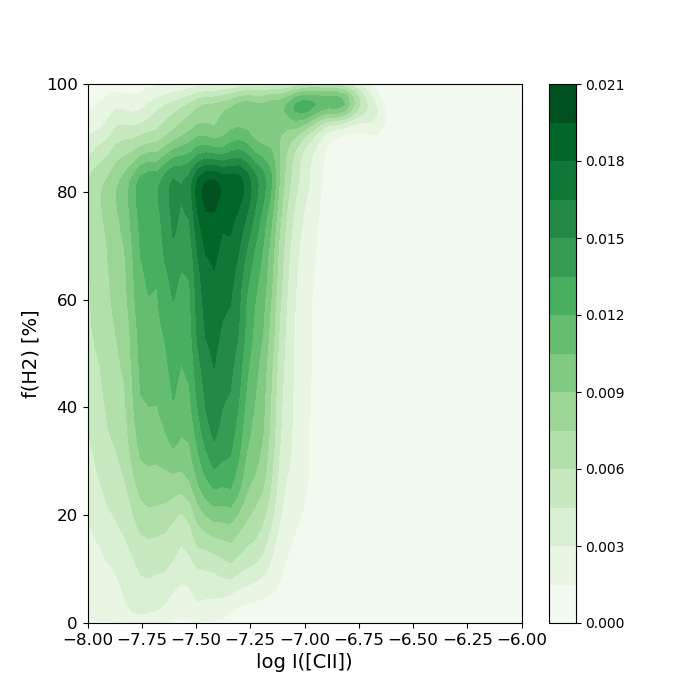}}; \node[right] at (4.,1.5) {\#9}; 
\end{tikzpicture}
\begin{tikzpicture}
  \node[anchor=south west,inner sep=0] (image) at (0,0) {\includegraphics[width=6cm,trim=20 20 20 20]{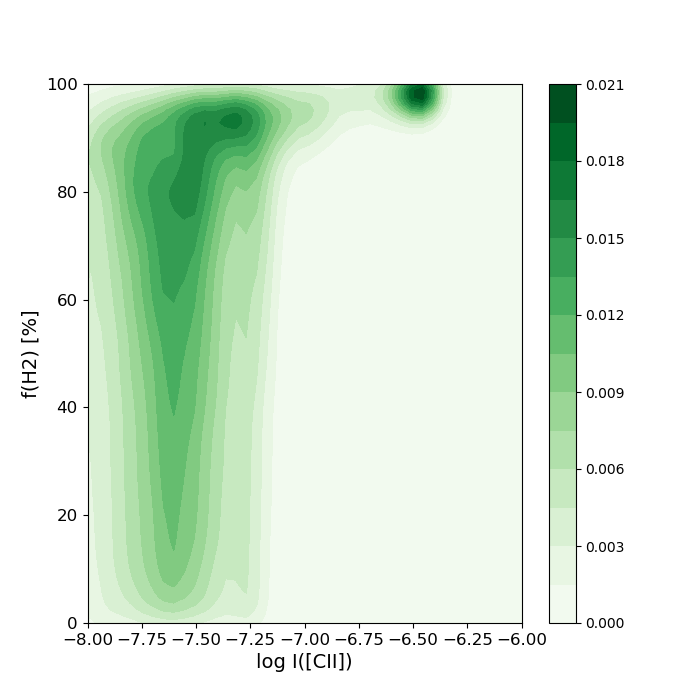}}; \node[right] at (4.,1.5) {\#10}; 
\end{tikzpicture}
\begin{tikzpicture}
  \node[anchor=south west,inner sep=0] (image) at (0,0) {\includegraphics[width=6cm,trim=20 20 20 20]{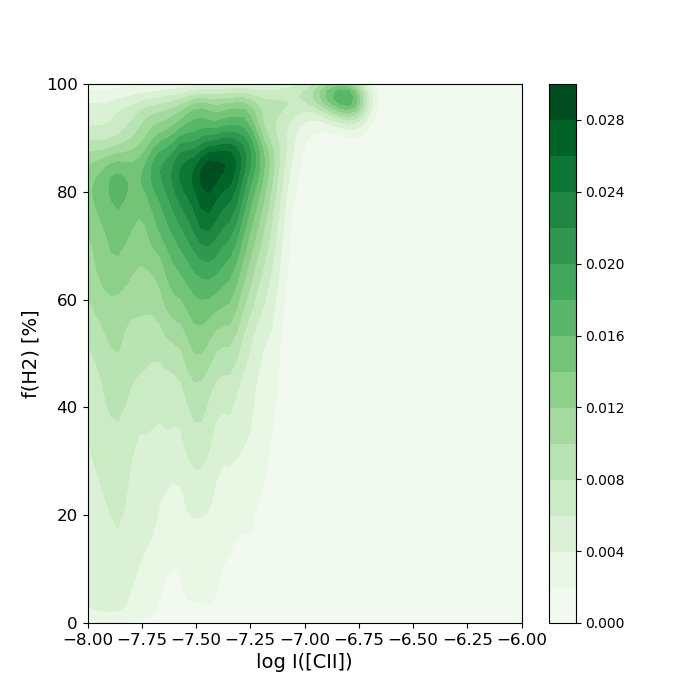}}; \node[right] at (4.,1.5) {\#11}; 
\end{tikzpicture}
\begin{tikzpicture}
  \node[anchor=south west,inner sep=0] (image) at (0,0) {\includegraphics[width=6cm,trim=20 20 20 20]{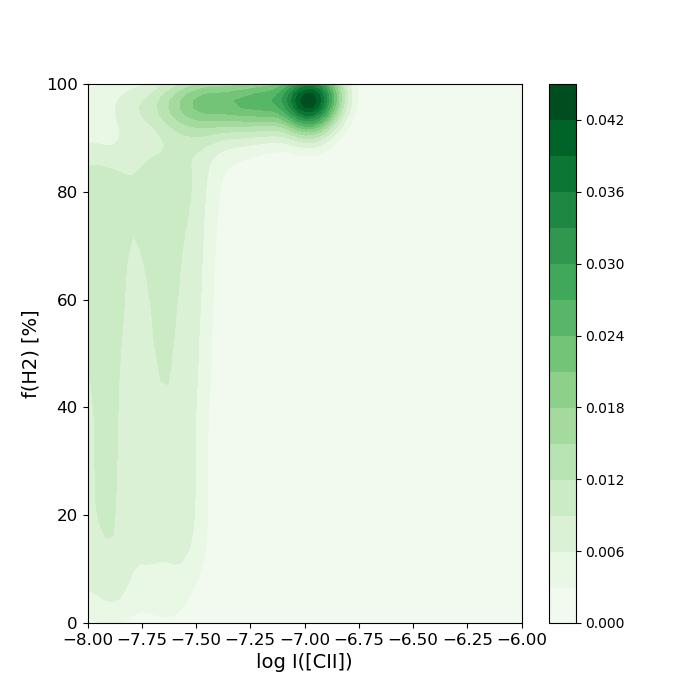}}; \node[right] at (4.,1.5) {\#12};
\end{tikzpicture}
\caption{Same as Fig.\,\ref{fig:resifco_each} but with the total molecular gas fraction (i.e., including the CO-dark H$_2$ gas). The vertical striping is due to the wide ranges for the molecular gas fraction determination for the faintest [C\2] components. The shade scales with the density of points.  }\label{fig:resiftot_each}
\end{figure*}

The [C\2] components with low $f({\rm H}_2|{\rm CO})$ values (Figure\,\ref{fig:resifco_each}) could a priori be either from CO-dark H$_2$ gas or atomic gas. When including the contribution from the CO-dark H$_2$ gas calculated by the models, most velocity components reach a large total (including CO-dark H$_2$) molecular gas fraction $f({\rm H}_2)\gtrsim80\%$ (Fig.\,\ref{fig:resiftot_each}), except for the faintest [C\2] components for which only an upper limit on $f({\rm H}_2)$ is often available. This indicates that [C\2]-bright regions are dominated by CO-dark H$_2$ gas while [C\2]-faint regions may include a significant contribution from neutral atomic gas.

The weak correlation between $f({\rm H}_2)$ and $I({\rm [CII]})$ in Figure\,\ref{fig:resiftot_each} indeed suggests that the contribution of the neutral atomic gas to the [C\2] emission is more significant toward faint [C\2] components, which is confirmed by Figure\,\ref{fig:resfcollh2icii_each} which shows the distribution of the fraction of [C\2] tracing CO-dark H$_2$ gas, $f_{\rm coll,H2}({\rm [CII]})$, versus\ $I([{\rm CII}])$, in particular for \#9 and \#11.

\begin{figure*} 
\begin{tikzpicture}
  \node[anchor=south west,inner sep=0] (image) at (0,0) {\includegraphics[width=6cm,trim=20 20 20 20]{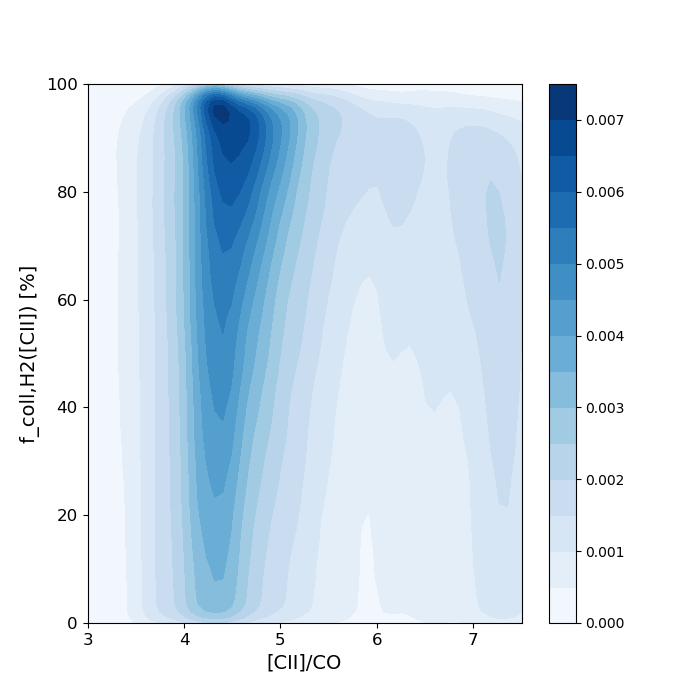}}; \node[right] at (4.,1) {\#1}; 
\end{tikzpicture}
\begin{tikzpicture}
  \node[anchor=south west,inner sep=0] (image) at (0,0) {\includegraphics[width=6cm,trim=20 20 20 20]{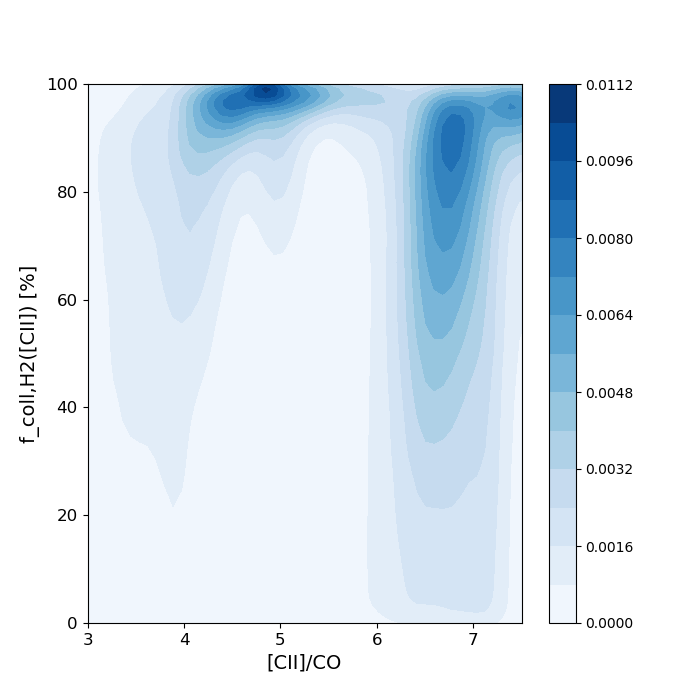}}; \node[right] at (4.,1) {\#2}; 
\end{tikzpicture}
\begin{tikzpicture}
  \node[anchor=south west,inner sep=0] (image) at (0,0) {\includegraphics[width=6cm,trim=20 20 20 20]{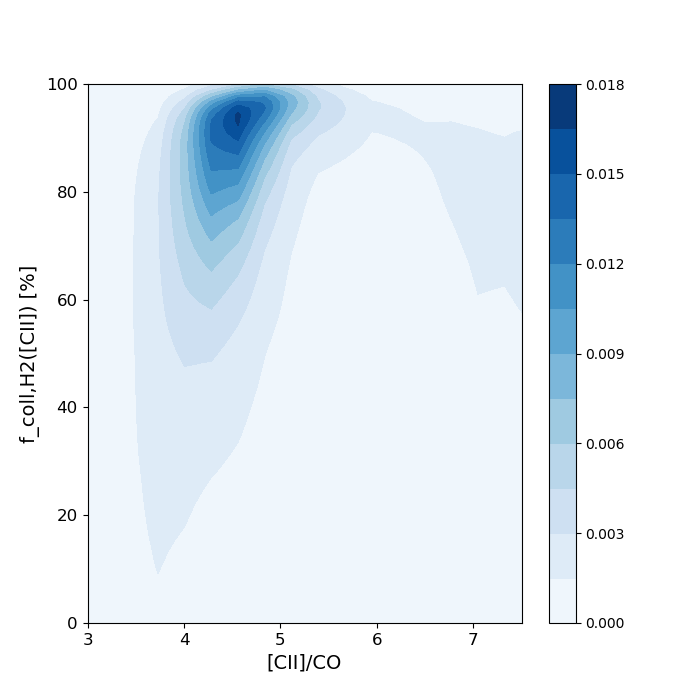}}; \node[right] at (4.,1) {\#3}; 
\end{tikzpicture}
\begin{tikzpicture}
  \node[anchor=south west,inner sep=0] (image) at (0,0) {\includegraphics[width=6cm,trim=20 20 20 20]{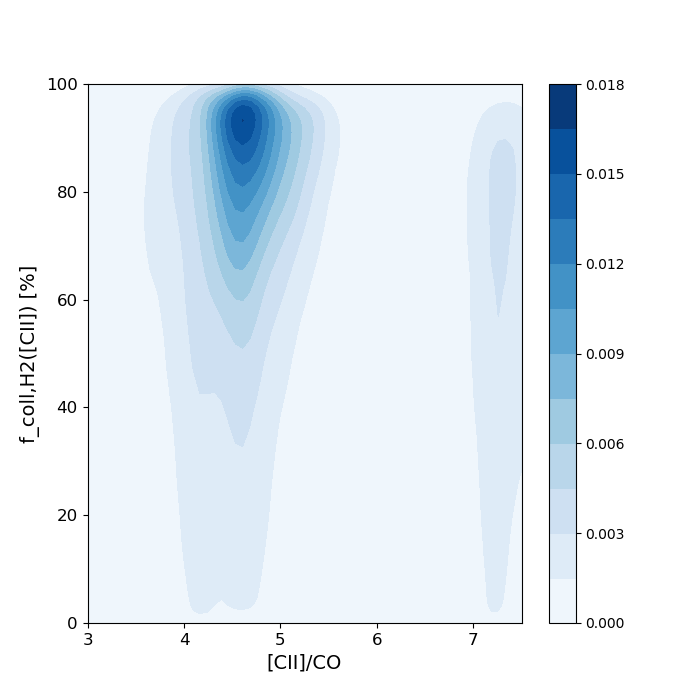}}; \node[right] at (4.,1) {\#4};
\end{tikzpicture}
\begin{tikzpicture}
  \node[anchor=south west,inner sep=0] (image) at (0,0) {\includegraphics[width=6cm,trim=20 20 20 20]{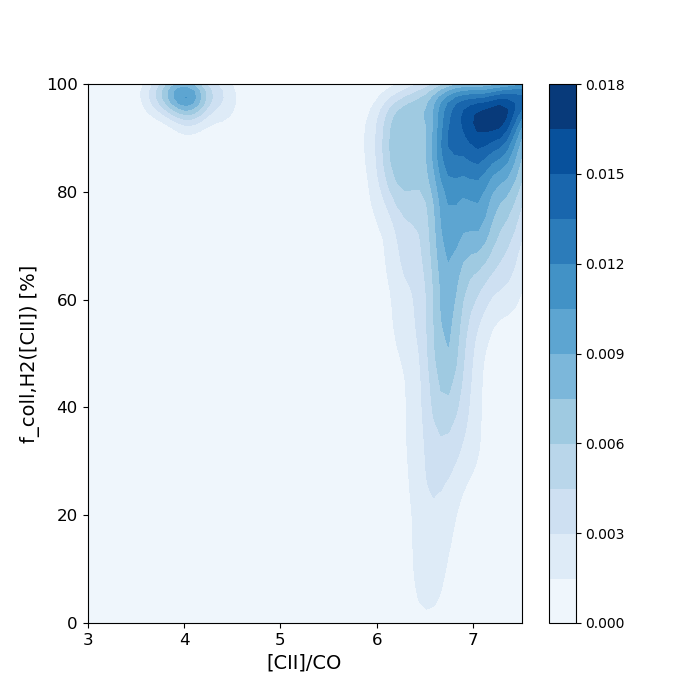}}; \node[right] at (4.,1) {\#5};
\end{tikzpicture}
\begin{tikzpicture}
  \node[anchor=south west,inner sep=0] (image) at (0,0) {\includegraphics[width=6cm,trim=20 20 20 20]{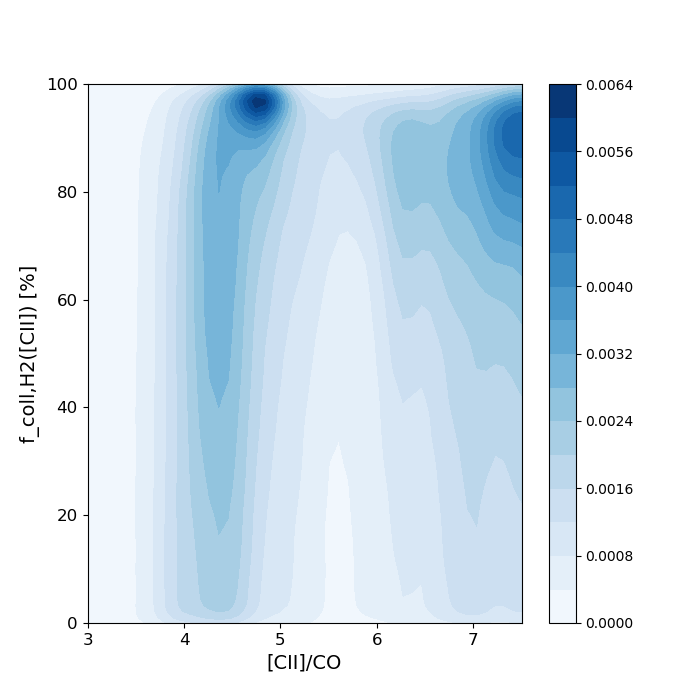}}; \node[right] at (4.,1) {\#6}; 
\end{tikzpicture}
\begin{tikzpicture}
  \node[anchor=south west,inner sep=0] (image) at (0,0) {\includegraphics[width=6cm,trim=20 20 20 20]{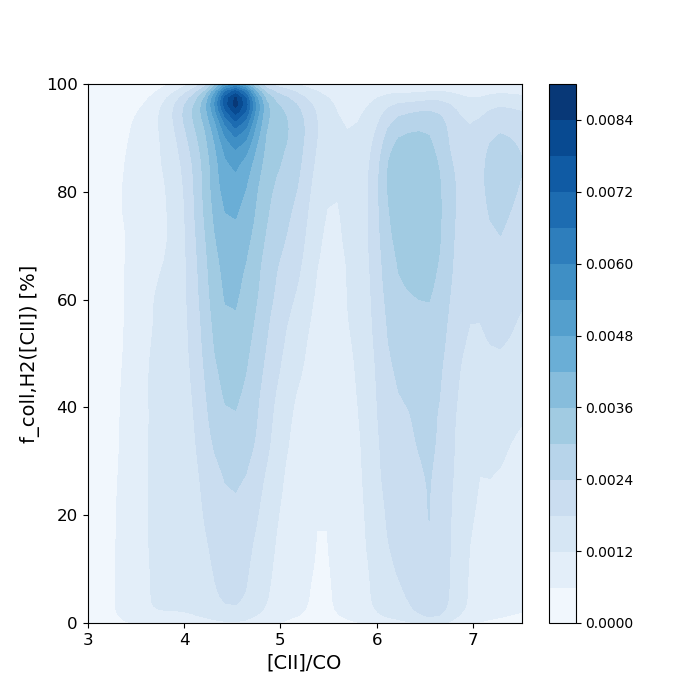}}; \node[right] at (4.,1) {\#7}; 
\end{tikzpicture}
\begin{tikzpicture}
  \node[anchor=south west,inner sep=0] (image) at (0,0) {\includegraphics[width=6cm,trim=20 20 20 20]{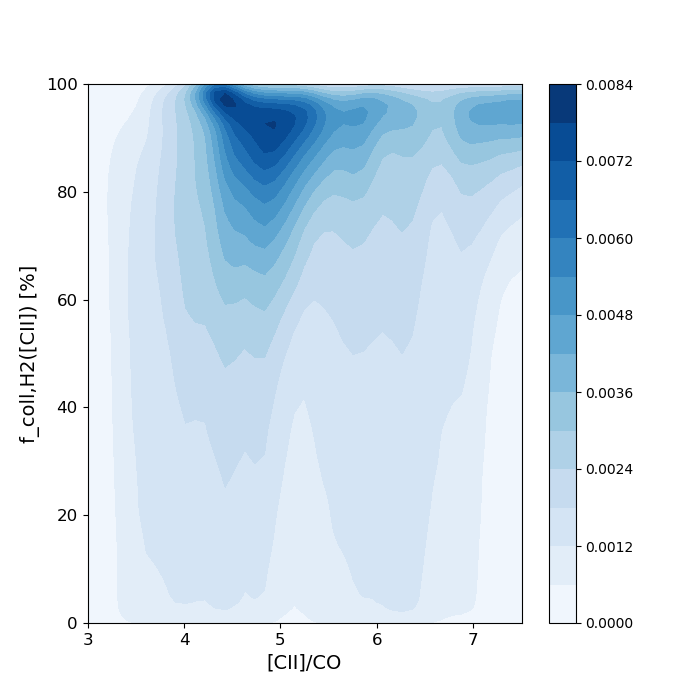}}; \node[right] at (4.,1) {\#8}; 
\end{tikzpicture}
\begin{tikzpicture}
  \node[anchor=south west,inner sep=0] (image) at (0,0) {\includegraphics[width=6cm,trim=20 20 20 20]{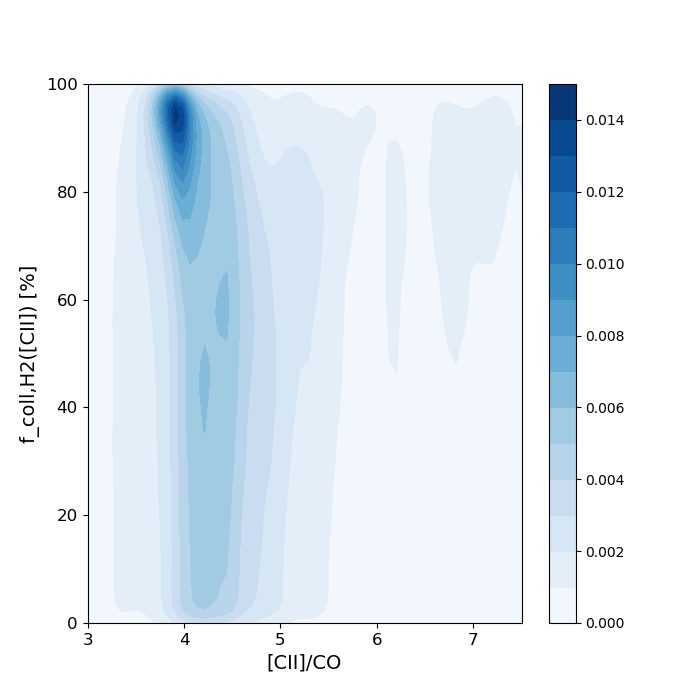}}; \node[right] at (4.,1) {\#9}; 
\end{tikzpicture}
\begin{tikzpicture}
  \node[anchor=south west,inner sep=0] (image) at (0,0) {\includegraphics[width=6cm,trim=20 20 20 20]{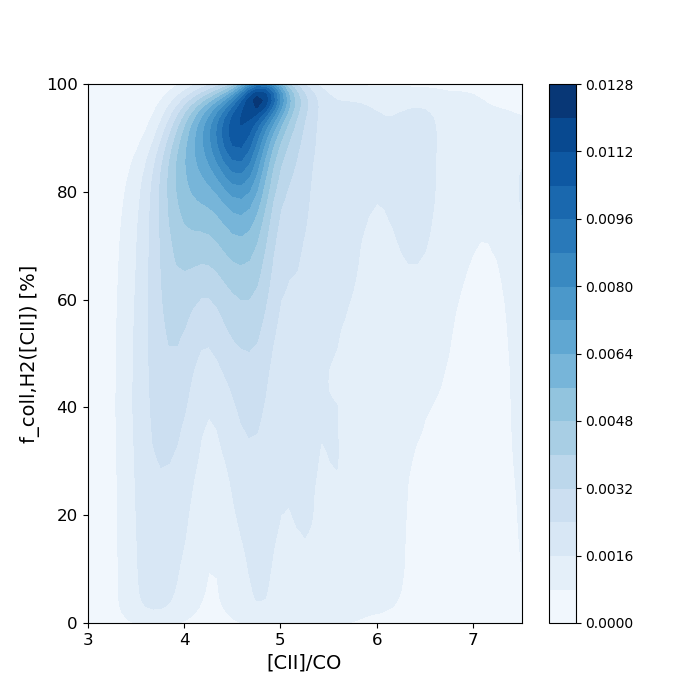}}; \node[right] at (4.,1) {\#10}; 
\end{tikzpicture}
\begin{tikzpicture}
  \node[anchor=south west,inner sep=0] (image) at (0,0) {\includegraphics[width=6cm,trim=20 20 20 20]{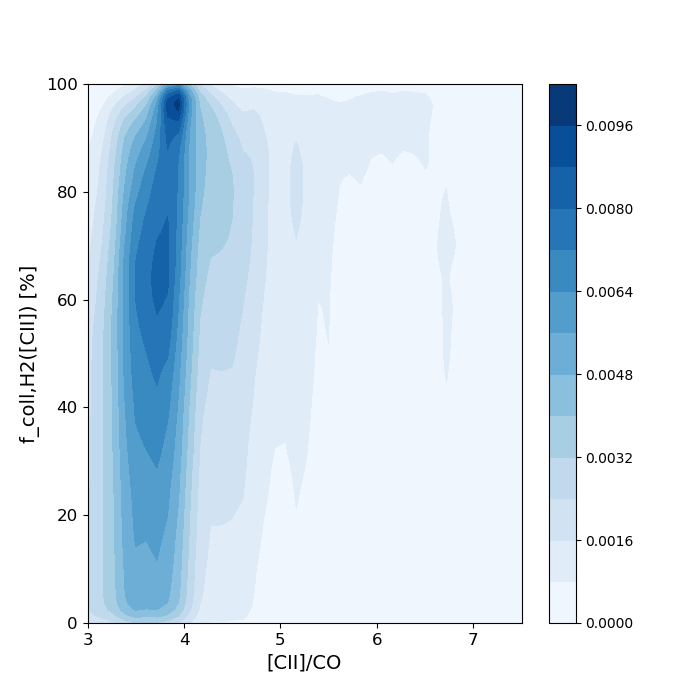}}; \node[right] at (4.,1) {\#11}; 
\end{tikzpicture}
\begin{tikzpicture}
  \node[anchor=south west,inner sep=0] (image) at (0,0) {\includegraphics[width=6cm,trim=20 20 20 20]{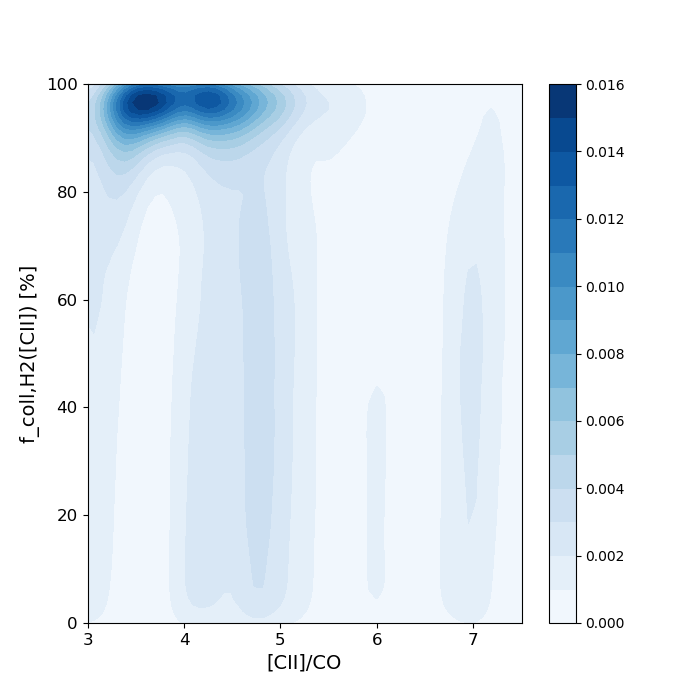}}; \node[right] at (4.,1) {\#12};
\end{tikzpicture}
\caption{Bivariate kernel density estimate of the fraction of [C\2] tracing CO-dark H$_2$ gas vs.\ [C\2]/CO for all pointings. The shade scales with the density of points. }\label{fig:resfcollh2ciico_each}
\end{figure*}

Figure\,\ref{fig:resfcollh2ciico_each} shows that the [C\2] contribution from the neutral atomic gas does not occur preferentially toward components with low [C\2]/CO. In this figure one can also clearly identify the two components in \#5, which are both dominated by CO-dark H$_2$ gas. Most pointings seem to show either a wide distribution of [C\2]/CO values or two main peaks, with the main peak at low [C\2]/CO being due to the velocity component bright in CO and [C\2], while the other peaks with large [C\2]/CO values correspond to [C\2] components with little associated CO emission.

\begin{figure}
\includegraphics[width=9cm,height=5cm]{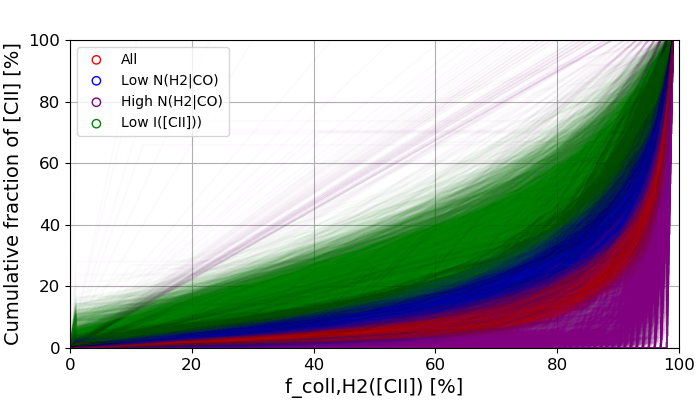}
\caption{Cumulative fraction of [C\2] plotted against the fraction of [C\2] emission arising in the CO-dark H$_2$ gas. Values are calculated with all pointings combined. Plots are shown for all components (red), for components with low CO column density ($<2\times10^{20}$\,cm$^{-2}$; dark blue), for components with high CO column density ($>1\times10^{21}$\,cm$^{-2}$; blue) , and for components with weak [C\2] ($<4\times10^{-8}$\,W\,m$^{-2}$\,sr$^{-1}$; green). }\label{fig:rescumul}
\end{figure}

Globally, we find that $>95$\%\ of the [C\2] emission across all pointings can be attributed to CO-dark H$_2$ gas (i.e., when $f_{\rm coll,H2}({\rm [CII]})>50\%$), as shown in Figure\,\ref{fig:rescumul}. This result does not depend on the CO column density and holds in particular for the brightest [C\2] components associated with CO. The [C\2] components that have the largest contribution from the neutral atomic gas (low $f_{\rm coll,H2}({\rm [CII]})$), correspond preferentially to the components with the faintest [C\2] surface brightness and not to components with low CO column density (Fig.\,\ref{fig:rescumul}). Even then, less than $\sim30$\%\ of the [C\2] emission in these low-[C\2]-surface-brightness components may be arising from gas that is mostly atomic (i.e., $f_{\rm coll,H2}({\rm [CII]})<50\%$). \cite{Okada2019a} find similar results toward several star-forming regions in the LMC, with $<15\%$ of [C\2] being attributed to neutral atomic gas. We wish to emphasize that the larger fraction of [C\2] in the neutral atomic gas in faint [C\2] components is driven by the variation of the [C\2] flux associated with the molecular phase. 
 
In summary, [C\2] mostly traces the CO-dark H$_2$ gas but there is evidence of a weak contribution from the neutral atomic gas in the faintest [C\2] components.
The thermal pressure in the CO-dark H$_2$ gas lies around $10^{3-5}$\,K\cc\ (Fig.\,\ref{fig:pressures}), which is similar to the range derived by \cite{Pineda2017a} for several diffuse (with no CO detection) LMC  regions. Considering the large filling factor of [C\2] emission observed in N\,11B with \textit{Herschel}/PACS (Fig.\,\ref{fig:spectracoalma}), it is therefore plausible that the CO-dark H$_2$ gas probed by our [C\2] observations corresponds to the diffuse ISM (median density $\sim200$\cc\ for all pointings) filling the space between CO clumps rather than thin CO-dark H$_2$ layers around CO clumps.

\subsubsection{Fraction of CO-dark H$_2$ gas}\label{sec:fdark}

\begin{figure*} 
\begin{tikzpicture}
  \node[anchor=south west,inner sep=0] (image) at (0,0) {\includegraphics[width=6cm,trim=20 20 20 20]{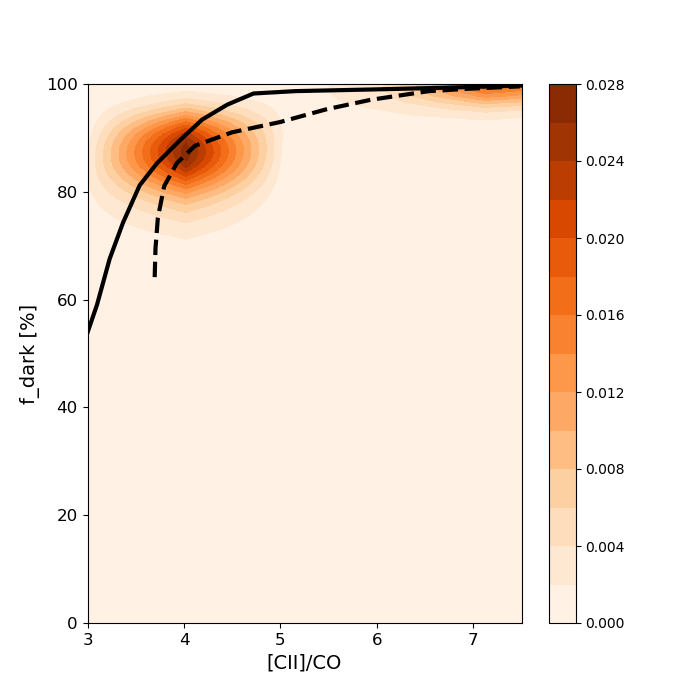}}; \node[right] at (1.6,1) {\#5};
 
\end{tikzpicture}\vspace{-0.05cm}
\begin{tikzpicture}
  \node[anchor=south west,inner sep=0] (image) at (0,0) {\includegraphics[width=6cm,trim=20 20 20 20]{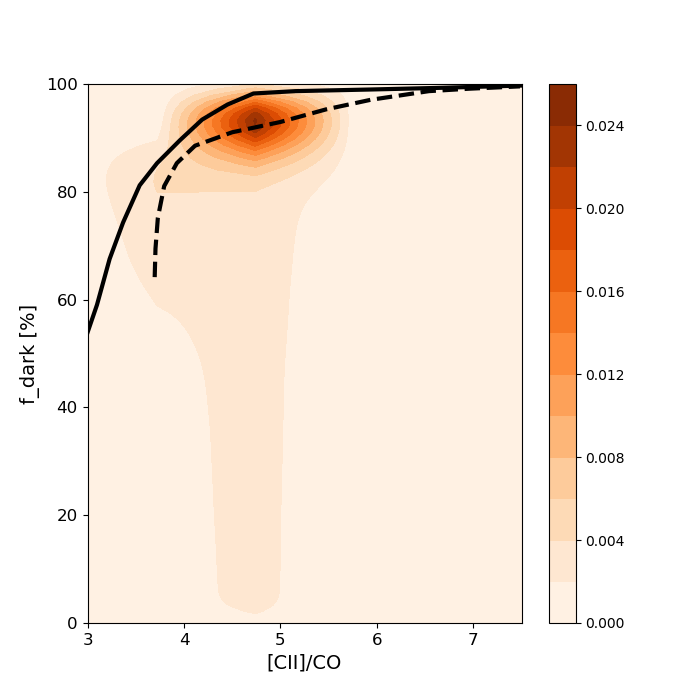}}; \node[right] at (1.6,1) {\#4};
\end{tikzpicture}\vspace{-0.05cm}
\begin{tikzpicture}
  \node[anchor=south west,inner sep=0] (image) at (0,0) {\includegraphics[width=6cm,trim=20 20 20 20]{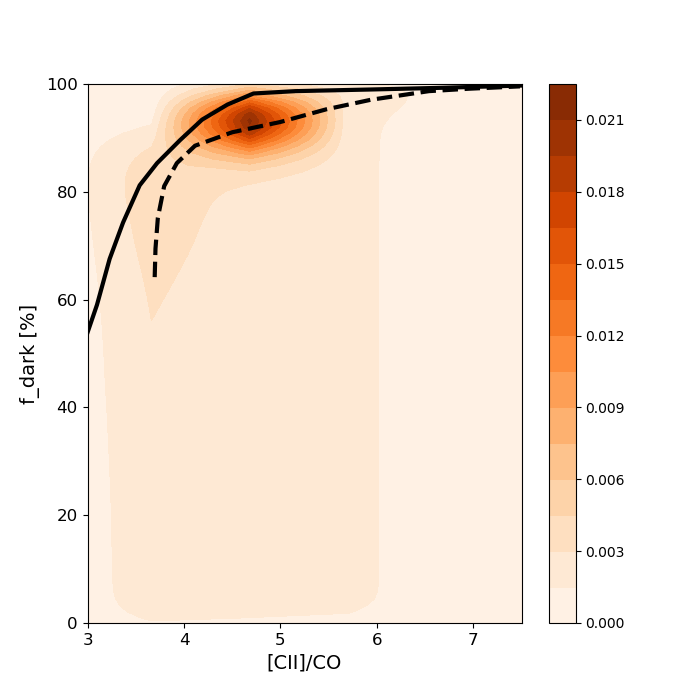}}; \node[right] at (1.6,1) {\#10}; 
\end{tikzpicture}\vspace{-0.05cm}
\begin{tikzpicture}
  \node[anchor=south west,inner sep=0] (image) at (0,0) {\includegraphics[width=6cm,trim=20 20 20 20]{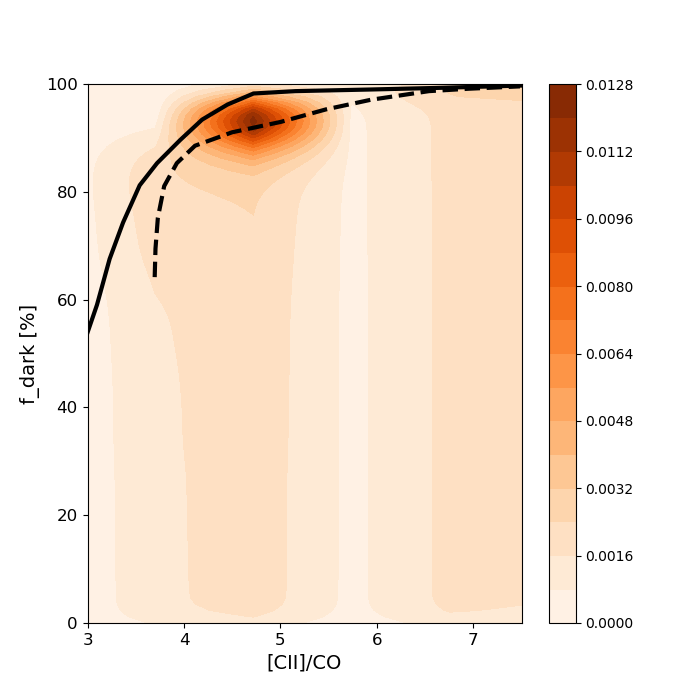}}; \node[right] at (1.6,1) {\#6}; 
\end{tikzpicture}\vspace{-0.05cm}
\begin{tikzpicture}
  \node[anchor=south west,inner sep=0] (image) at (0,0) {\includegraphics[width=6cm,trim=20 20 20 20]{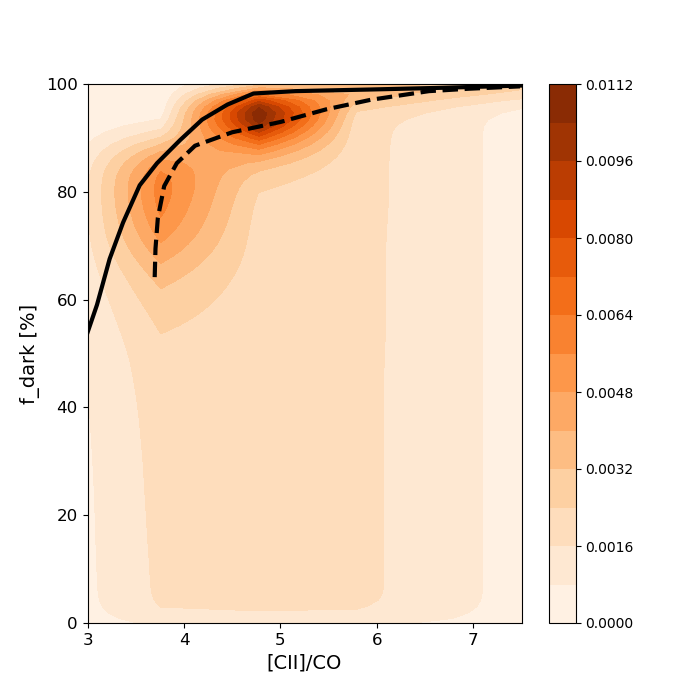}}; \node[right] at (1.6,1) {\#8}; 
\end{tikzpicture}\vspace{-0.05cm}
\begin{tikzpicture}
  \node[anchor=south west,inner sep=0] (image) at (0,0) {\includegraphics[width=6cm,trim=20 20 20 20]{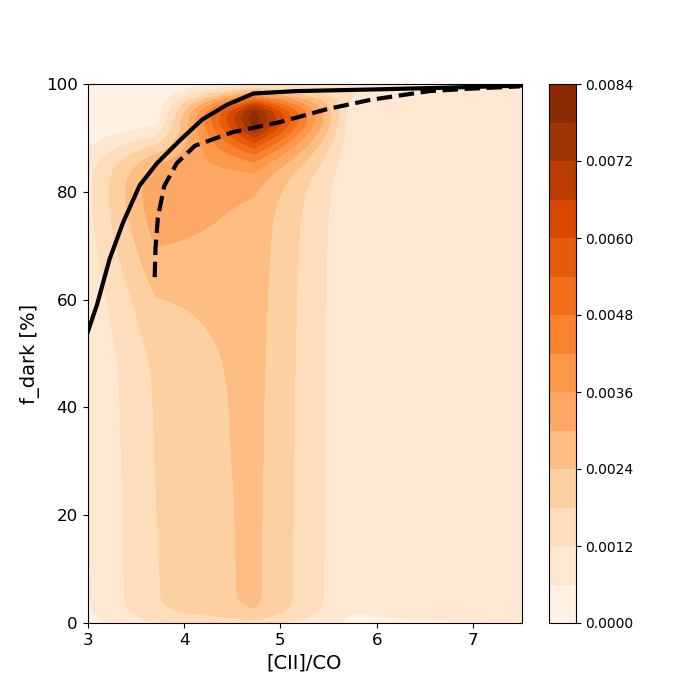}}; \node[right] at (1.6,1) {\#1}; 
\end{tikzpicture}\vspace{-0.05cm}
\begin{tikzpicture}
  \node[anchor=south west,inner sep=0] (image) at (0,0) {\includegraphics[width=6cm,trim=20 20 20 20]{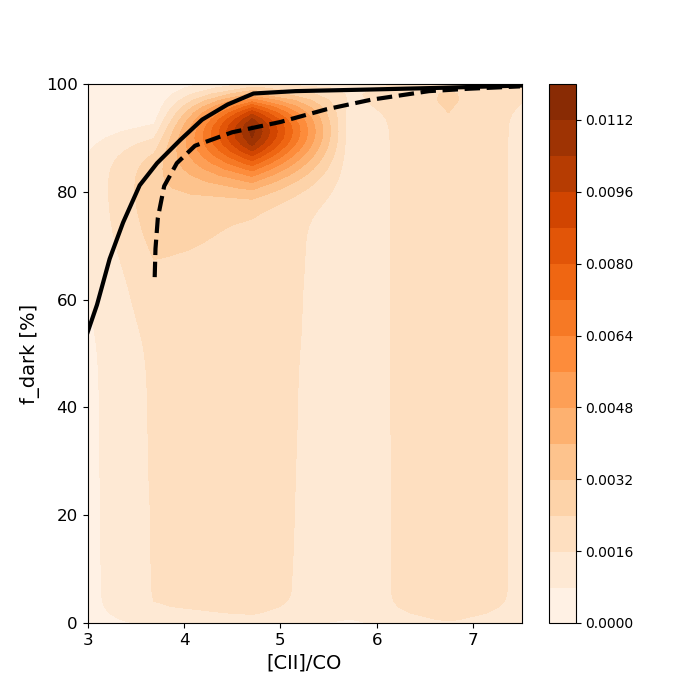}}; \node[right] at (1.6,1) {\#7}; 
\end{tikzpicture}\vspace{-0.05cm}
\begin{tikzpicture}
  \node[anchor=south west,inner sep=0] (image) at (0,0) {\includegraphics[width=6cm,trim=20 20 20 20]{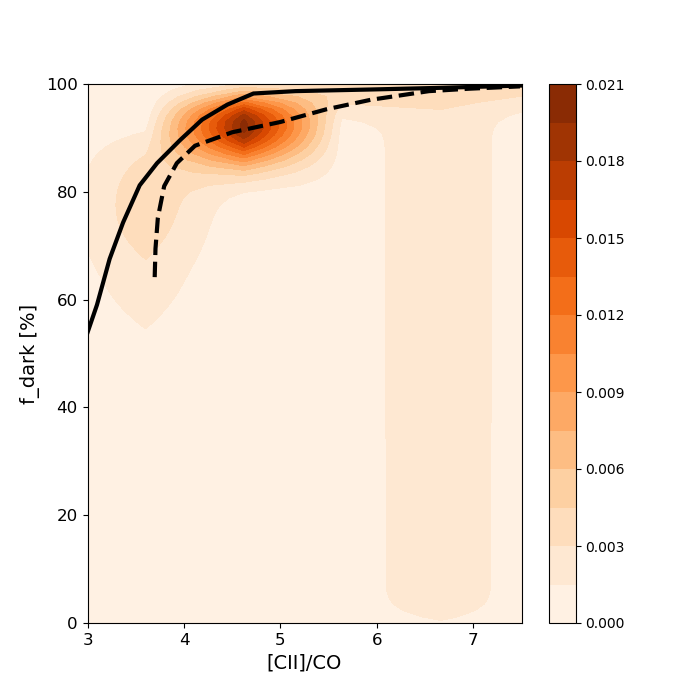}}; \node[right] at (1.6,1) {\#2}; 
\end{tikzpicture}\vspace{-0.05cm}
\begin{tikzpicture}
  \node[anchor=south west,inner sep=0] (image) at (0,0) {\includegraphics[width=6cm,trim=20 20 20 20]{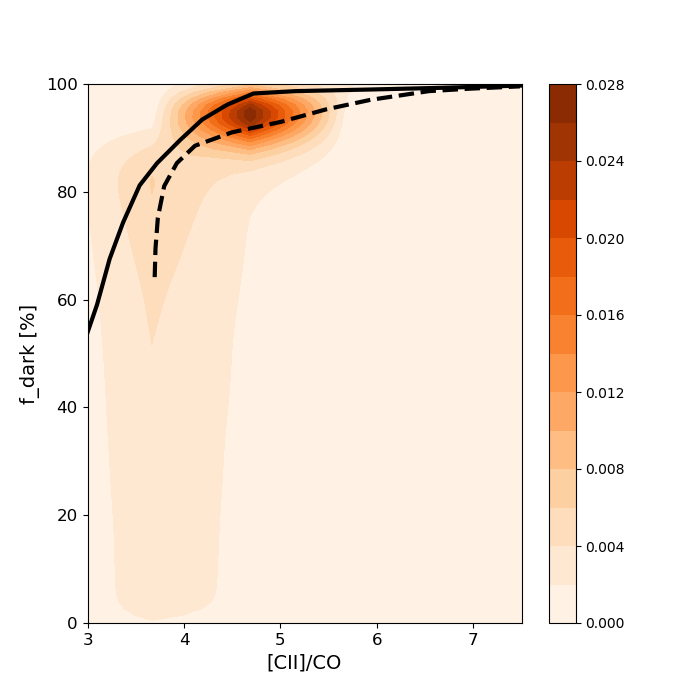}}; \node[right] at (1.6,1) {\#3}; 
\end{tikzpicture}\vspace{-0.05cm}
\begin{tikzpicture}
  \node[anchor=south west,inner sep=0] (image) at (0,0) {\includegraphics[width=6cm,trim=20 20 20 20]{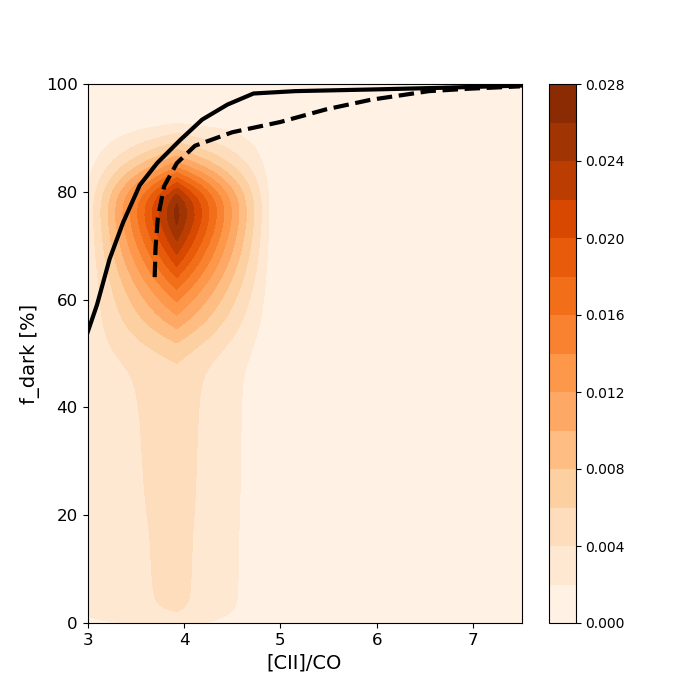}}; \node[right] at (1.6,1) {\#11}; 
\end{tikzpicture}\vspace{-0.05cm}
\begin{tikzpicture}
  \node[anchor=south west,inner sep=0] (image) at (0,0) {\includegraphics[width=6cm,trim=20 20 20 20]{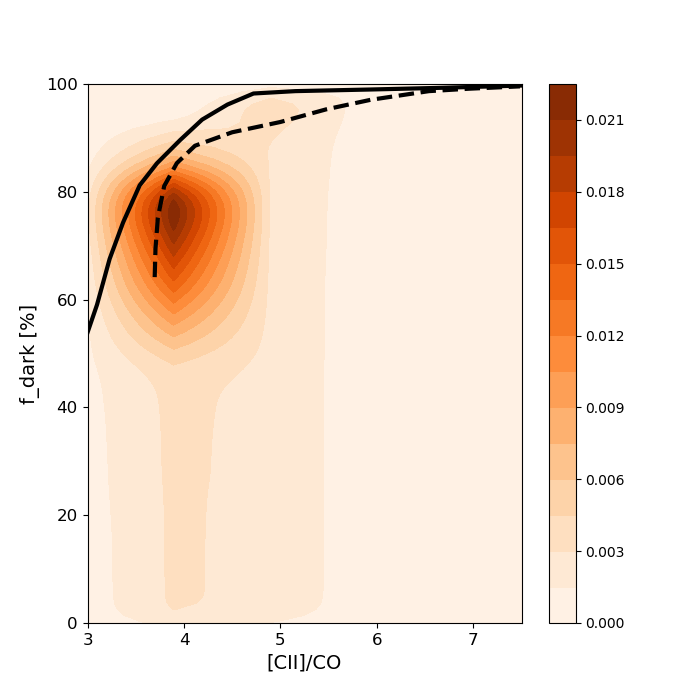}}; \node[right] at (1.6,1) {\#9}; 
\end{tikzpicture}\vspace{-0.05cm}
\begin{tikzpicture}
  \node[anchor=south west,inner sep=0] (image) at (0,0) {\includegraphics[width=6cm,trim=20 20 20 20]{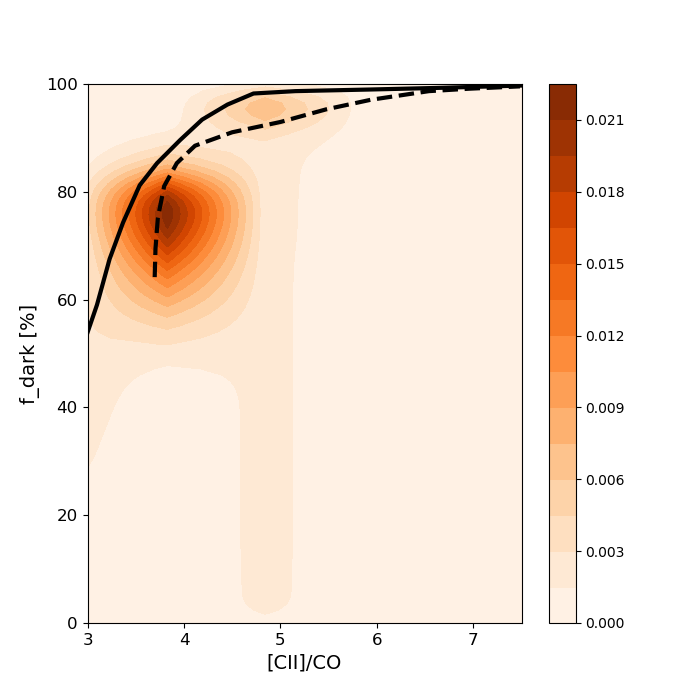}}; \node[right] at (1.6,1) {\#12};
\end{tikzpicture}\vspace{-0.05cm}
\caption{Bivariate kernel density estimate of the fraction of CO-dark H$_2$ gas $f_{\rm dark}$ vs.\ the [C\2]/CO ratio for all pointings. The shade scales with the density of points. The dashed curve is an attempt to connect most of the distribution peaks from different pointings (with the plots ordered according the location of probability peaks along the curve) while the solid curve shows the expected relation between $f_{\rm dark}$ and [C\2]/CO at low density ($\sim30$\cc; Fig.\,\ref{fig:model_plots}). }\label{fig:resdarkciico_each}
\end{figure*}

We show in Figure\,\ref{fig:resdarkciico_each} the distribution of the fraction of CO-dark H$_2$ gas, $f_{\rm dark}$, versus\ [C\2]/CO for each pointing. The fraction of CO-dark H$_2$ gas is also a proxy for the fraction of  DNM (i.e., CO-dark H$_2$ gas and optically-thick H\1) since the CO-dark H$_2$ gas traced by [C\2] is the main contribution ($\gtrsim95\%$) to the DNM according to our models. The weak contribution of optically thick H\1\ to the DNM is also found in the local and diffuse ISM \citep{Murray2018a,Liszt2018a}.

Most pointings show a well-defined peak in Figure\,\ref{fig:resdarkciico_each} corresponding to the brightest [C\2] component. Fainter [C\2] components often result in a wide range of $f_{\rm dark}$ values because they are more likely to arise from the neutral atomic phase (with relatively large uncertainty; Sect.\,\ref{sec:ciiorigin2}).
From Figure\,\ref{fig:resdarkciico_each} we see that most of the molecular gas is CO-dark overall, with $f_{\rm dark}\gtrsim60\%$.
Since most of the molecular gas is CO-dark, we conclude that the velocity-integrated ([C\2]+[O\1])/PAH ratio, which suggests a constant photoelectric-effect heating efficiency across N\,11 (Sect.\,\ref{sec:ciioipah}), corresponds in fact to the physical conditions of the CO-dark H$_2$ gas rather than the neutral atomic medium. 

The $f_{\rm dark}$ values we find are in good agreement with the estimates of \cite{Galametz2016a} in N\,11 using the dust-to-gas mass ratio. Our results are also in general agreement with other studies in nearby low-metallicity environments. \cite{Fahrion2017a} used SOFIA/GREAT observations the dwarf galaxy NGC\,4214 at $\approx200$\,pc spatial scale and found that only $\approx10$\%\ of [C\2] could be attributed to the CNM and that $\approx80$\%\ of the H$_2$ mass  is not traced by CO. Other studies in the Magellanic Clouds also show significant CO-dark H$_2$ gas fractions. \cite{RequenaTorres2016a} examined several star-forming regions in the SMC, and found that most of the [C\2] emission originates from CO-dark H$_2$ gas. \cite{Chevance2016b} observed LMC-30\,Dor and determined from PDR models that $\gtrsim80$\%\ of the molecular gas is CO-dark H$_2$ gas.

In N\,11, there is a clear correlation between $f_{\rm dark}$ and [C\2]/CO (Fig.\,\ref{fig:resdarkciico_each}), which is not surprising because the CO-dark H$_2$ gas in the model is by construction traced by C$^+$ and because the contribution of [C\2] in the neutral atomic gas is relatively small (Sect.\,\ref{sec:modelstrat}). For this reason, the [C\2]/CO ratio we use in Figure\,\ref{fig:resdarkciico_each} is the observed value, i.e., with no correction for the [C\2] in the atomic gas. For a given [C\2]/CO ratio, the range of $f_{\rm dark}$ values is driven by the H$^0$ column density which sets the density in the atomic gas and consequently in the molecular gas (Sect.\,\ref{sec:modelstrat}).

Most points seem to follow the same relationship between $f_{\rm dark}$ and [C\2]/CO (dashed curve in Fig.\,\ref{fig:resdarkciico_each}). For [C\2]/CO$\gtrsim10^{4.5}$, this is because $f_{\rm dark}$ depends little on the gas density (Sect.\,\ref{sec:modelstrat}). For lower [C\2]/CO values, $f_{\rm dark}$ depends more on density
and components with similar [C\2]/CO ratios may have significantly different $f_{\rm dark}$ values (e.g., from $\approx75\%$ for \#9 to $\approx90\%$ for \#5).

\begin{figure*}
\begin{tikzpicture}
  \node[anchor=south west,inner sep=0] (image) at (0,0) {\includegraphics[width=6cm]{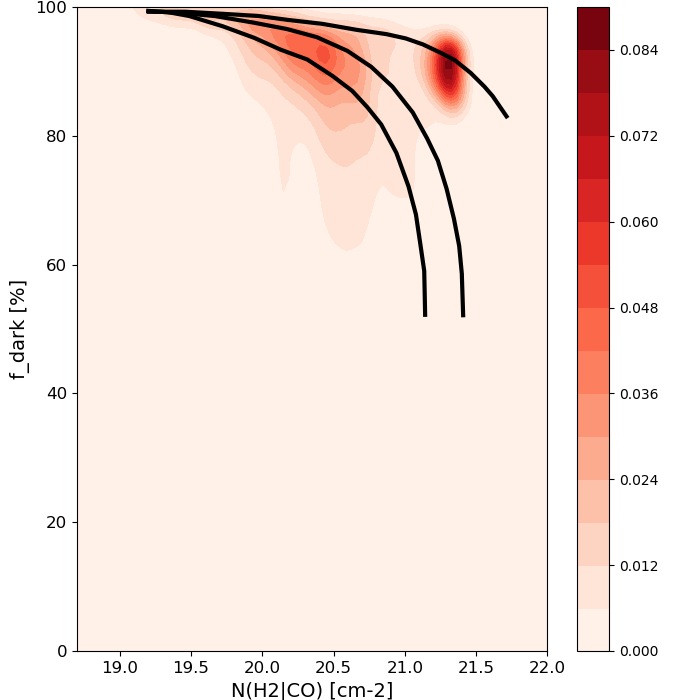}}; \node[right] at (1.05,5.4) {\#2};
  \end{tikzpicture}\vspace{-0.07cm}
\begin{tikzpicture}
  \node[anchor=south west,inner sep=0] (image) at (0,0) {\includegraphics[width=6cm]{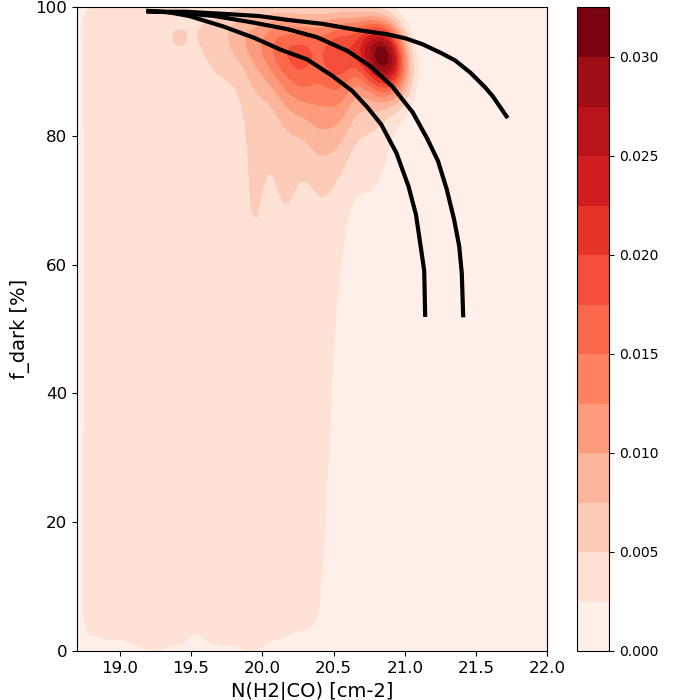}}; \node[right] at (1.05,5.4) {\#10};
  \end{tikzpicture}\vspace{-0.07cm}
\begin{tikzpicture}
  \node[anchor=south west,inner sep=0] (image) at (0,0) {\includegraphics[width=6cm]{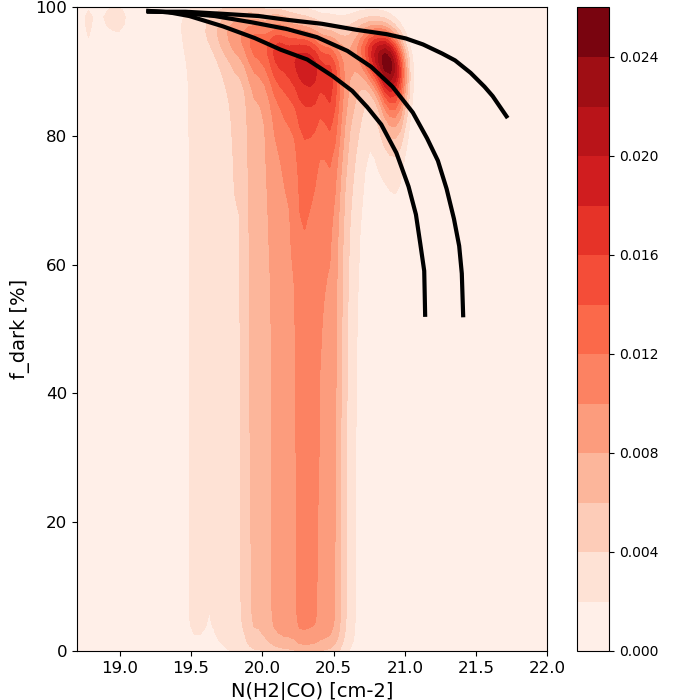}}; \node[right] at (0.912,5.4) {\#6}; 
  \end{tikzpicture}\vspace{-0.07cm}
\begin{tikzpicture}
  \node[anchor=south west,inner sep=0] (image) at (0,0) {\includegraphics[width=6cm]{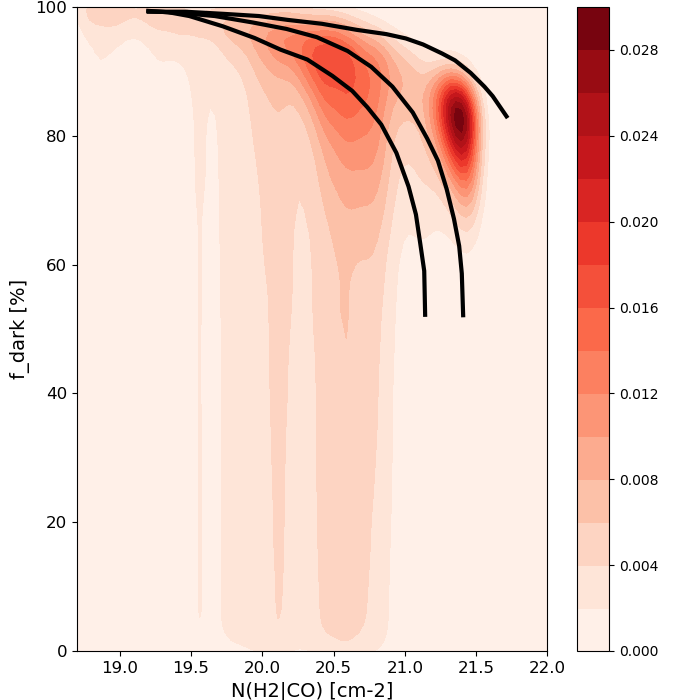}}; \node[right] at (1.05,5.4) {\#8};
  \end{tikzpicture}\vspace{-0.07cm}
\begin{tikzpicture}
  \node[anchor=south west,inner sep=0] (image) at (0,0) {\includegraphics[width=6cm]{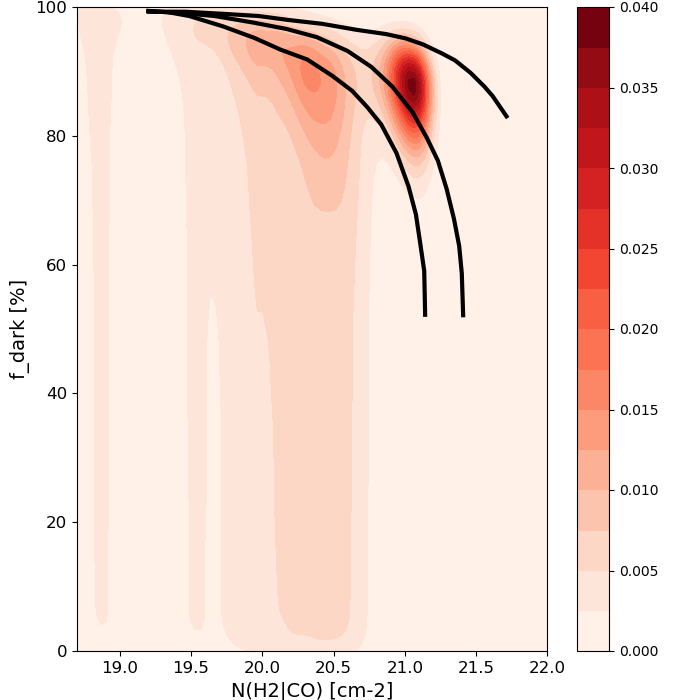}}; \node[right] at (1.05,5.4) {\#7};
  \end{tikzpicture}\vspace{-0.07cm}
\begin{tikzpicture}
  \node[anchor=south west,inner sep=0] (image) at (0,0) {\includegraphics[width=6cm]{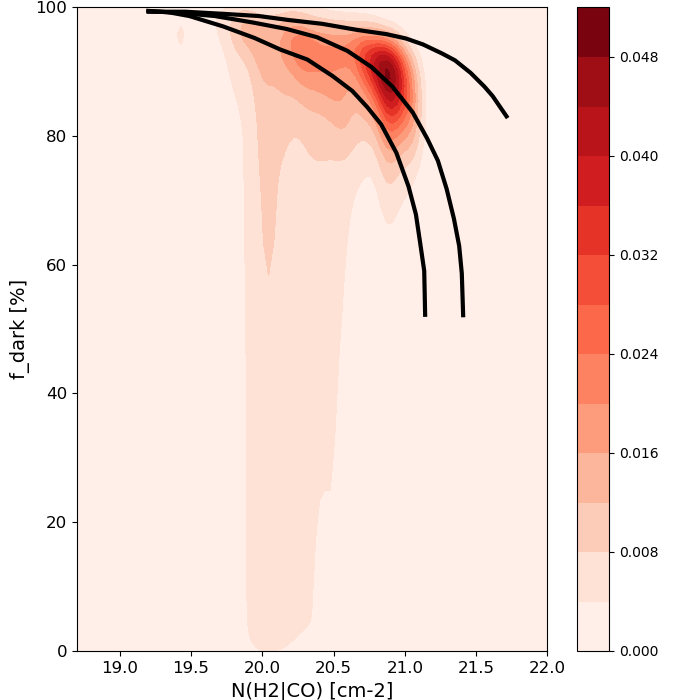}}; \node[right] at (1.05,5.4) {\#4};
  \end{tikzpicture}\vspace{-0.07cm}
\begin{tikzpicture}
  \node[anchor=south west,inner sep=0] (image) at (0,0) {\includegraphics[width=6cm]{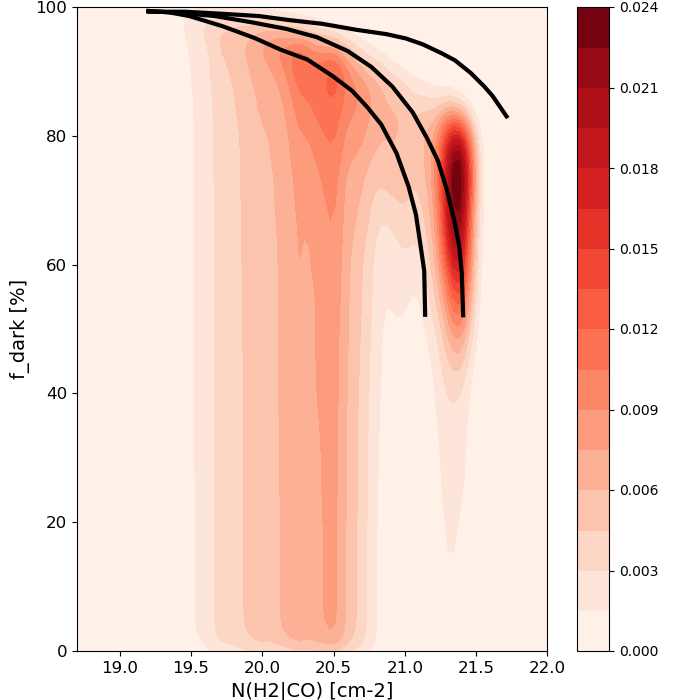}}; \node[right] at (1.05,5.4) {\#9};
  \end{tikzpicture}\vspace{-0.07cm}
\begin{tikzpicture}
  \node[anchor=south west,inner sep=0] (image) at (0,0) {\includegraphics[width=6cm]{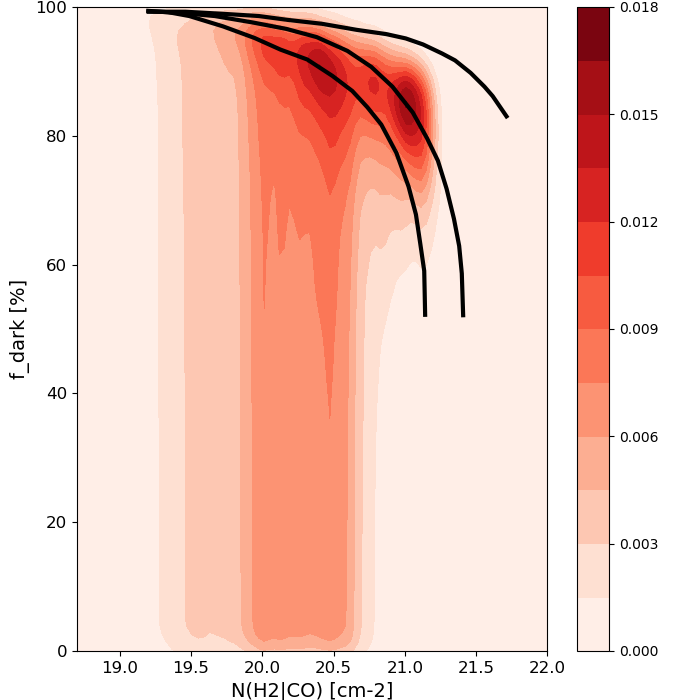}}; \node[right] at (1.05,5.4) {\#1};
  \end{tikzpicture}\vspace{-0.07cm}
\begin{tikzpicture}
  \node[anchor=south west,inner sep=0] (image) at (0,0) {\includegraphics[width=6cm]{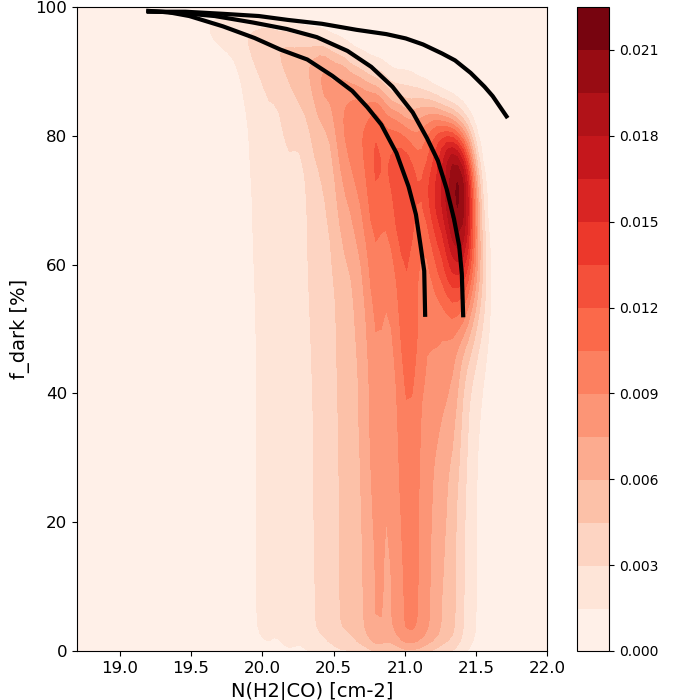}}; \node[right] at (1.05,5.4) {\#11};
  \end{tikzpicture}\vspace{-0.07cm}
\begin{tikzpicture}
  \node[anchor=south west,inner sep=0] (image) at (0,0) {\includegraphics[width=6cm]{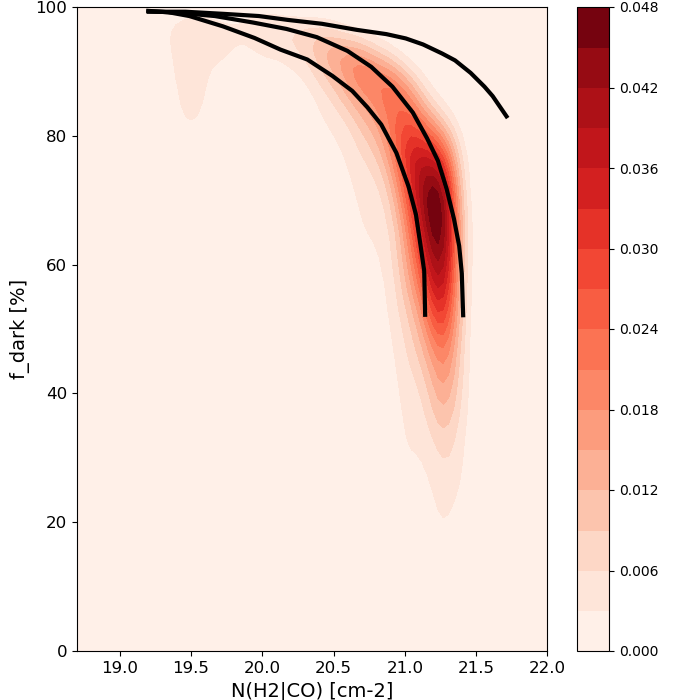}}; \node[right] at (1.05,5.4) {\#12}; 
  \end{tikzpicture}\vspace{-0.07cm}
\begin{tikzpicture}
  \node[anchor=south west,inner sep=0] (image) at (0,0) {\includegraphics[width=6cm]{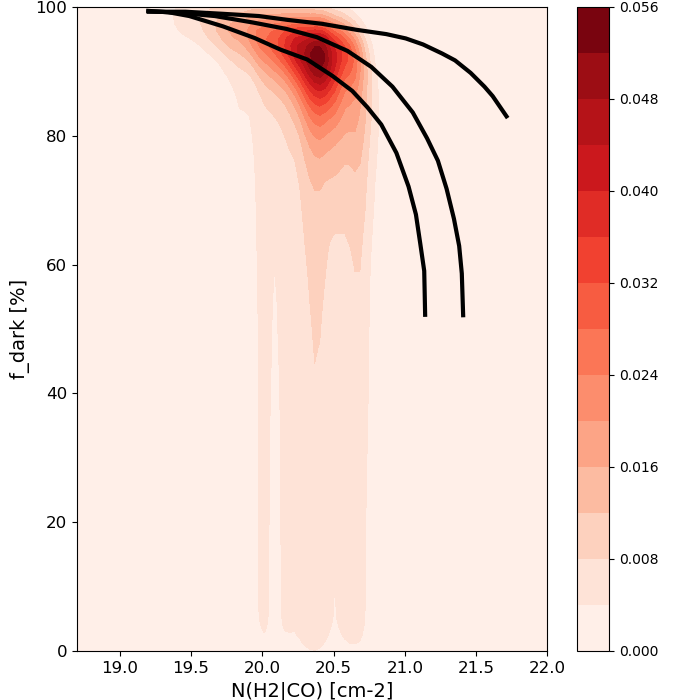}}; \node[right] at (1.05,5.4) {\#3};
  \end{tikzpicture}\vspace{-0.07cm}
\begin{tikzpicture}
  \node[anchor=south west,inner sep=0] (image) at (0,0) {\includegraphics[width=6cm]{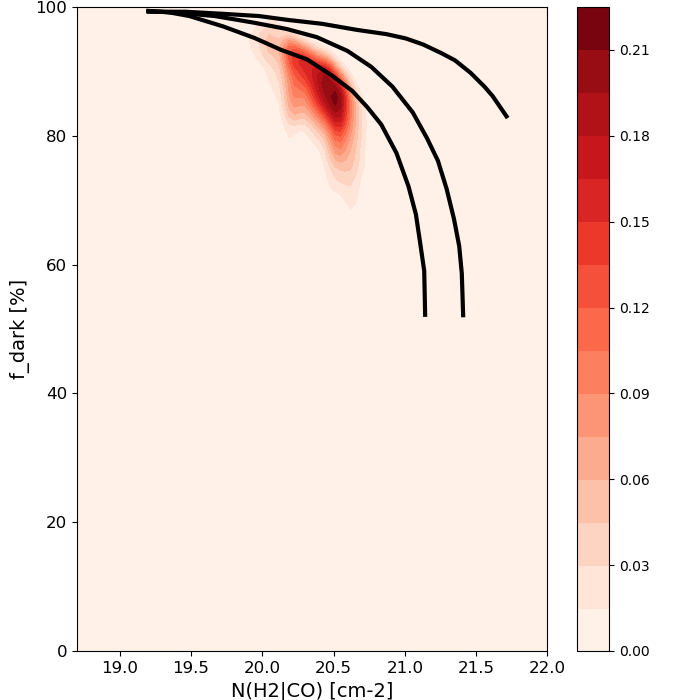}}; \node[right] at (1.05,5.4) {\#5};
  \end{tikzpicture}\vspace{-0.07cm}
\caption{Bivariate kernel density estimate of the fraction of CO-dark H$_2$ gas vs.\ the H$_2$ column density measured from CO (i.e., ignoring CO-dark H$_2$ gas) for all pointings. The shade scales with the density of points. The black curves serve as guides to compare the location of the probability peaks in different pointings. Plots are ordered according to the location of the main peak with respect to the black curves.  }\label{fig:resdarknco_each}
\end{figure*}

We also find that $f_{\rm dark}$ is anti-correlated with the CO column density (Fig.\,\ref{fig:resdarknco_each}). Most of the molecular gas is therefore CO-dark, but there is more CO-dark gas in CO-faint regions ($\sim100$\%\ for $N({\rm H_2|CO})\lesssim10^{20}$\,cm$^{-2}$) as compared to CO peaks ($\sim70-90$\%\ for $N({\rm H_2|CO})\sim10^{21}$\,cm$^{-2}$). This result is reminiscent of the findings of \cite{Okada2015a}  that the amount of [C\2] that cannot be attributed to the gas traced by CO is larger between CO peaks in another LMC star-forming region, N\,159.

The effective $X_{\rm CO}$ factor including the contribution of the CO-dark H$_2$ gas lies in the range $10^{21-22}$\,(K\,\kms)$^{-1}$ for most of the bright velocity components (Fig.\,\ref{fig:resxco_ciico_each}), in good agreement with values obtained in \cite{Israel1997a,Galliano2011a,Chevance2016b}.

\subsubsection{Physical parameters controlling $f_{\rm dark}$}\label{sec:stellarfeedback}

\begin{figure}
\includegraphics[width=10cm]{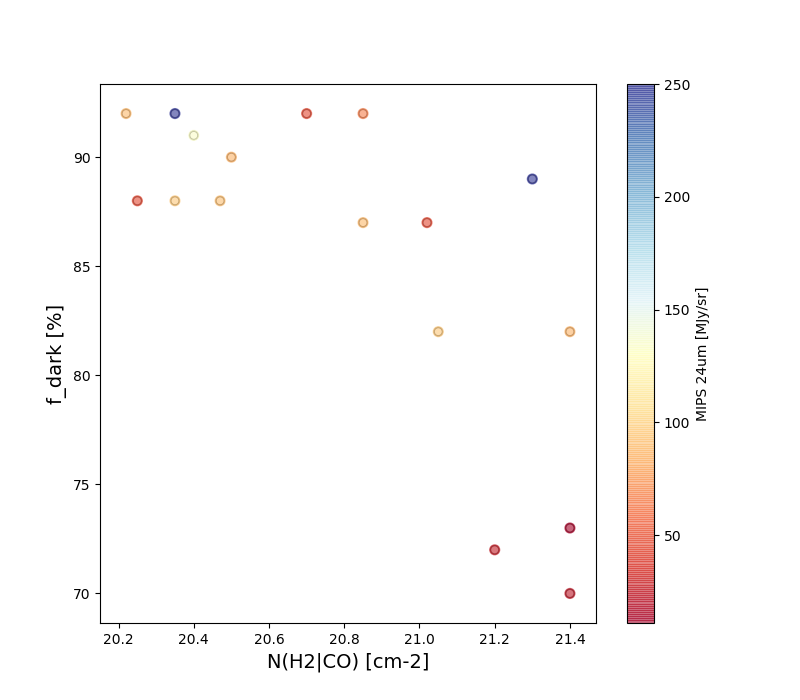}
\caption{Fraction of CO-dark H$_2$ gas plotted against the CO column density for the peaks identified by eye in Figure\,\ref{fig:resdarknco_each}. The shade scales with the \textit{Spitzer}/MIPS $24$\mic\ surface brightness. }\label{fig:resdarknco_mips}
\end{figure}

We find that $f_{\rm dark}$ is somewhat larger for components with a large [C\2]/CO ratio (Fig.\,\ref{fig:resdarkciico_each}), which should correspond to molecular clouds illuminated by the UV radiation from massive stars, providing some evidence that stellar feedback may play a role. In Figure\,\ref{fig:resdarknco_each}, the pointings are ordered based on the shape of the $f_{\rm dark}$ with $N({\rm H_2|CO})$ distribution, from pointings with a high and almost flat $f_{\rm dark}$ distribution until column densities of $\approx10^{21.3}$\,cm$^{-2}$ to pointings showing a steep decrease of $f_{\rm dark}$ for column densities above $\approx10^{20}$\,cm$^{-2}$. The difference between the various pointings is most pronounced for large column densities $N({\rm H_2|CO})>10^{21}$\,cm$^{-2}$. We note that the larger $f_{\rm dark}$ values are not a direct consequence of the fact that a relatively dense molecular phase might be required, since in such cases a somewhat lower column density of CO-dark H$_2$ gas will be determined (Fig.\,\ref{fig:modelstrat}).
Interestingly, the sequence of pointings in Figure\,\ref{fig:resdarknco_each} also correlates with the presence of bright H$\alpha$ and $24$\mic\ emission near molecular clouds (Fig.\,\ref{fig:hicoha}). For instance pointings \#9, \#11, and \#12 show particularly low $f_{\rm dark}$ values and high CO column densities and they are also the pointings with the faintest H$\alpha$ (or $24$\mic) emission. This result is also shown in Figure\,\ref{fig:resdarknco_mips} where we report the $24$\mic\ emission for each peak in the distribution of each pointing.

Stellar feedback could have an impact on the fraction of CO-dark H$_2$ gas, either through intense radiation field (ionization, radiation pressure) or through dynamical/mechanical effects from winds and supernovae shocks resulting in disruption/dispersal of molecular clouds (e.g., \citealt{Dale2012a}) and therefore to a lower beam-averaged extinction $A_V$. The average extinction toward a given pointing is driven by the selective disruption and photodissociation of the most diffuse clouds (thereby selecting out the clouds with the highest extinction) and the overall eroding of all clouds. From a theoretical perspective, \cite{Wolfire2010a} find that for fixed $A_V$ the fraction of CO-dark H$_2$ gas in a molecular cloud is insensitive to the incident radiation field, although with the latter not exceeding $30$ times the local Galactic radiation field. In contrast, these latter authors find that $f_{\rm dark}$ increases with decreasing $A_V$. On the other hand, kiloparsec-scale simulations of spiral galaxies show that $f_{\rm dark}$ is a function of the radiation field and of the surface density \citep{Smith2014a}. \cite{Fahrion2017a} modeled the star-forming dwarf galaxy NGC\,4214 and found that the fraction of CO-dark H$_2$ gas mass depends on the evolutionary stage of the star-forming regions, with a larger fraction towards the naked cluster as compared to compact embedded regions. Madden et al.\ (in preparation) modeled PDRs in galaxies from the \textit{Herschel} Dwarf Galaxy Survey (DGS) and find that the effective extinction derived in the model is the main parameter controlling $f_{\rm dark}$.

Unfortunately, our models do not allow us to directly examine parameters such as $A_V$ or the radiation field strength $G_0$, for lack of other transitions or dust measurements corresponding to individual velocity components, but we may hope to disentangle both parameters from $N({\rm H_2|CO})$ or [C\2]/CO. Since $N({\rm H_2|CO})$ is proportional to the CO intensity which correlates with extinction on parsec scales \citep{Lee2018a}, the trends in Figure\,\ref{fig:resdarknco_each} can be understood to first order as the variations of $f_{\rm dark}$ with $A_V$ for a single cloud hypothesis. With this in mind, the decrease of $f_{\rm dark}$ for $N({\rm H_2|CO})\gtrsim10^{20}$\,cm$^{-2}$ would be due to the increase of CO column density with the cloud size while the mass in the CO-dark H$_2$ gas layer saturates, and $A_V$ should then be the main parameter controlling $f_{\rm dark}$. However, the fact that we observe different $f_{\rm dark}$ values for a given $N({\rm H_2|CO})$ value implies either that a parameter other than $A_V$ controls $f_{\rm dark}$ (a parameter correlating with [C\2]/CO since $f_{\rm dark}$ correlates best with [C\2]/CO; Fig.\,\ref{fig:resdarkciico_each}) or that $N({\rm H_2|CO})$ does not accurately trace $A_V$  (for instance because our observations combine multiple clouds).

\subsection{Influence of metallicity}\label{secapp:pineda}

In this section we apply our models to the data from \cite{Pineda2017a} for LMC and SMC pointings (lines of sight toward H\1-bright, CO-bright, and/or $160$\mic-bright regions). The SMC data points are useful probes of a lower-metallicity environment ($\approx0.2$\,Z$_\odot$) as compared to the LMC ($\approx0.5$\,Z$_\odot$). The spectral profiles of H\1\ and CO were used in \cite{Pineda2017a} but individual velocity components were not adjusted to mitigate the different angular resolution in each tracer. We use the H\1\ CNM value (integrated over the [C\2] line FWHM) in \cite{Pineda2017a}. For our models, the CO-to-H$_2$ conversion factor for the SMC is taken as $X'_{\rm CO}=1\times10^{21}$\,cm$^{-2}$\,(K\,\kms)$^{-1}$ \citep{RomanDuval2014a}. We selected data points with detections in H\1, [C\2], and CO, and with $I([{\rm CII}])>2\times10^{-8}$\,W\,m$^{-2}$\,sr$^{-1}$ in order to be coherent with the present study (Sect.\,\ref{sec:selcomps}). 

\begin{figure*}
\includegraphics[width=6cm]{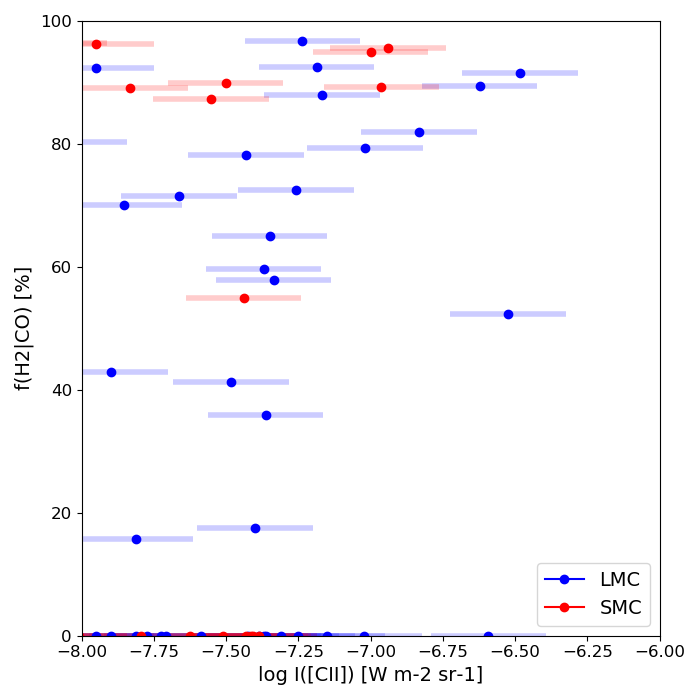}
\includegraphics[width=6cm]{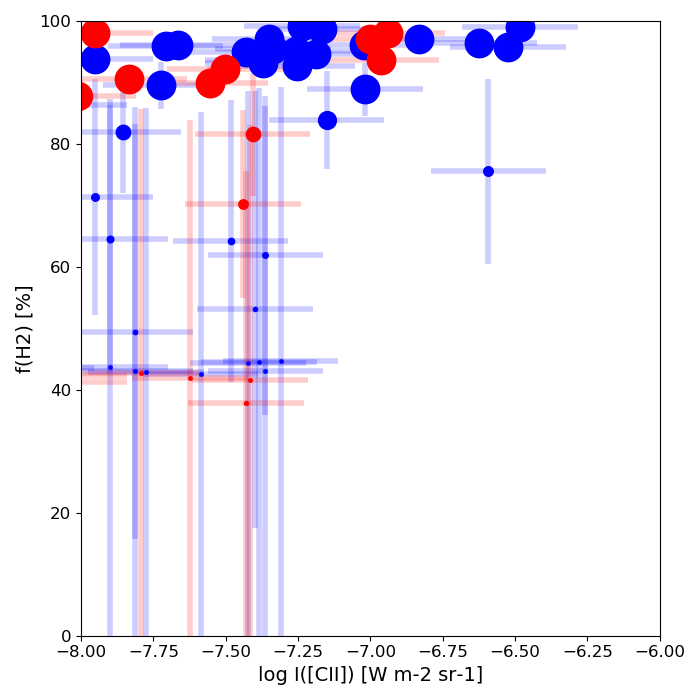}
\includegraphics[width=6cm]{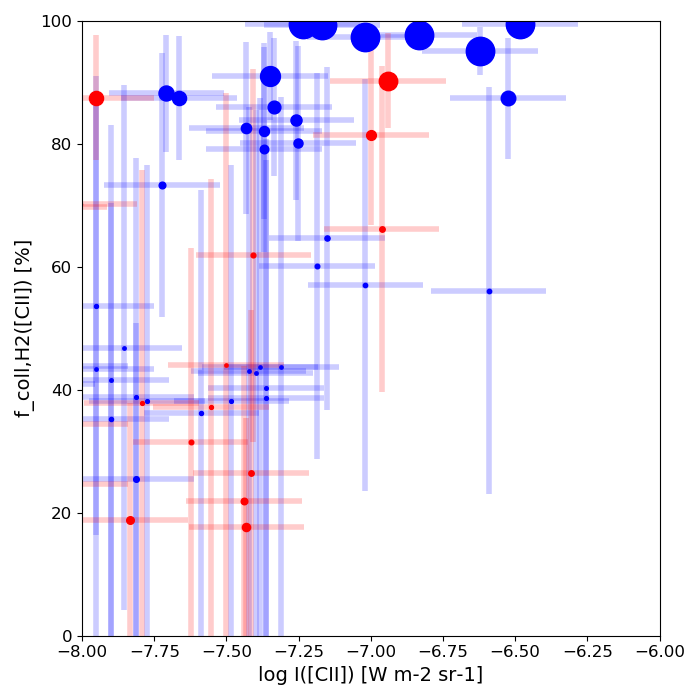}
\begin{tikzpicture}
  \node[anchor=south west,inner sep=0] (image) at (0,0) {\includegraphics[width=6cm]{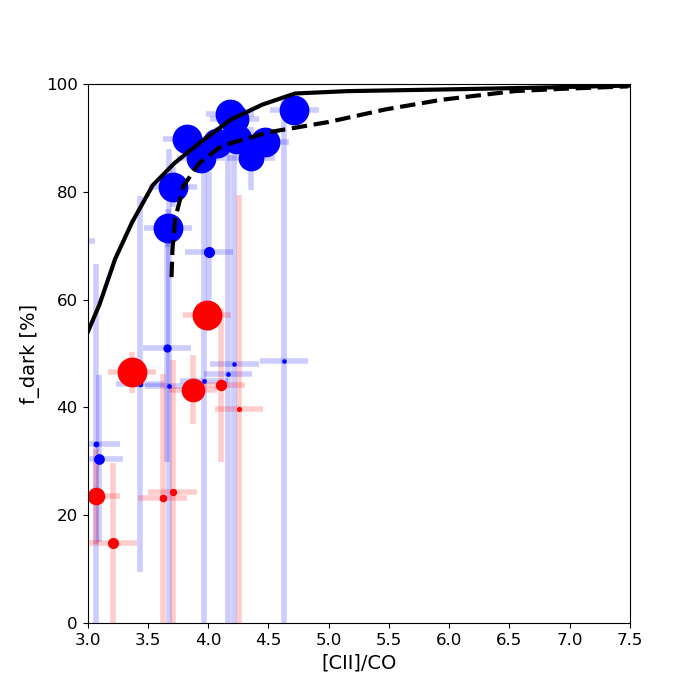}}; 
\end{tikzpicture}
\begin{tikzpicture}
  \node[anchor=south west,inner sep=0] (image) at (0,0) {\includegraphics[width=6cm]{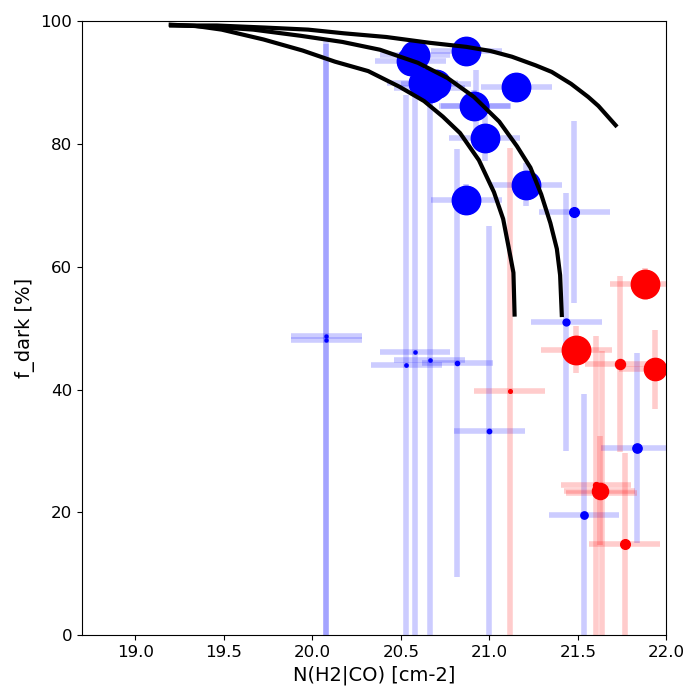}}; 
  \end{tikzpicture}
\caption{Results from \cite{Pineda2017a} for LMC (blue) and SMC (red) regions. The symbol size is inversely proportional to the error bar of the y-axis parameter. The curves are the same as in Figs.\,\ref{fig:model_plots} and \ref{fig:resdarknco_each}. }\label{fig:respineda}
\end{figure*}

Results are shown in Figure\,\ref{fig:respineda}. Overall, the results agree with our study for LMC data points, that is, bright [C\2] components have a molecular gas fraction near unity, the contribution of neutral atomic gas to the [C\2] emission is larger for faint [C\2] components ($\lesssim6\times10^{-8}$\,W\,m$^{-2}$\,sr$^{-1}$), and the fraction of CO-dark H$_2$ gas decreases with the CO column density. The comparison with our results illustrates some advantages related to the profile decomposition. First, while our study was performed with fewer pointings,  by accounting for multiple velocity components corresponding to various physical conditions, statistically significant results could be obtained. Furthermore, we were able to some extent to probe components with large [C\2]/CO values which may correspond to relatively quiescent clouds (Fig.\,\ref{fig:resdarkciico_each}).

Toward SMC regions, the fraction of [C\2] emission corresponding to CO-dark H$_2$ gas, $f_{\rm coll,H2}({\rm [CII]})$, is somewhat lower, while $f_{\rm dark}$ is somewhat larger {for a given CO column density} as compared to the LMC. In other words, more [C\2] can be explained by atomic gas in the SMC, but the fraction of CO-dark H$_2$ gas is larger. It is possible that the larger $f_{\rm dark}$ values obtained for the SMC are related to a larger incident radiation field but it is also possible that the gas temperature we assume for the SMC (the same as for the LMC) is not applicable. As noted by \cite{Glover2016a}, the detectability of the CO-dark H$_2$ gas with [C\2] or [O\1] in fact improves greatly with the ISRF strength $G_0$ (which increases the photoelectric-effect heating rate and hence the gas temperature). While we have chosen the same temperatures for the LMC and SMC regions, $f_{\rm dark}$ in the SMC would be compatible with the LMC values if the temperature was scaled up by a factor of only two to four. 
Results for the DGS using integrated measurements show that there is no simple relationship between metallicity and $f_{\rm dark}$ and that the latter mostly depends on extinction $A_V$ (Madden et al.\ in prep.).

\section{Conclusions}

We present a study of the CO-dark H$_2$ gas in the giant H\2\ region N\,11 in the LMC, with a half-solar metallicity. Twelve pointings were observed with SOFIA/GREAT in [C\2]. The [C\2] velocity profile is compared to that of CO(1-0) observed with MOPRA and ALMA, and to that of H\1\ observed with ATCA+Parkes. The objectives are to measure the total molecular gas content, the fraction of CO-dark H$_2$ gas, and to probe the influence of the environment (in particular stellar feedback). First we investigated velocity-integrated properties. The main results obtained are as follows.
\begin{itemize}
\item The [C\2] emission originates mostly from the neutral gas, except for one velocity component toward one of the pointings (\#5) for which the result is unclear. This result is found through various methods using the [C\2] line width, the [N\2]/[C\2] ratio in the ionized gas, the [N\2] $205$\mic\ velocity profile observed with GREAT, and the photoelectric-effect heating efficiency proxy ([C\2]+[O\1])/PAH.
\item The photoelectric-effect heating efficiency proxy ([C\2]+[O\1])/PAH is found to be constant throughout the regions in N\,11 observed with \textit{Herschel}/PACS, generalizing somewhat the results previously obtained in N\,11B by \cite{Lebouteiller2012b} that the sum [C\2]+[O\1] traces the total cooling, that PAH emission traces the gas heating, and that [C\2] mostly originates from the neutral gas. Our results suggest that this gas is CO-dark H$_2$ and not atomic. 
\item The total profile width of [C\2] is found to be between that of CO(1-0) and H\1, but the [C\2] profile resembles more that of CO(1-0). 
\end{itemize}

We then decomposed the profiles using a Bayesian method and a statistical approach that makes use of the many pointings together with a range of input parameters concerning the number of velocity components, the minimum individual component line width, and the minimum separation between components. A simple model was used to compute the [C\2] line intensity accounting for collisions with H$^0$, H$_2$, and $e^-$ in a two-phase medium (neutral atomic and molecular) as a function of gas temperature and density. The main results obtained are as follows:
\begin{itemize}
\item The variations of $I({\rm [CII]})$ are driven by the [C\2] emission in the molecular phase. As a result, the [C\2] components with the largest contribution from H$^0$ gas are preferentially those with a low [C\2] surface brightness rather than those with low [C\2]/CO or low CO column density. However, the contribution from H$^0$ gas to [C\2] is never dominant. 
\item There is a sharp transition between CO-bright and CO-dark H$_2$ gas, with the latter quickly becoming the dominant H$_2$ reservoir. 
\item Overall (combining all pointings and all velocity components), more than $90$\%\ of the [C\2] emission arises in the (CO-dark) H$_2$ gas.
\item Most of the molecular gas is CO-dark (fraction between $40$\ and $100$\%), in particular toward the brightest [C\2] components.
\item The CO-dark H$_2$ gas traced by [C\2] is rather diffuse with $\sim200$\cc\ on average for all pointings. We identify in particular a specific [C\2] velocity component toward \#5 with a density around $\sim100$\cc. 
\item The contribution of optically thick H\1\ to the dark neutral medium is not significant.
\item Most components follow the same trend of the fraction of CO-dark H$_2$ versus\ [C\2]/CO, with some deviations driven by the gas density.
\item The effective $X_{\rm CO}$ factor including the CO-dark H$_2$ gas lies in the range $10^{21-22}$\,(K\,\kms)$^{-1}$ for most of the bright velocity components. 
\item The fraction of CO-dark H$_2$ gas decreases with increasing CO column density, but while it is rather constant for CO column density $<10^{20.5}$\,cm$^{-2}$, it shows a large dispersion above this value. We argue that, for a given CO column density, the larger fractions of CO-dark H$_2$ gas are found toward CO clouds at the interface with H$\alpha$-bright regions. It is plausible that stellar feedback (either through radiation or dynamical/mechanical effects from stellar winds and supernovae shocks) results in the disruption and dispersal of molecular clouds and in turn a lower extinction on average.
\item Our simple models were applied to the LMC and SMC pointings in \cite{Pineda2017a}. We find circumstancial evidence that the fraction of CO-dark H$_2$ gas is larger for a given CO column density in the SMC as compared to the LMC, but this conclusion is weakened by the uncertain gas temperature. 
\end{itemize}

The main caveat concerns the derived column density and number density in the atomic medium, with limited spatial resolution, even though the velocity decomposition somewhat mitigates the correspondence between components in the different tracers. Further observations at larger spatial resolution in H\1\ would greatly improve our knowledge of the origin of [C\2] and its performance as a CO-dark H$_2$ gas tracer. Moreover, observations of [O\1] would shed light on the physical conditions of the few components where [C\2] arises in the neutral atomic medium. In particular, more constraints are needed to examine the incident radiation field and extinction in individual clouds, which would then help the understanding of the nature of stellar feedback responsible for the variations of the fraction of CO-dark H$_2$ gas. Finally, we are aware that the assumption that the velocity components from different tracers arise from a given cloud is increasingly problematic at increasingly large scales, but we cannot unfortunately thoroughly test this hypothesis with the present dataset.

\begin{acknowledgements}
Based on observations made with the NASA/DLR Stratospheric Observatory for Infrared Astronomy (SOFIA). SOFIA is jointly operated by the Universities Space Research Association, Inc. (USRA), under NASA contract NAS2-97001, and the Deutsches SOFIA Institut (DSI) under DLR contract 50 OK 0901 to the University of Stuttgart. This paper makes use of the following ALMA data: ADS/JAO.ALMA\#2012.1.00532.S and ADS/JAO.ALMA\#2013.1.00556.S. ALMA is a partnership of ESO (representing its member states), NSF (USA) and NINS (Japan), together with NRC (Canada) and NSC and ASIAA (Taiwan) and KASI (Republic of Korea), in cooperation with the Republic of Chile. The Joint ALMA Observatory is operated by ESO, AUI/NRAO and NAOJ. V.\ Lebouteiller wishes to thank S.\ Kannappan and C.\ Iliadis for making it possible to do part of this work at UNC Chapel Hill. MC gratefully acknowledges funding from the Deutsche Forschungsgemeinschaft (DFG) through an Emmy Noether Research Group, grant number KR4801/1-1. M.-Y.L. was partially funded through the sub-project A6 of the Collaborative Research Council 956, funded by the Deutsche Forschungsgemeinschaft (DFG). DC is supported by the European Union's Horizon 2020 research and innovation programme under the Marie Sk\l{}odowska-Curie grant agreement No 702622. FLP is supported by the ANR grant LYRICS (ANR-16-CE31-0011). This work was supported by the Programme National ``Physique et Chimie du Milieu Interstellaire'' (PCMI) of CNRS/INSU with INC/INP co-funded by CEA and CNES. 
\end{acknowledgements}

\pagebreak
\bibliography{/local/home/vleboute/ownCloud/bibtexendum/bibtexendum}

\pagebreak

\appendix

\section{Profile decomposition for each pointing}\label{secapp:profiles}

\begin{figure*}
  \includegraphics[width=4cm,clip,trim=0 0 0 0,valign=t]{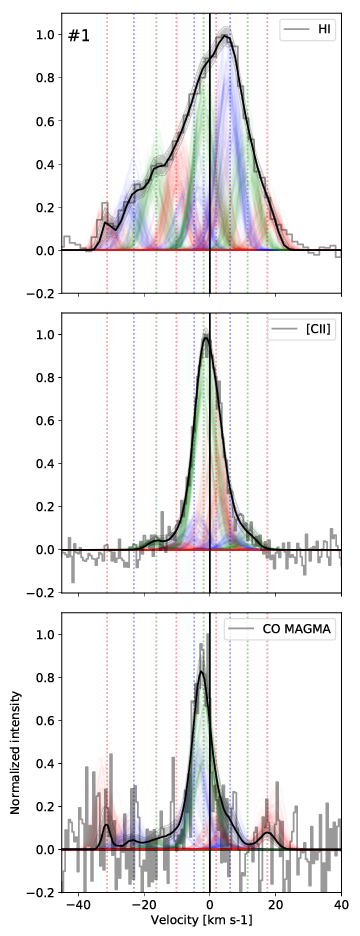}
  \includegraphics[width=4cm,clip,trim=0 0 0 0,valign=t]{{Figures/Decomposition/ptgfit_2/ncomps10_minsig1.0_mindeltav1.0/Fit.pdf}.png}
  \includegraphics[width=4cm,clip,trim=0 0 0 0,valign=t]{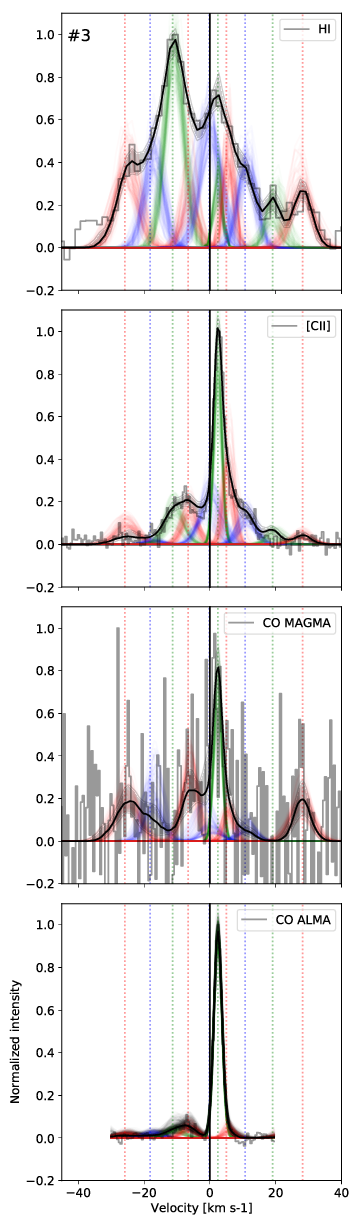}
\caption{Profile decomposition for pointings \#1, \#2, and \#3.  }\label{fig:decompose_pointing123_illustration}
\end{figure*}

\begin{figure*}
  \includegraphics[width=4cm,clip,trim=0 0 0 0,valign=t]{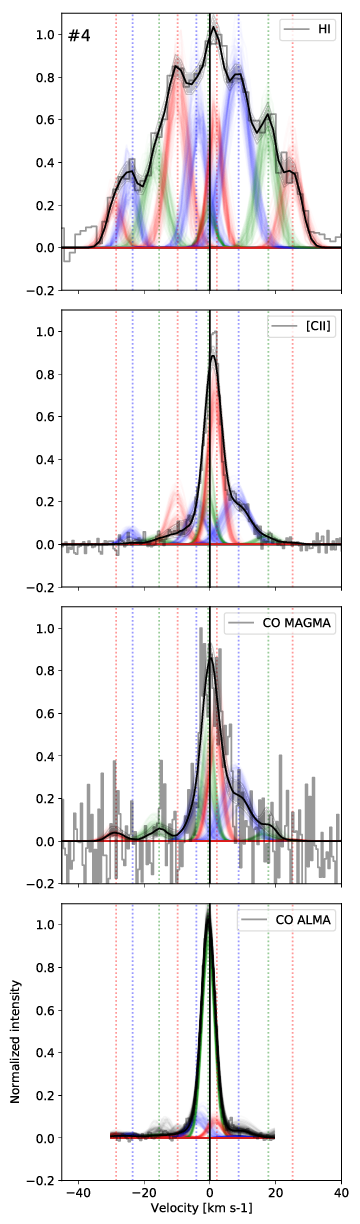}
  \includegraphics[width=4cm,clip,trim=0 0 0 0,valign=t]{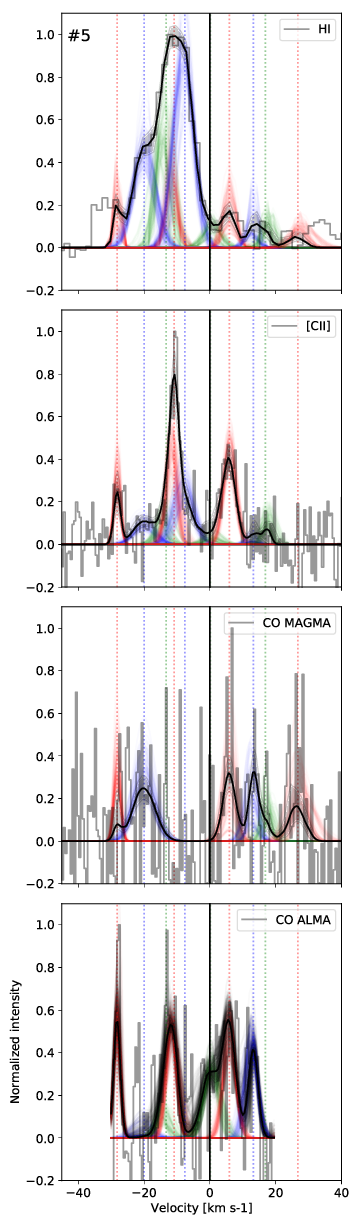}
  \includegraphics[width=4cm,clip,trim=0 0 0 0,valign=t]{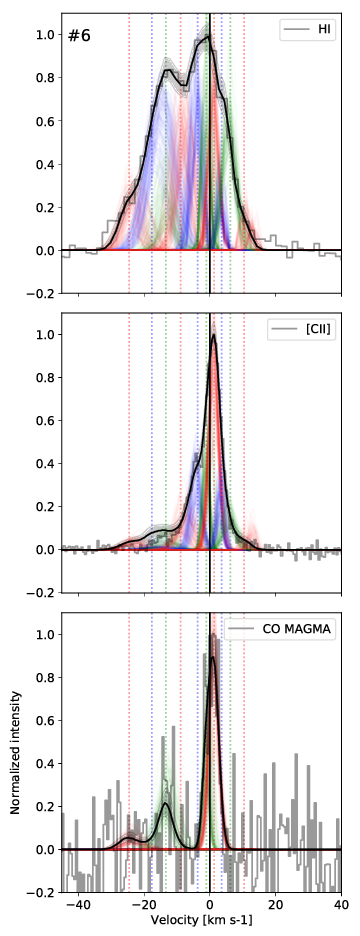}
\caption{Profile decomposition for pointings \#4, \#5, and \#6.  }\label{fig:decompose_pointing456_illustration}
\end{figure*}

\begin{figure*}
  \includegraphics[width=4cm,clip,trim=0 0 0 0,valign=t]{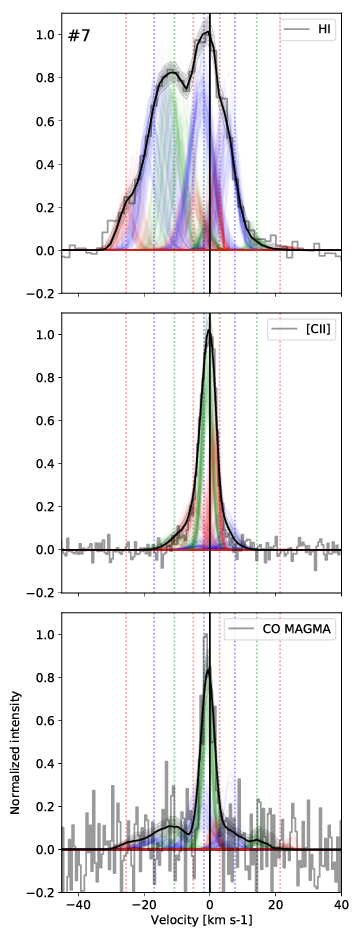}
  \includegraphics[width=4cm,clip,trim=0 0 0 0,valign=t]{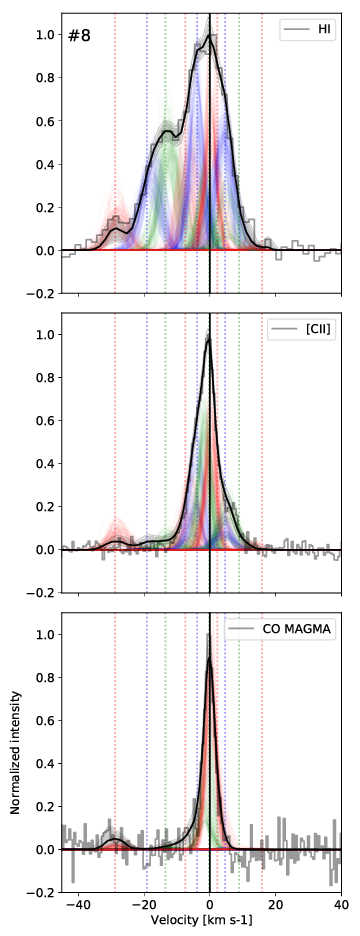}
  \includegraphics[width=4cm,clip,trim=0 0 0 0,valign=t]{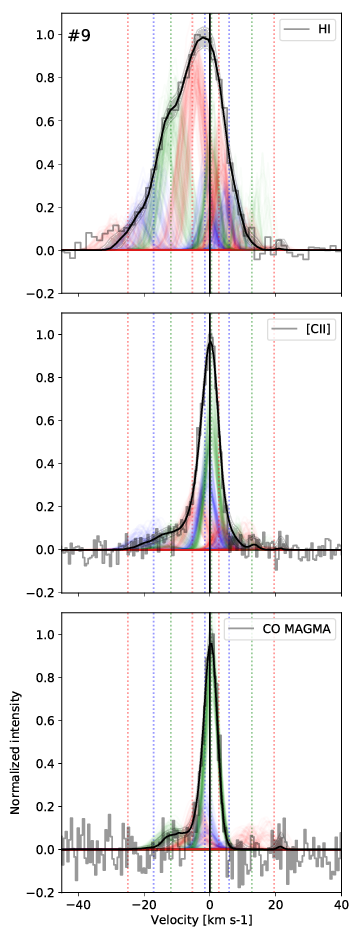}
\caption{Profile decomposition for pointings \#7, \#8, and \#9.  }\label{fig:decompose_pointing789_illustration}
\end{figure*}

\begin{figure*}
  \includegraphics[width=4cm,clip,trim=0 0 0 0,valign=t]{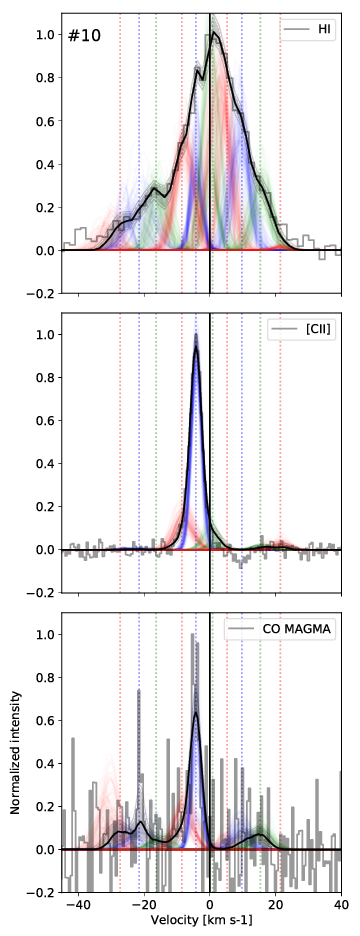}
  \includegraphics[width=4cm,clip,trim=0 0 0 0,valign=t]{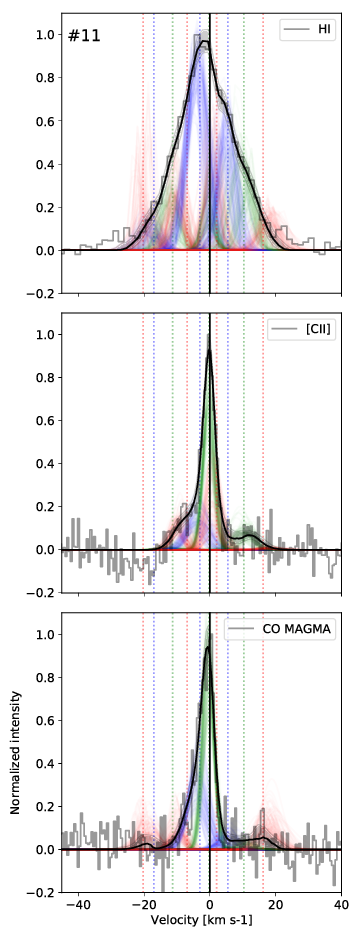}
  \includegraphics[width=4cm,clip,trim=0 0 0 0,valign=t]{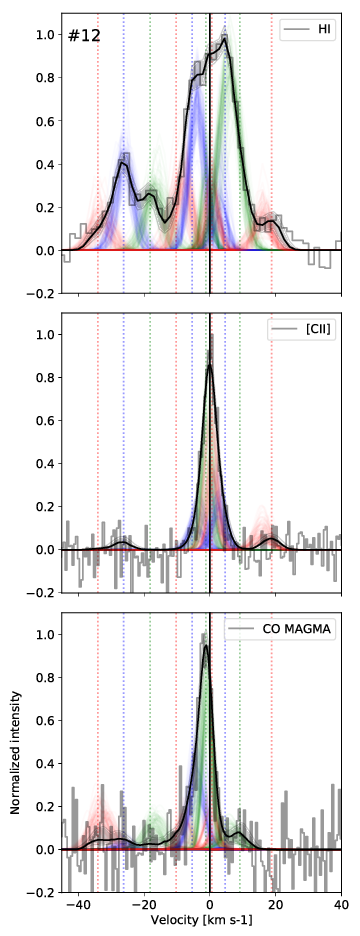}
\caption{Profile decomposition for pointings \#10, \#11, and \#12.  }\label{fig:decompose_pointing101112_illustration}
\end{figure*}

The spectral decomposition of the H\1\, CO, and [C\2] spectra obtained for the $12$ GREAT pointings is shown in Figures\,\ref{fig:decompose_pointing123_illustration}, \ref{fig:decompose_pointing456_illustration}, \ref{fig:decompose_pointing789_illustration}, \ref{fig:decompose_pointing101112_illustration}. The decomposition is only shown for the model with ten components, a minimum line width of $1$\kms, and a minimum component separation of $1$\kms. 

\section{Kernel density estimates}\label{secapp:kdes}

\begin{figure*} 
\begin{tikzpicture}
  \node[anchor=south west,inner sep=0] (image) at (0,0) {\includegraphics[width=6cm,trim=20 20 20 20]{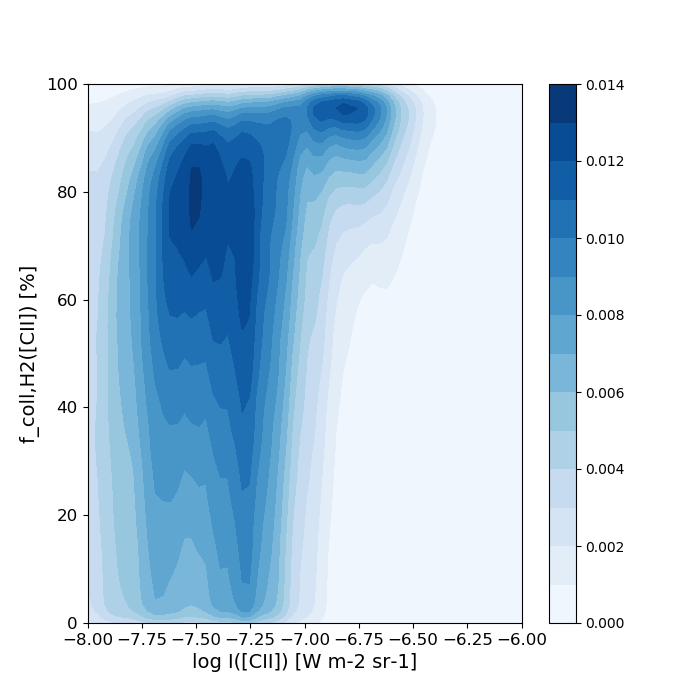}}; \node[right] at (4.,1) {\#1}; 
\end{tikzpicture}
\begin{tikzpicture}
  \node[anchor=south west,inner sep=0] (image) at (0,0) {\includegraphics[width=6cm,trim=20 20 20 20]{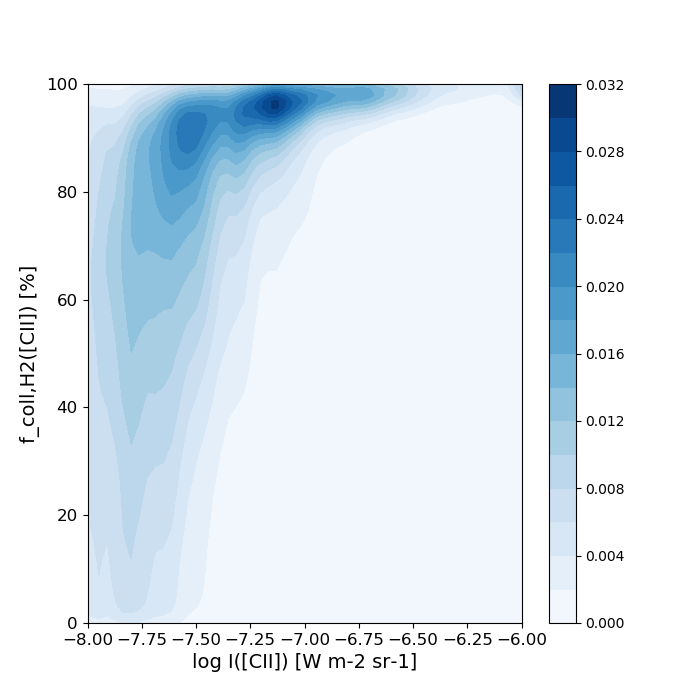}}; \node[right] at (4.,1) {\#2}; 
\end{tikzpicture}
\begin{tikzpicture}
  \node[anchor=south west,inner sep=0] (image) at (0,0) {\includegraphics[width=6cm,trim=20 20 20 20]{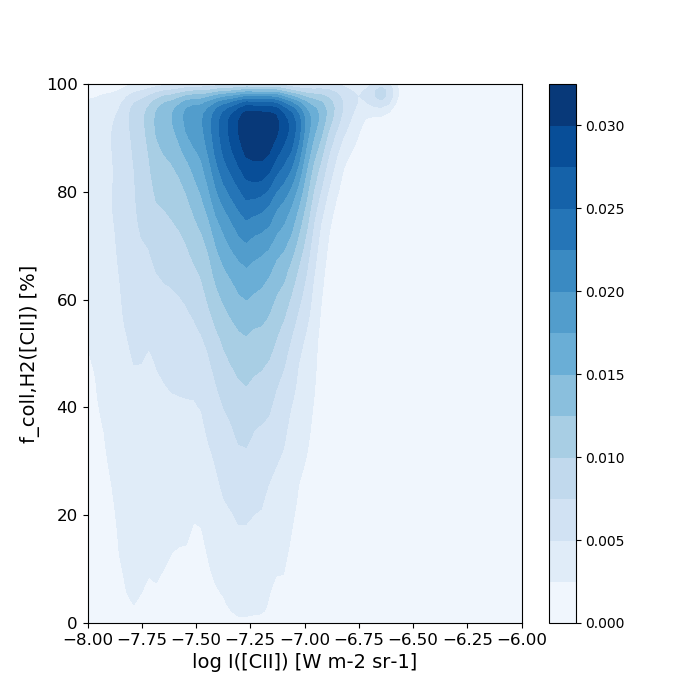}}; \node[right] at (4.,1) {\#3}; 
\end{tikzpicture}
\begin{tikzpicture}
  \node[anchor=south west,inner sep=0] (image) at (0,0) {\includegraphics[width=6cm,trim=20 20 20 20]{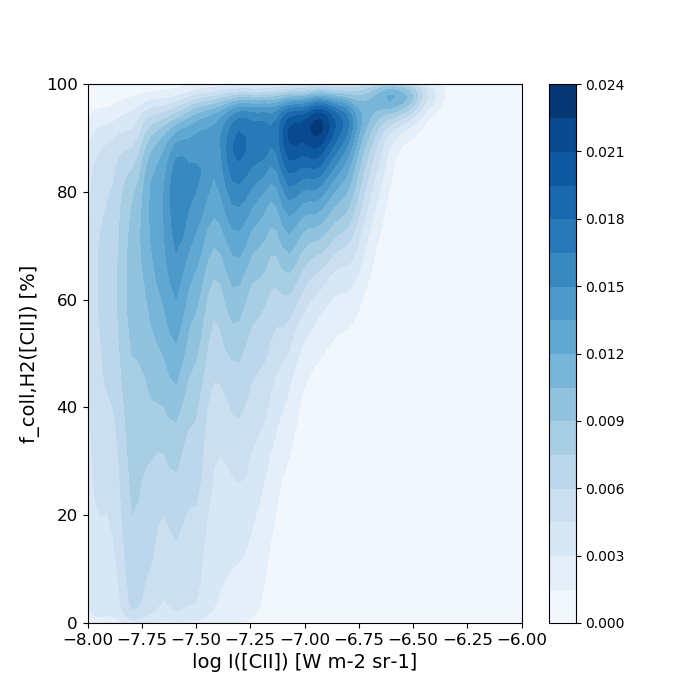}}; \node[right] at (4.,1) {\#4};
\end{tikzpicture}
\begin{tikzpicture}
  \node[anchor=south west,inner sep=0] (image) at (0,0) {\includegraphics[width=6cm,trim=20 20 20 20]{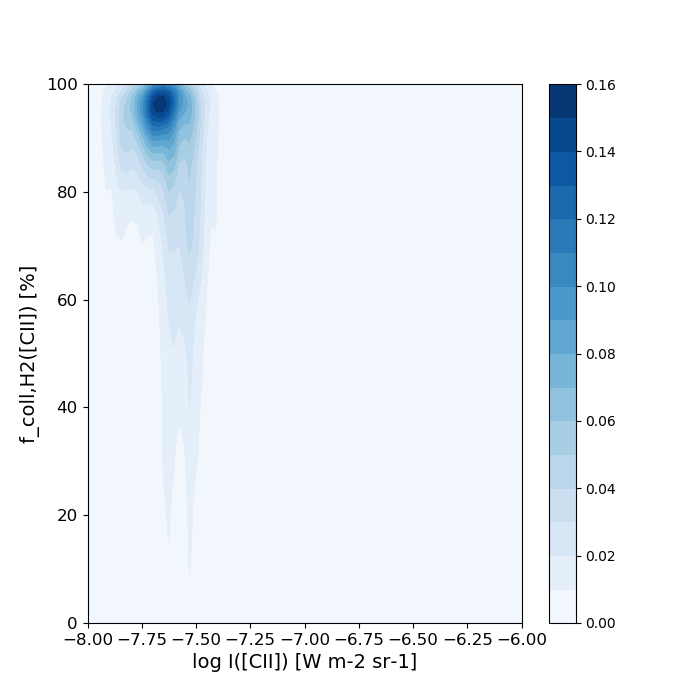}}; \node[right] at (4.,1) {\#5};
\end{tikzpicture}
\begin{tikzpicture}
  \node[anchor=south west,inner sep=0] (image) at (0,0) {\includegraphics[width=6cm,trim=20 20 20 20]{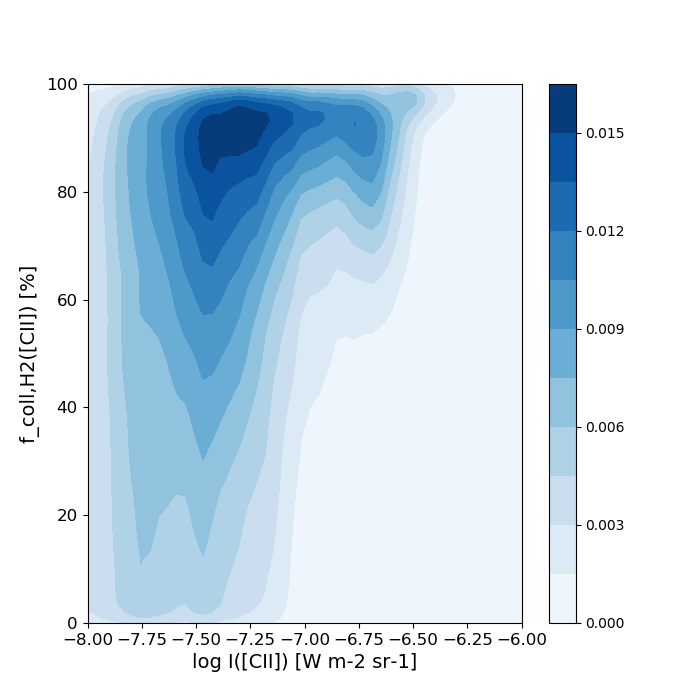}}; \node[right] at (4.,1) {\#6}; 
\end{tikzpicture}
\begin{tikzpicture}
  \node[anchor=south west,inner sep=0] (image) at (0,0) {\includegraphics[width=6cm,trim=20 20 20 20]{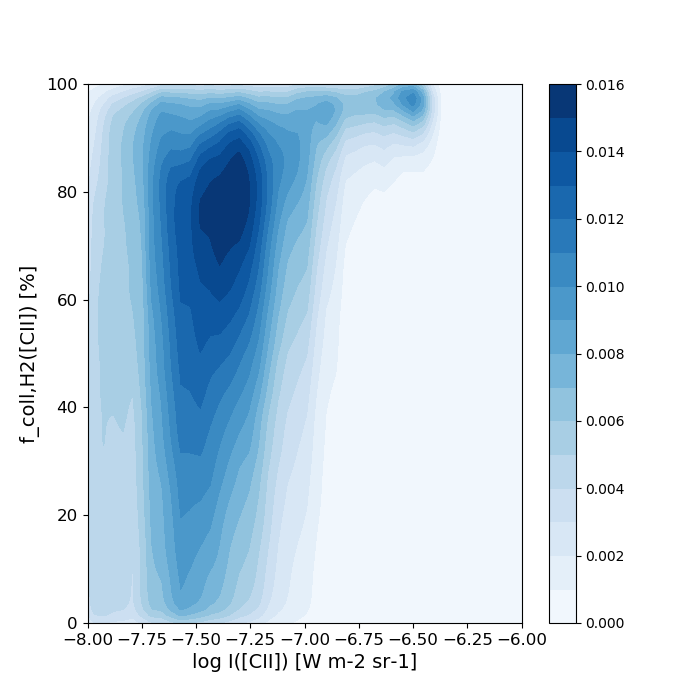}}; \node[right] at (4.,1) {\#7}; 
\end{tikzpicture}
\begin{tikzpicture}
  \node[anchor=south west,inner sep=0] (image) at (0,0) {\includegraphics[width=6cm,trim=20 20 20 20]{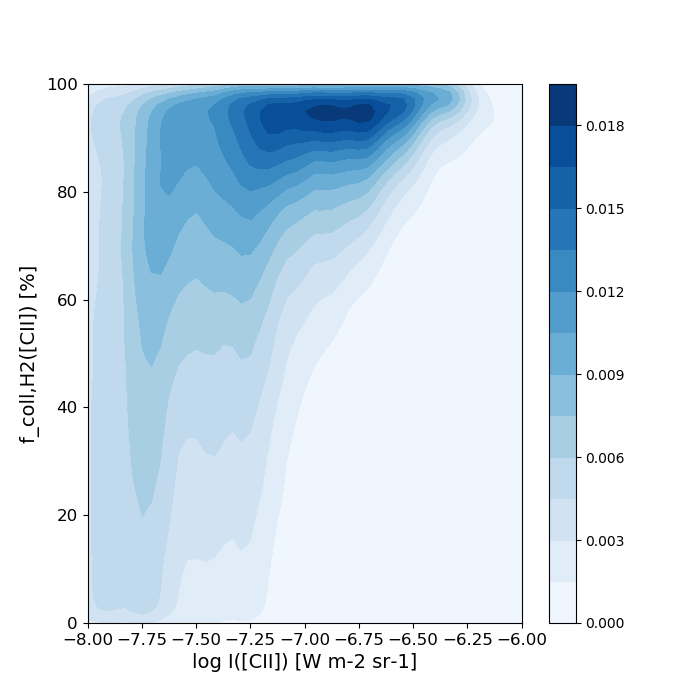}}; \node[right] at (4.,1) {\#8}; 
\end{tikzpicture}
\begin{tikzpicture}
  \node[anchor=south west,inner sep=0] (image) at (0,0) {\includegraphics[width=6cm,trim=20 20 20 20]{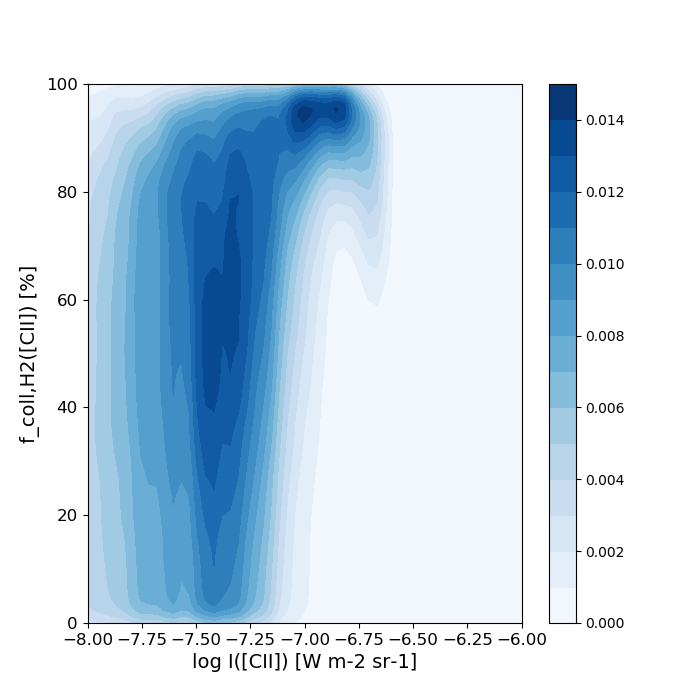}}; \node[right] at (4.,1) {\#9}; 
\end{tikzpicture}
\begin{tikzpicture}
  \node[anchor=south west,inner sep=0] (image) at (0,0) {\includegraphics[width=6cm,trim=20 20 20 20]{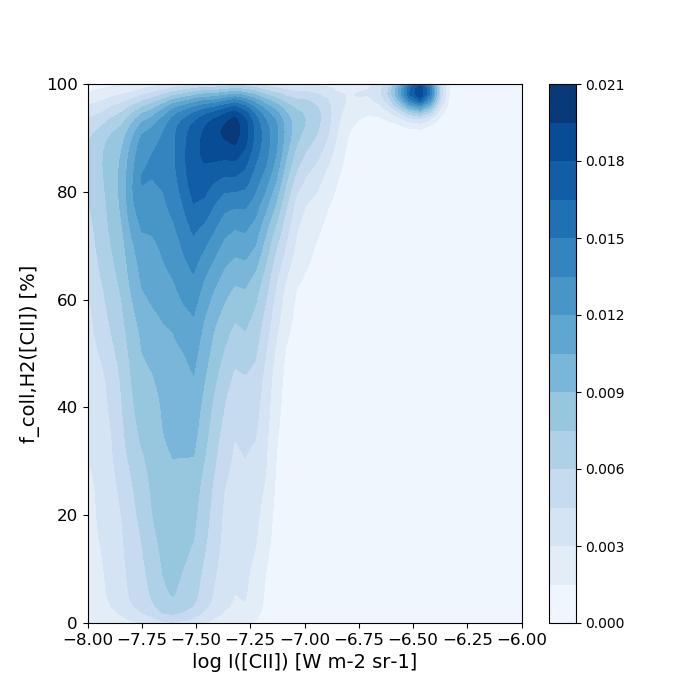}}; \node[right] at (4.,1) {\#10}; 
\end{tikzpicture}
\begin{tikzpicture}
  \node[anchor=south west,inner sep=0] (image) at (0,0) {\includegraphics[width=6cm,trim=20 20 20 20]{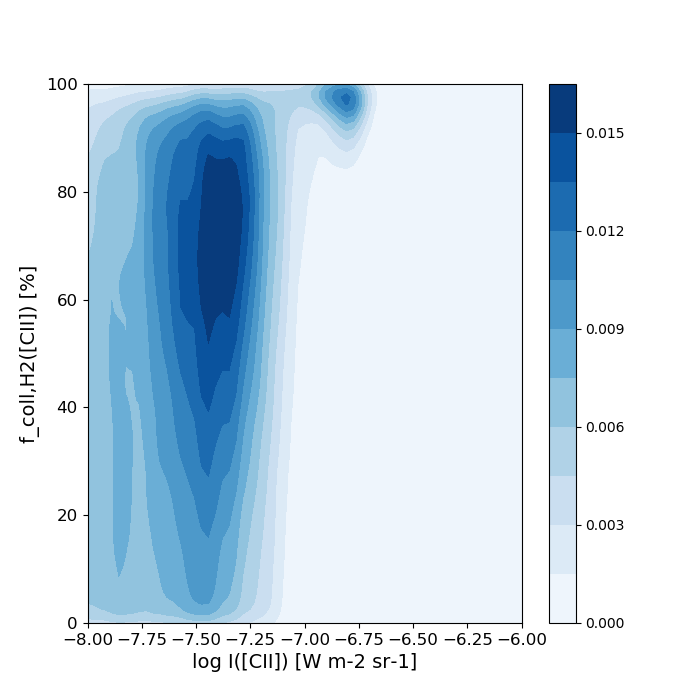}}; \node[right] at (4.,1) {\#11}; 
\end{tikzpicture}
\begin{tikzpicture}
  \node[anchor=south west,inner sep=0] (image) at (0,0) {\includegraphics[width=6cm,trim=20 20 20 20]{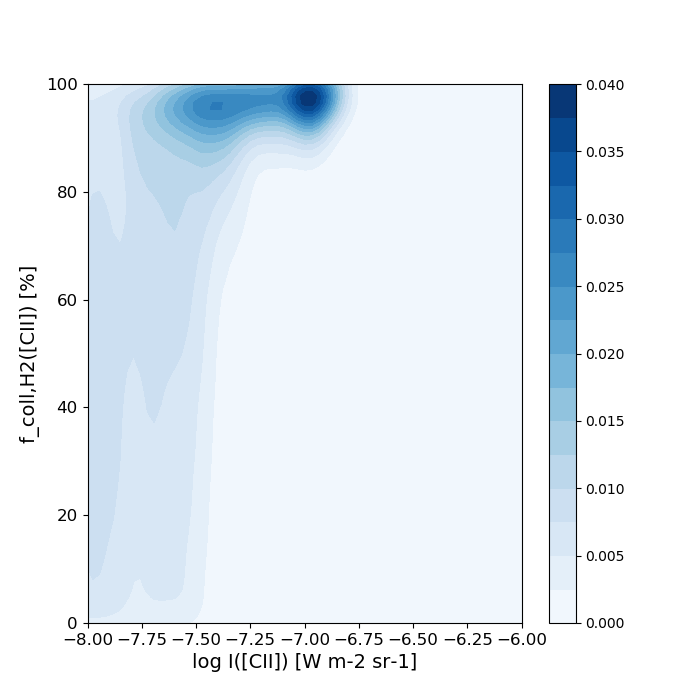}}; \node[right] at (4.,1) {\#12};
\end{tikzpicture}
\caption{Bivariate kernel density estimate of the fraction of [C\2] tracing CO-dark H$_2$ gas vs.\ $I([{\rm CII}])$ for all pointings. The shade scales with the density of points. }\label{fig:resfcollh2icii_each}
\end{figure*}

The kernel density estimate of the fraction of [C\2] tracing CO-dark H$_2$ gas, $f_{\rm coll,H2}({\rm [CII]})$, and $I([{\rm CII}])$ is shown in Figure\,\ref{fig:resfcollh2icii_each}. The distribution is globally similar to that of $f({\rm H}_2)$ versus\ $I([{\rm CII}])$ (Fig.\,\ref{fig:resiftot_each}) because most H$_2$ is CO-dark and because the CO-dark H$_2$ gas is traced by [C\2].

\begin{figure*} 
\begin{tikzpicture}
  \node[anchor=south west,inner sep=0] (image) at (0,0) {\includegraphics[width=6cm,trim=20 20 20 20]{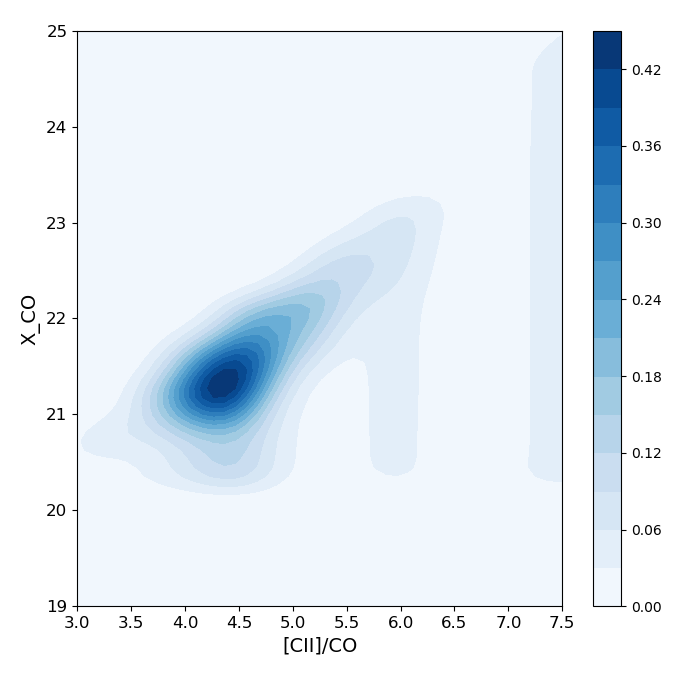}}; \node[right] at (4.2,1) {\#1}; 
\end{tikzpicture}
\begin{tikzpicture}
  \node[anchor=south west,inner sep=0] (image) at (0,0) {\includegraphics[width=6cm,trim=20 20 20 20]{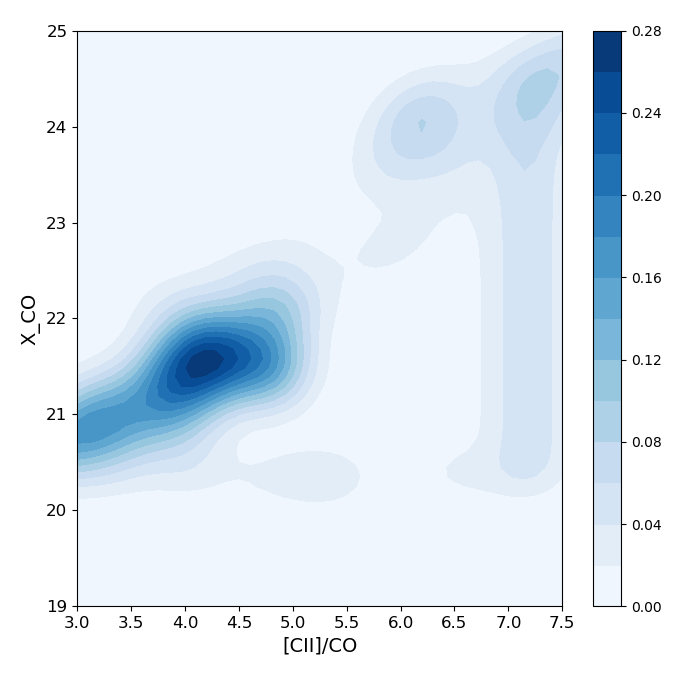}}; \node[right] at (4.2,1) {\#2}; 
\end{tikzpicture}
\begin{tikzpicture}
  \node[anchor=south west,inner sep=0] (image) at (0,0) {\includegraphics[width=6cm,trim=20 20 20 20]{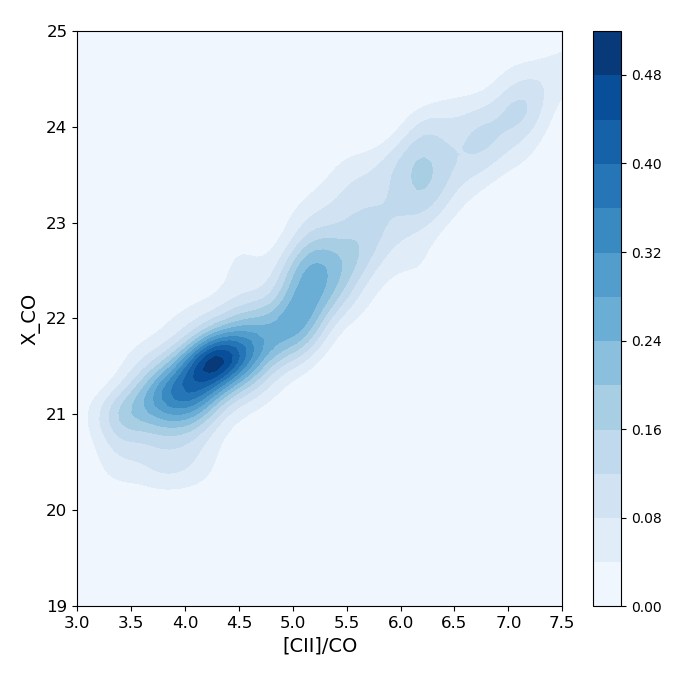}}; \node[right] at (4.2,1) {\#3}; 
\end{tikzpicture}
\begin{tikzpicture}
  \node[anchor=south west,inner sep=0] (image) at (0,0) {\includegraphics[width=6cm,trim=20 20 20 20]{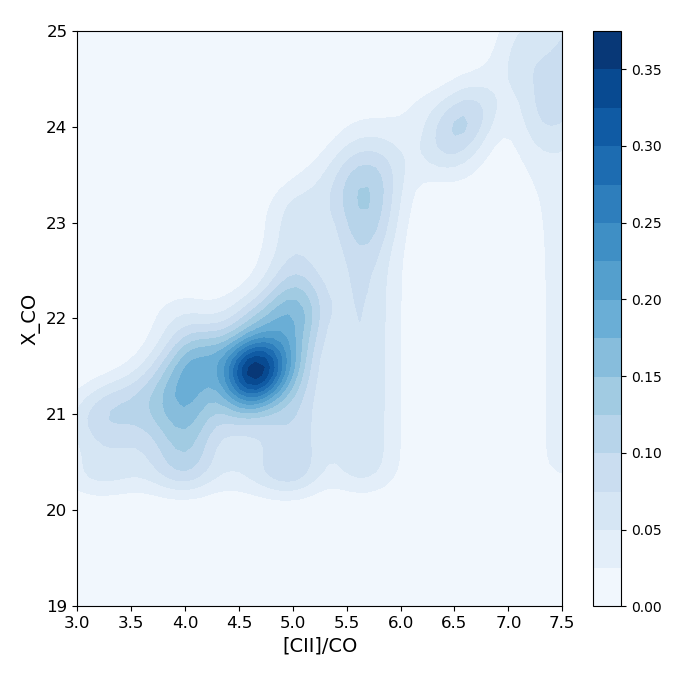}}; \node[right] at (4.2,1) {\#4};
\end{tikzpicture}
\begin{tikzpicture}
  \node[anchor=south west,inner sep=0] (image) at (0,0) {\includegraphics[width=6cm,trim=20 20 20 20]{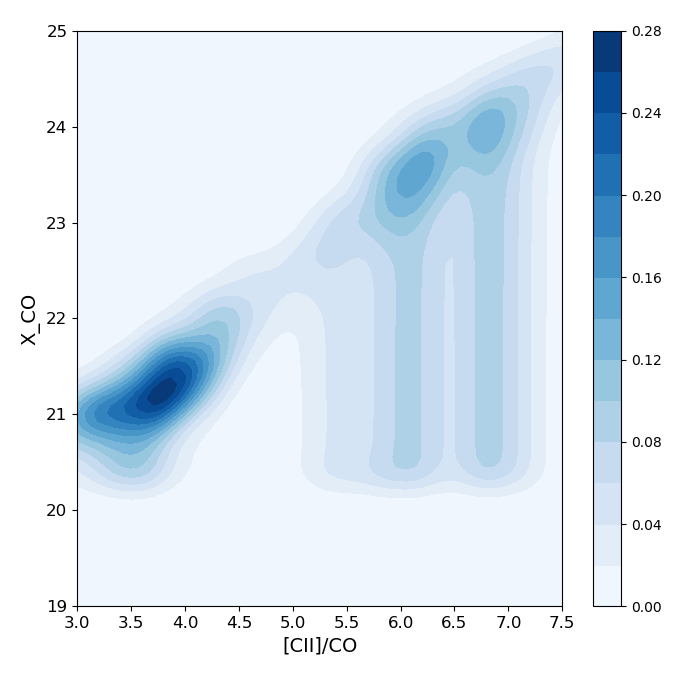}}; \node[right] at (4.2,1) {\#5};
\end{tikzpicture}
\begin{tikzpicture}
  \node[anchor=south west,inner sep=0] (image) at (0,0) {\includegraphics[width=6cm,trim=20 20 20 20]{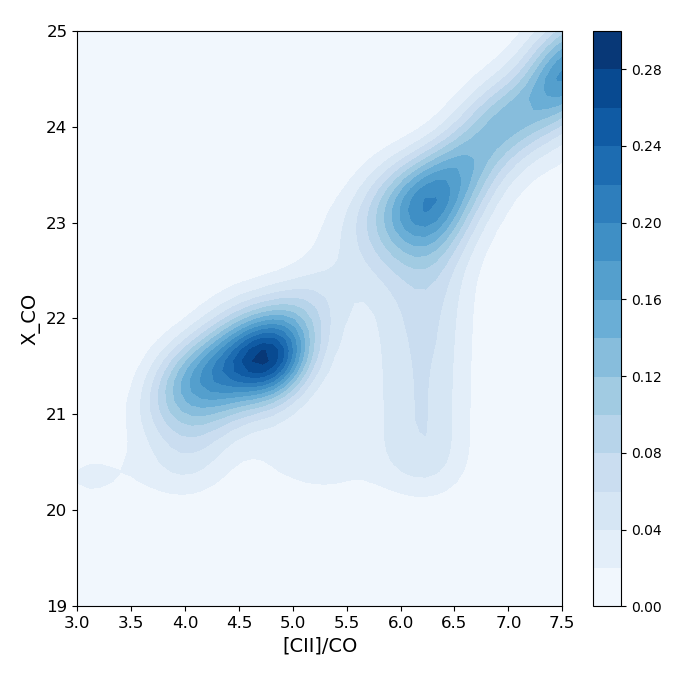}}; \node[right] at (4.2,1) {\#6}; 
\end{tikzpicture}
\begin{tikzpicture}
  \node[anchor=south west,inner sep=0] (image) at (0,0) {\includegraphics[width=6cm,trim=20 20 20 20]{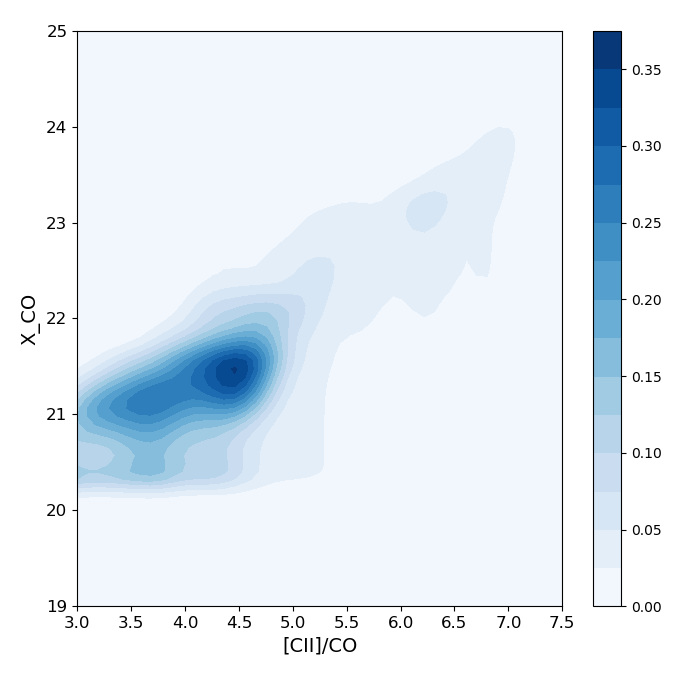}}; \node[right] at (4.2,1) {\#7}; 
\end{tikzpicture}
\begin{tikzpicture}
  \node[anchor=south west,inner sep=0] (image) at (0,0) {\includegraphics[width=6cm,trim=20 20 20 20]{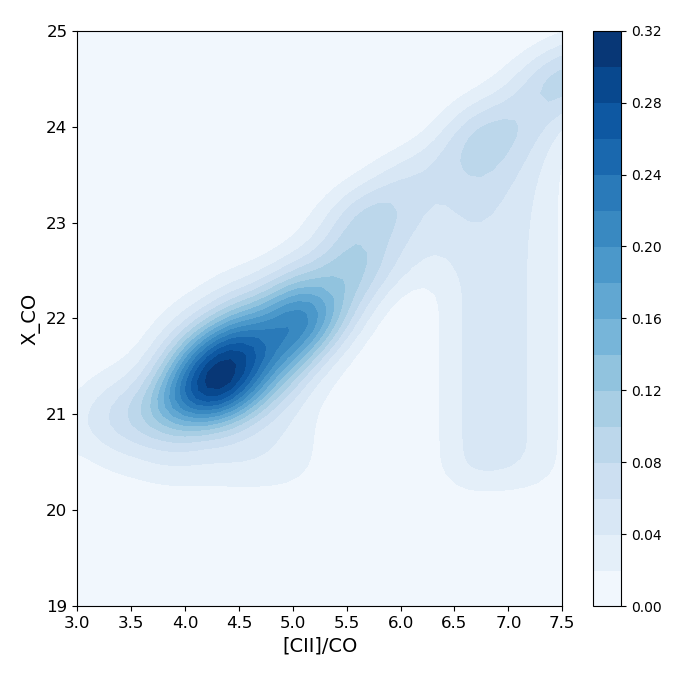}}; \node[right] at (4.2,1) {\#8}; 
\end{tikzpicture}
\begin{tikzpicture}
  \node[anchor=south west,inner sep=0] (image) at (0,0) {\includegraphics[width=6cm,trim=20 20 20 20]{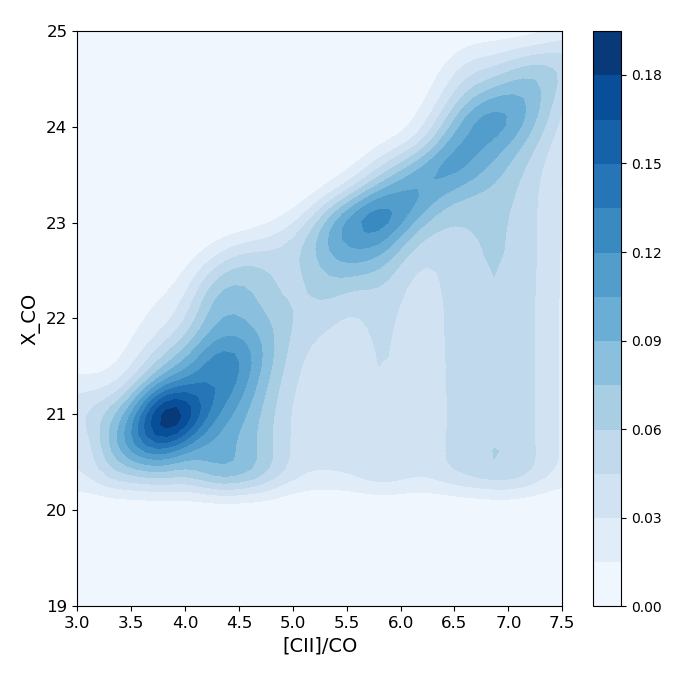}}; \node[right] at (4.2,1) {\#9}; 
\end{tikzpicture}
\begin{tikzpicture}
  \node[anchor=south west,inner sep=0] (image) at (0,0) {\includegraphics[width=6cm,trim=20 20 20 20]{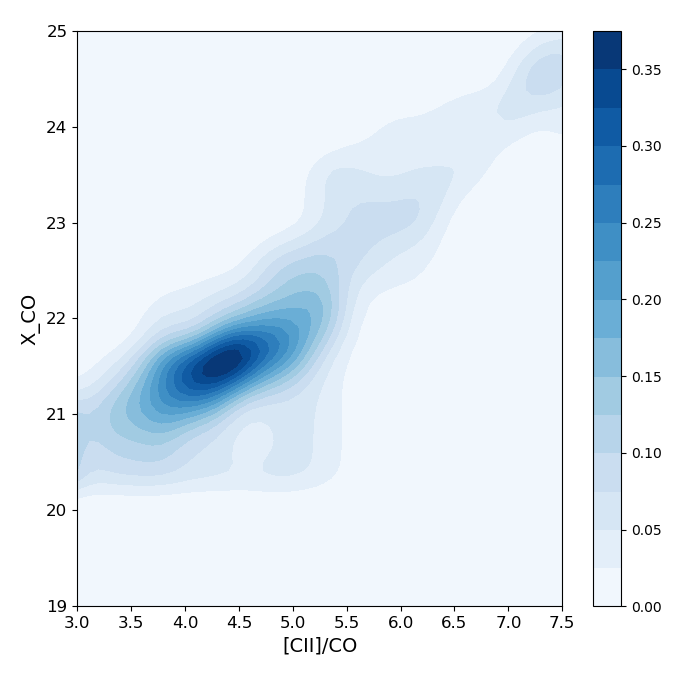}}; \node[right] at (4.2,1) {\#10}; 
\end{tikzpicture}
\begin{tikzpicture}
  \node[anchor=south west,inner sep=0] (image) at (0,0) {\includegraphics[width=6cm,trim=20 20 20 20]{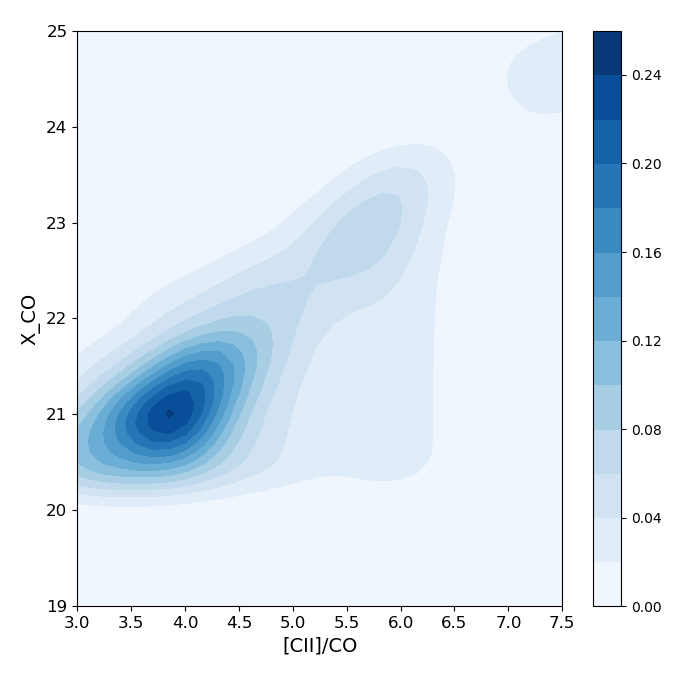}}; \node[right] at (4.2,1) {\#11}; 
\end{tikzpicture}
\begin{tikzpicture}
  \node[anchor=south west,inner sep=0] (image) at (0,0) {\includegraphics[width=6cm,trim=20 20 20 20]{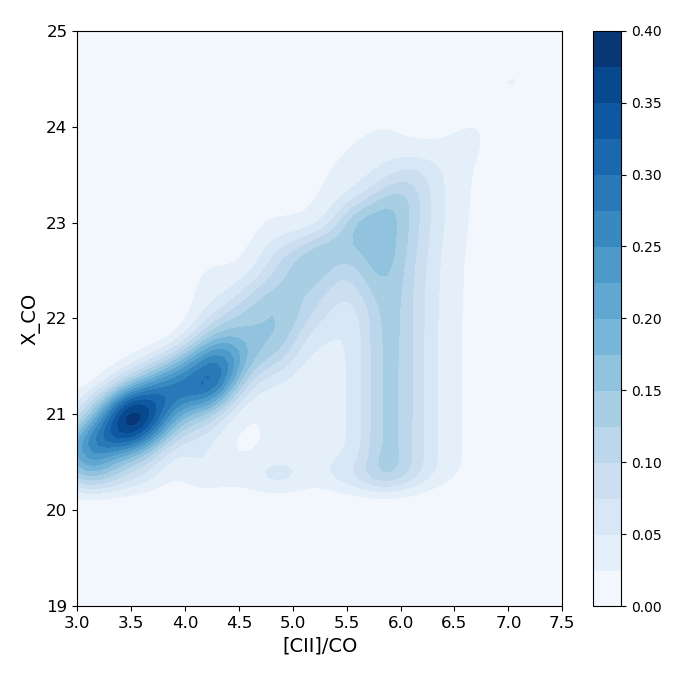}}; \node[right] at (4.2,1) {\#12};
\end{tikzpicture}
\caption{Bivariate kernel density estimate of $X_{\rm CO}$ vs.\ [C\2]/CO for all pointings. The shade scales with the density of points.}\label{fig:resxco_ciico_each}
\end{figure*}

Figure\,\ref{fig:resxco_ciico_each} shows the kernel density estimate of the effective $X_{\rm CO}$ conversion factor and the [C\2]/CO ratio. Since the $X_{\rm CO}$ conversion factor includes the contribution of the CO-dark H$_2$ gas, it is larger than the fiducial value $X'_{\rm CO}=2\times10^{20}$\,cm$^{-2}$\,(K\,\kms)$^{-1}$ (Sect.\,\ref{sec:obs_co}). We find an effective $X_{\rm CO}$ in the range $10^{21-22}$\,(K\,\kms)$^{-1}$ for most of the bright velocity components, in good agreement with values obtained in \cite{Israel1997a,Galliano2011a,Chevance2016b}. 

\section{Gas temperature determination}\label{secapp:oicii}

\begin{figure*}
  \includegraphics[width=9cm,height=8cm]{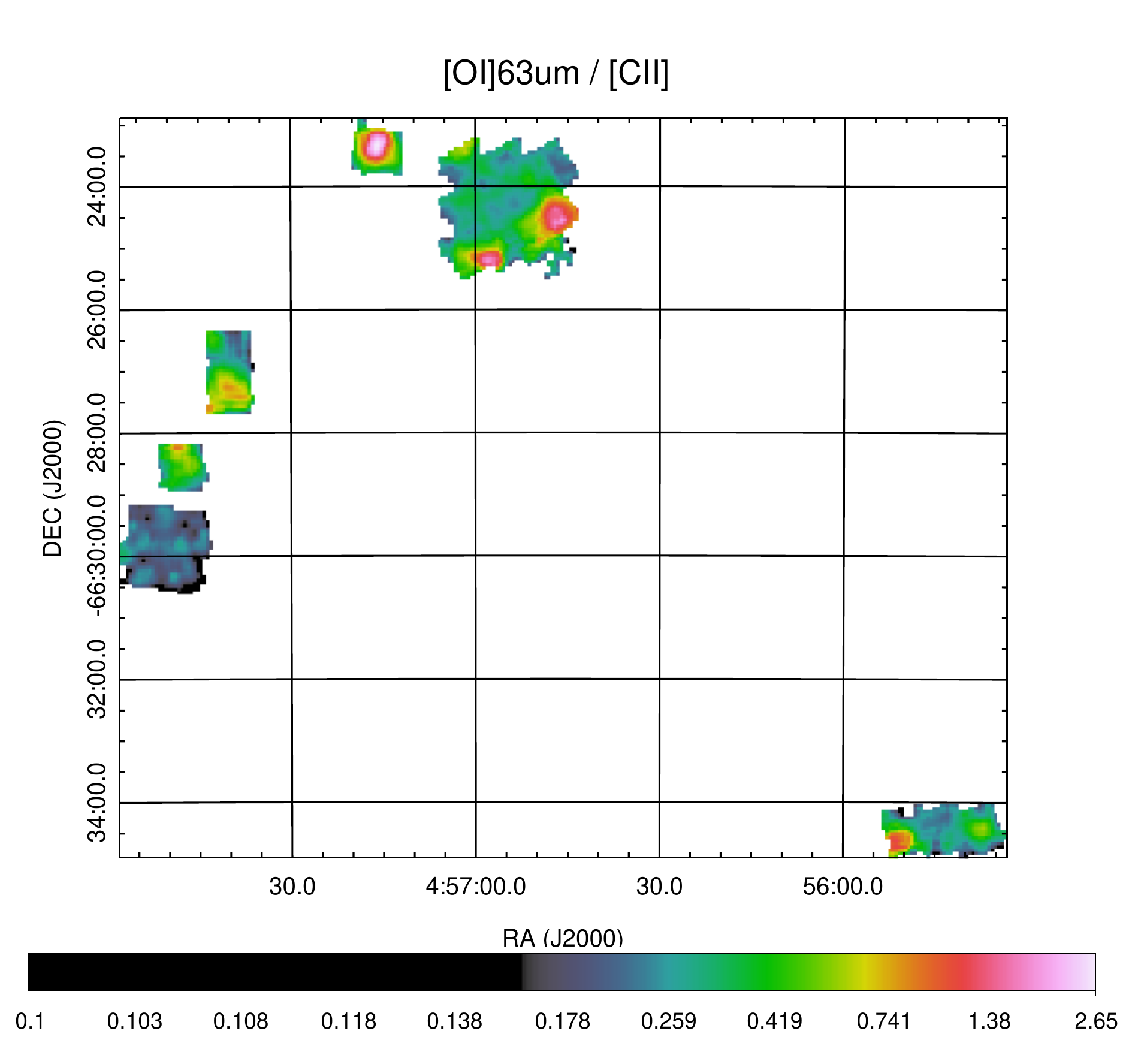}
\includegraphics[width=9cm,height=8cm]{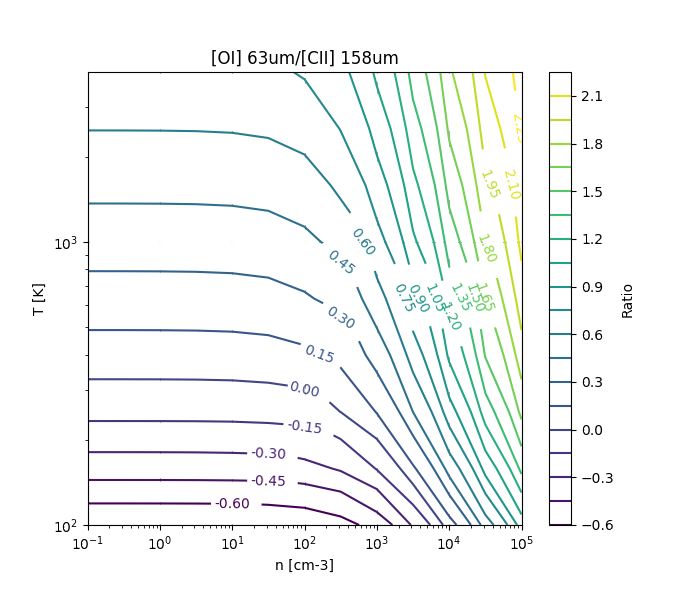}
\caption{\textit{Left} $-$ Map of [O\1] $63$\mic/[C\2] with PACS with values between $\approx-0.7$ and $\approx0.4$ in $\log$ unit. \textit{Right} $-$ Theoretical ratio [O\1]/[C\2] (in $\log$ units) as a function of the gas density and temperature for a purely atomic gas. We use the N\,11B chemical abundances from \cite{ToribioSanCipriano2017a}. }\label{fig:oicii}
\end{figure*}

The temperature in the neutral atomic gas can in principle be estimated from the [O\1] $63$\mic\//[C\2] ratio observed with \textit{Herschel}/PACS \citep{Lebouteiller2012b}. Apart from optical depth effects for the [O\1] $63$\mic\ line (see \citealt{Lebouteiller2012b}), another problem resides in the fact that [C\2], and presumably [O\1], mostly trace the molecular phase rather than the neutral atomic phase, as our present results suggest. For this reason we use the extended emission seen in the PACS maps and assume it to be dominated by atomic gas. For extended emission, the [O\1] $63$\mic/[C\2] ratio is in the range $\sim0.2-0.5$ (Fig.\,\ref{fig:oicii} \textit{left}), corresponding to a temperature of a few hundred Kelvin for densities $\lesssim10^3$\cc\ (Fig.\,\ref{fig:oicii} \textit{right}), while somewhat lower temperature values would be found if the medium were partly molecular. 

\end{document}